\documentclass{elsarticle}
\usepackage[title]{appendix}
\usepackage{soul}
\usepackage{amsmath}
\numberwithin{equation}{section}
\usepackage{geometry}
\usepackage{amsfonts}
\usepackage{bm}
\usepackage{adjustbox}
\usepackage{multirow}
\usepackage{color}
\usepackage{booktabs}
\usepackage{cellspace}
\usepackage{makecell}
\setcellgapes{3pt}
\usepackage{graphicx}
\usepackage{subfigure}
\usepackage[table ]{xcolor}
\usepackage[utf8]{inputenc}
\usepackage[colorlinks,linkcolor=blue]{hyperref}
\usepackage[font=small,labelfont={bf,sf},tableposition=top]{caption}
\usepackage{amsmath}
\numberwithin{equation}{section}
\newcommand{\tabincell}[2]{\begin{tabular}{@{}#1@{}}#2\end{tabular}}
\renewcommand\appendix{\setcounter{secnumdepth}{-2}}

\biboptions{numbers,sort&compress}

\begin{document}

\begin{frontmatter}

\title{Recommender systems based on graph embedding techniques: \\A comprehensive review}
\author[label1]{Yue Deng \corref{cor1}}
\cortext[cor1]{Corresponding author}
\ead{201921210214@std.uestc.edu.cn}
\address[label1]{Institute of Fundamental and Frontier Sciences, University of Electronic Science and Technology of China, \\Chengdu 611731, China}

\begin{abstract}
As a pivotal tool to alleviate the information overload problem, recommender systems aim to predict user's preferred items from millions of candidates by analyzing observed user-item relations. As for alleviating the sparsity and cold start problems encountered by recommender systems, researchers generally resort to employing side information or knowledge in recommendation as a strategy for uncovering hidden (indirect) user-item relations, aiming to enrich observed information (or data) for recommendation. However, in the face of the high complexity and large scale of side information and knowledge, this strategy largely relies for efficient implementation on the scalability of recommendation models. Not until after the prevalence of machine learning did graph embedding techniques be a recent concentration, which can efficiently utilize complex and large-scale data. In light of that, equipping recommender systems with graph embedding techniques has been widely studied these years, appearing to outperform conventional recommendation implemented directly based on graph topological analysis (or resolution). As the focus, this article systematically retrospects graph embedding-based recommendation from embedding techniques for bipartite graphs, general graphs and knowledge graphs, and proposes a general design pipeline of that. In addition, after comparing several representative graph embedding-based recommendation models with the most common-used conventional recommendation models on simulations, this article manifests that the conventional models can still overall outperform the graph embedding-based ones in predicting implicit user-item interactions, revealing the comparative weakness of graph embedding-based recommendation in these tasks. To foster future research, this article proposes constructive suggestions on making a trade-off between graph embedding-based recommendation and conventional recommendation in different tasks, and puts forward some open questions.
\end{abstract}


\end{frontmatter}

\tableofcontents

\clearpage

\section{Introduction}
\label{introduction}
Does big data \cite{cai2015challenges,fang2015survey} benefit people's lives? On its face, the question seems absurd. It is true that, for example, the traffic flow big data helps to quantify the potential infectious crowds during the pandemic \cite{zhao2020tracking}, the scientific research big data could facilitate academic-industry collaboration \cite{jain2014big}, or the multimedia social big data usually entertains consumers \cite{kumar2020multimedia}. But meanwhile, the high-volume, high-velocity and high-variety, also called the three “V” features \cite{laney20013d}, of big data bring problems. Information overload \cite{jacoby1984perspectives, bontcheva2013social} is a case in point, referring to the excess of big data available to a person when making a decision, say, which articles are relevant to a researcher's focus, which products meet a consumer's demand, or which movies pique an audience's interest. Consequently, it would discount one's information retrieval \cite{hersh2021information} efficiency. Counterbalancing these pros and cons of big data to maximize its benefits requires the development of big data mining techniques \cite{wu2013data}, among which recommender systems \cite{resnick1997recommender, ricci2011introduction, lu2012recommender} have turned out to be a pivotal tool to alleviate the information overload problem, aiming to predict a user's (\textit{e.g.}, researcher, consumer or audience) preferred items (\textit{e.g.}, articles, products or movies) from millions of candidates. Apart from this, recommender systems have seen commercial practices ranging from startup-investors matching \cite{xu2020recommending} to energy efficiency in buildings \cite{himeur2021survey}.

Developing recommender systems requires surmounting the sparsity problem \cite{singh2020scalability, idrissi2020systematic} and cold start problem \cite{lika2014facing, gope2017survey, gondaliya2019solutions, sethi2021cold} encountered by recommendation models, the core component of recommender systems. The rationale for recommendation models lies in the accurate inference for user's preferences for items, the prerequisite for well recommendation performance, by analyzing observed user-item relations, among which user-item interactions (Sec.~\ref{information} gives details) are primary resources. However, user-item interactions are usually sparse as a result of only a few of the total number of items that were interacted by a user, called the sparsity problem. When coming to a new user, that no interaction between the user and items yet has been observed leads to the cold start problem, and the same is true of a new item analogically. Given these problems, inadequate user-item interactions as input for recommendation models will lower the accuracy of inference for user's preferences and eventually weaken recommendation performance. As for tackling the sparsity and cold start problems, employing side information \cite{sun2019research, guo2019exploiting} or knowledge \cite{weikum2020machine, gao2020deep, guo2020survey} (Sec.~\ref{information} gives details) as a supplement to user-item interactions has been proven promising recently, aiming to uncover hidden (indirect) user-item relations to enrich observed information for recommendation models.

Concerning the ability to efficiently employ side information or knowledge to promote recommendation performance, the discussion about whether graph embedding-based recommendation (Sec.~\ref{graph_embedding} gives details) usually outperforms conventional recommendation implemented based on graph topological analysis (Sec.~\ref{conventional_models} gives details) is an ongoing controversy. With regard to the scalability \cite{sarwar2001sparsity, singh2020scalability}, graph embedding-based recommendation outperforms conventional recommendation, which can efficiently implement recommendation per second for millions of users and items when data is highly complex and large-scale as a result of the three “V” features of side information and knowledge inherited from big data. This result is brought from the two's distinctive rationales: after organizing data (or information) into graph representations (Sec.~\ref{graph_representation} gives details), conventional recommendation runs by analyzing a graph's topological characteristics such as users' co-interactions with common items \cite{sarwar2001item} or global topological diffusion \cite{zhang2007heat, zhang2007recommendation}. In contrast, graph embedding-based recommendation runs by using nodal embedding vectors, which preserve graph topological features once learned from the graph representations by embedding techniques \cite{goyal2018graph} (Secs.~\ref{Fundamental-Technologies} and \ref{LEMGE} give retrospects). In view of that, when employing side information or knowledge, graph embedding-based recommendation can directly reuse the learned nodal embedding vectors rather than repeating the analysis of graph topological characteristics as conventional recommendation does. Therefore, the scalability of it can be substantially improved. Besides, the storability of embedding vectors can support downstream machine learning tasks \cite{yang2021network}, which require feature vectors of data instances as inputs, like classification \cite{bhagat2011node, khattak2020tweets,verma2017hunt, narayanan2017graph2vec, galland2019invariant, chen2019gl2vec, wang2021hyperparameter}, link prediction \cite{liben2007link, lu2011link, rossi2021knowledge, kumar2020link} or clustering \cite{wang2020commerce}. Such a property of embedding vectors enables graph embedding-based recommendation to outperform conventional recommendation in terms of model extensibility.

Nevertheless, as for model explainability (or interpretability) \cite{zhang2018explainable}: why did models return such recommendations to a user, graph embedding-based recommendation substantially underperforms conventional recommendation as a result of its general adoption of machine learning methodology \cite{jordan2015machine}, almost a black box, whose idea lies on the input-output data fitting for underlying pattern discovery by numerical or analytic optimization methods \cite{sun2019survey}, whereas conventional recommendation can directly realize the explainability through resolving the graph topological characteristics pertaining to a user-item node pair. Although some recent studies argued that by employing knowledge in recommendation\cite{zhang2018explainable, gao2019explainable, palmonari2020knowledge, xie2021explainable} (Sec.~\ref{KGERS} gives details) or by causal learning (causal inference) \cite{gopnik2007causal, liang2016causal, bonner2018causal, yao2020survey, wang2020causal, xu2021causal, xu2021learning} to reason and understand user's preferences the explainability of recommendation results can also be indirectly realized, the explainability of recommendation models still faces fundamental limits. In addition, controversies over graph embedding-based recommendation and conventional recommendation are also embodied in recommendation accuracy. To be sure, by employing side information and knowledge, graph embedding-based recommendation can achieve distinctive improvement in recommendation accuracy beyond conventional recommendation \cite{duwairi2016enhanced,yu2020privacy,zhang2020doccit2vec,mao2021application}. However, this has been cast into doubt by its comparative weakness in some recommendation tasks for predicting implicit user-item interactions compared with conventional recommendation, proved in Sec.~\ref{evaluation} on simulations. Similar results were unraveled by Dacrema et al. \cite{dacrema2019we} too.

Faced with these ongoing controversies, the current lack of unified evaluation criterion on graph embedding-based recommendation and conventional recommendation will lead to longstanding discussions in the future, involving extended perspectives from accuracy, scalability, extensibility and explainability, and participated by interdisciplinary researchers ranging from mathematicians to data scientists. In fact, developing both graph embedding-based recommendation and conventional recommendation is not contradictory. The methods of analyzing graph topological characteristics behind conventional recommendation can inspire graph embedding-based recommendation in the utilization of such as subgraphs \cite{zhao2021heterogeneous}, motifs \cite{lim2016motif, peng2020motif, zhang2020motif}, and neighborhood \cite{lin2015learning, ji2016knowledge, grover2016node2vec} to promote its explainability \cite{wang2021hyperparameter}. On the other hand, graph embedding-based recommendation has pioneered novel recommendation scenarios, like conversational recommender system (CRS) \cite{Sun2018Conversational} or news recommendation \cite{okura2017embedding}, providing more promising application prospects for conventional recommendation. It seems that developing both of them to complement each other could improve recommender systems larger than only focusing on one side.

Unlike all-around review articles about conventional recommendation \cite{resnick1997recommender, ricci2011introduction, lu2012recommender}, newly published reviews on graph embedding-based recommendation \cite{sun2019research, zhu2020knowledge, dai2020survey, guo2020survey,wang2020graph, gao2020deep, wu2021survey, wang2021graph, choudhary2021survey} generally lack a systematic structure and an in-depth description, which seems to be insufficient for researchers focused on conventional recommendation before or interdisciplinary researchers to apprehend this novel prevalent field. To bridge this gap, this article builds an all-around perspective on recommender systems involving both graph embedding-based and conventional methods, and proposes a general design pipeline of graph embedding-based recommendation. As the focus, this article systematically retrospects graph embedding-based recommendation from embedding techniques for bipartite graphs, general graphs and knowledge graphs. In addition, this article further compares the strengths and weaknesses of representative graph embedding-based recommendation models with those of the most common-used conventional recommendation models, on simulations, in different recommendation tasks, revealing that the conventional recommendation models can outperform the graph embedding-based recommendation models in predicting implicit user-item interactions. By analyzing these experimental results, this article provides constructive suggestions on making a trade-off between graph embedding-based recommendation and conventional recommendation, and proposes some open questions for future research.

The rest of this article is organized as follows. Sec.~\ref{S2} covers basic definitions of subjects and problems, building an all-around perspective on recommender systems and proposing a general design pipeline of graph embedding-based recommendation. Secs.~\ref{BiGE}, \ref{GGE}, and \ref{KGBE} retrospect embedding techniques for bipartite graphs, general graphs and knowledge graphs, respectively, and then retrospect the graph embedding-based recommendation models based on them, correspondingly. Tabs.~\ref{a_comparison_of_graph_embedding_based_recommendation} and \ref{Framework_and_literature_technologies_and_models} provide an overview of the reviewed models. Sec.~\ref{evaluation} presents the experimental results on evaluating representative graph embedding-based recommendation models and the most common-used conventional recommendation models in different recommendation scenarios with distinctive data scales. After analyzing these experimental results, Sec.~\ref{evaluation} provides several constructive trade-off suggestions as well as open questions for future research. Finally, Sec.~\ref{outlook} puts forward some prospects on graph embedding-based recommendation, ranging from current challenges to potential solutions.

\section{Preliminaries}
\label{S2}

After introducing several major controversies between graph embedding-based recommendation and conventional recommendation in Sec.~\ref{introduction}, in this section Sec.~\ref{recommender_systems} devotes to an all-around perspective of recommender systems and illuminates the rationale behind conventional recommendation. Then, Sec.~\ref{graph_embedding_and_RS} illustrates what is graph embedding as well as what is the rationale for it to be applied in recommendation, preparing for Secs. \ref{BiGE}, \ref{GGE} and \ref{KGBE}. Following that, Sec.~\ref{pipeline} proposes a general design pipeline of graph embedding-based recommendation. Finally, the notations used in this article are presented in Tab.~\ref{Notation} at the end of this section.

\subsection{Recommender systems}
\label{recommender_systems}

In general, the target of recommendation is to infer user's preferences for items by analyzing user-item relations with observed information (or data) related to users and items, aiming to predict unobserved (or never happened) user-item interactions. This section divides the observed information into three categories: user-item interactions, side information and knowledge, according to their respective distinguishable complexity. Before being employed in recommendation, the three kinds of information should be represented by graph representations, including bipartite graphs, general graphs and knowledge graphs, correspondingly, which are the bedrock of measuring the k-order proximity between users and items in order to predict unobserved user-item interactions. To clarify the above process, Sec.~\ref{conventional_models} takes two common-used conventional recommendation models as instances for illustration.

\subsubsection{Information (or data) for recommendation}
\label{information}

In engineering, observed information in recommender systems can be recorded by tuples. For example, consider an event that \textit{a 24-year-old male student named Tom watched Iron Man on Netflix on January 28, 2021, and rated this movie with five points}, also called observed information. In engineering, it can be recorded by a tuple like \{Tom, male, 24, student, watched, Iron Man, 5, 2021-1-28, Netflix\}, in which user-item interactions (\textit{i.e.}, \{Tom, watched, Iron Man, 5, 2021-1-28\}), side information (\textit{i.e.}, \{Tom, male, 24, student\}) and knowledge (\textit{i.e.}, \{Iron Man, Netflix\}) can be further split out. Tab.~\ref{information_for_recommendation} briefly compares the three kinds of information.

\begin{table}[!ht]
\footnotesize
\setlength{\abovecaptionskip}{0.3cm}
\setlength{\belowcaptionskip}{0.3cm}
\caption{\textbf{A brief comparison between user-item interactions, side information and knowledge.} Taking side information or knowledge as a supplement to user-item interactions contributes to a higher recommendation accuracy \cite{duwairi2016enhanced,yu2020privacy,zhang2020doccit2vec,mao2021application}, where richer user-item relations are uncovered as the ground of preference inference, which helps to alleviate the sparsity and cold start problems.}
\label{information_for_recommendation}
\begin{adjustbox}{center}
\begin{tabular}{@{}cccc@{}}
\toprule
\textbf{Information}                                              & \textbf{Instances}                                                                                           & \textbf{Functions}                                                                                     & \textbf{Comparative complexity} \\ \midrule
user-item interactions                                                     & \begin{tabular}[c]{@{}c@{}}\textit{e.g.}, user's clicks, browses, \\ or ratings on items\end{tabular}                & \begin{tabular}[c]{@{}c@{}}can reflect user's preferences \\ for items\end{tabular}                & low                 \\
side information        & \begin{tabular}[c]{@{}c@{}}\textit{e.g.}, users' social relationships \\ and locations or item's profiles\end{tabular} & \begin{tabular}[c]{@{}c@{}}provides diverse properties \\ of users and items\end{tabular}                 & middle              \\
knowledge                                                                                & \textit{e.g.} encyclopedias of items                                                                                 & \begin{tabular}[c]{@{}c@{}}provides abundant direct or indirect  \\ relations between items \end{tabular} & high                \\ \bottomrule
\end{tabular}
\end{adjustbox}
\end{table}

As the primary resources for recommendation, \textbf{user-item interactions} can be divided into two categories: explicit ones and implicit ones, according to whether the interactions explicitly carry user's affection degree on items or not. Specifically, \textbf{explicit user-item interactions} can be defined as the ratings of items given by users, used to quantify user's affection degree on items based on the assumption that one tends to rate higher on items those he prefers than those he may show indifference. Under this definition, user's rating biases termed \textbf{user biases} \cite{koren2009matrix, adomavicius2014biasing, manjur2021exploring} can be dug out from explicit user-item interactions. For example, consider two users: when calculated on a five-point scale, one is used to rate by at least three points on items which he even showed indifference while the other is so extremely strict that never rated by more than three points on items which he even loved. Consequently, the rating biases between the two users obviously exist. In the same way, \textbf{item biases} \cite{koren2009matrix, park2014uncovering} could be brought from the user biases. For example, when it comes to one specific item, its average ratings given by tolerant users or critical users could be different. In this regard, in order to remove the distorted view of user's preferences or item's popularity, taking both the user biases and item biases into account can promote a high-quality recommendation \cite{koren2009matrix, adomavicius2014biasing}. However, despite their advantages in being able to directly reflect user's preferences for items, the limitation of explicit user-item interactions are clear for the two reasons: (1) In practice, since when surfing online one would prefer to browse, click or watch than rate, accessible explicit user-item interactions are usually sparse, which are insufficient to be as the input of models, not to mention an accurate recommendation. (2) As playing an enhanced role for user's privacy protection \cite{wang2018toward, huang2019privacy} in data security, explicit user-item interactions even could be inaccessible in some recommendation scenarios. To tackle these issues, \textbf{implicit user-item interactions} have become additional resources, which are defined as a binary state using value $1$ to indicate the existence of a user-item interaction (such as click, browse or watch) and value $0$ otherwise. Compared to the explicit ones, implicit user-item interactions don't require user's extra operations like ratings or comments on items; so they generally occur more frequently and can be more easily accessed. However, a minor criticism of implicit user-item interactions is that they can't directly carry user's preferences for items, bringing the so-called one-class problem \cite{liu2020cofigan}, which could resort to converting implicit user-item interactions to explicit ones \cite{hu2008collaborative, pan2008one} as a strategy.

In general, user-item interactions are occurring constantly, not merely at a specific time or in a constrained period. Correspondingly, inferred from newly occurring interactions, user's preferences for new items could sprout and those for old items could fade, which means that user's preferences for items could vary over time, usually in long-term and short-term termed as user's \textbf{long-term preferences} and \textbf{short-term preferences}, respectively. Specifically, user's long-term preferences could vary over time in ways large and small, like the changes of user's personal hobbies (one may prefer comedy movies in his childhood while finding science fiction movies more interesting after entering college), special events like seasonal changes and holidays or even the changes of one's family status. At the same time, user's short-term preferences could be affected by one's latest interacted items. For example, one's interest in comedy movies may decline after watching lots of comedy movies within a short period. In view of that, capturing both user's long-term and short-term preferences for items contributes to a higher recommendation accuracy \cite{xiang2010temporal, yu2015combining, wu2019long}. For that purpose, by utilizing deep learning methods \cite{fang2020deep} to mine the underlying patterns involved in user behavior, sequential recommendation \cite{xu2019survey} has been the recent focus of research into employing user's short-term preferences for items in recommendation. Others include the methods of matrix factorization and Markov processes (Sec.~\ref{with:dynamics} gives details). Since not all the reviewed recommendation models in this article take the changes of user's preferences into account, this article defines terms \textbf{temporal user-item interactions} and \textbf{static user-item interactions} to distinguish recommendation models considering that changes or not, respectively.

To alleviate the cold start and sparsity problems of recommendation, \textbf{side information} \cite{sun2019research, guo2019exploiting}, which is generally characterized by the properties of users and items, is utilized to uncover more hidden (indirect) user-item relations. Back to the first example of this section, the side information \{Tom, male, 24, student\} records Tom's properties including his gender, age and occupation. In light of that, it could be uncovered that some indirect relations between \textit{Iron Man} and audiences who are close to Tom in these personal properties might exist. In practice, side information usually refers to user's social information \cite{chen2016incorporate} and locations \cite{xie2016learning} or item's profiles \cite{liao2018attributed}, labels \cite{li2017attributed, huang2017label} and textual content \cite{wang2017ice}, to name a few. Among them, as for user's social information, microblogging \cite{mirkovski2018understanding} is a primary resource, which uncovers user's interpersonal relationships (like following or friends) by tweets on social platforms or user's individual profiles. Side information of microblogging has two virtues dear to researchers: abundant and almost real-time \cite{zhao2014we}. In detail, the abundant information can carry multitudes of diverse relationships between users as well as user's preferences for things that were directly expressed. Such abundance is far beyond that of user-item interactions, which definitely contributes to inferring user's references for items more accurately by, for example, taking one's friends' preferences as a reference. Besides, side information of microblogging is usually considered real-time, benefits from user's tweeting habits that one is inclined to share his daily events or feelings in microblogging or other social platforms. In this way, it provides more opportunities to trace user's latest preferences, which user-item interactions couldn't be that real-time because user's latest preferences can be reflected by these interactions only when new ones occur.

As the most complex one, \textbf{knowledge} \cite{weikum2020machine, gao2020deep, guo2020survey} can be displayed or expressed in a language form logically organized by subjects, predicates and objects, related to objective facts of the world. Among the three elements, subjects and objects are usually termed as \textbf{entities}, abstract or concrete things of fiction or reality with specific types or attributes. The connections (\textit{i.e.}, predicates) between entities are usually termed as \textbf{relations}. With these terms, knowledge can be defined as a collection of entities with different types or attributes and the relations between them. Take the first example of this section one step further, suppose some knowledge about the movie \textit{Iron Man} was excerpted from Wikipedia, displaying that \textit{Iron Man is a 2008 American superhero film based on the Marvel Comics character of the same name. Produced by Marvel Studios and distributed by Paramount Pictures, it is the first film in the Marvel Cinematic Universe (MCU).} Based on that, entities can be split out, like from the fact that \textit{the movie Iron Man belonging to the class of American superhero film the movie} the movie \textit{Iron Man} can be split out as an entity with an attribute of \textit{the first film in the MCU}. Meanwhile, the relation \textit{produced} between the entities \textit{Marvel Studios} and \textit{Iron Man} as well as the relation \textit{distributed} between the entities \textit{Paramount Pictures} and \textit{Iron Man} also can be split out. By employing this knowledge represented by entities and relations in recommendation, hidden  (or indirect) relations between Tom and other movies could be uncovered, like that Tom might be inclined to the productions by \textit{Paramount Pictures}, which reflect Tom's underlying preferences for other movies. In practice, the category information on e-commerce platforms, which is organized in a form by tree logic, is also a common-used resource of knowledge, providing a better understanding of user's preferences for items on multiple levels.

To be sure, when speaking of knowledge, one is tempted to argue that, throughout much of recent research, common scientific sense seemed to dictate that knowledge is supposed to be a subcategory of side information, commonly being as a supplement to user-item interactions. Despite general acknowledgement of that, the accepted classification belied the essential distinctions between side information and knowledge by making specious continuity of them. In the first place, the resources of side information and knowledge are different. Side information is usually requested from one's personal information forwardly. For example, when starting to use an APP installed on a cell phone, one might be requested to share his personal information with the APP, at best his name or gender, and at worst his address book or geographical location, which are private. Besides, as for side information of item's profiles, user descriptions of items about such as one's usage experiences are also requested from consumers forwardly. In sharp contrast, since always existed in the real world, knowledge can be naturally perceived and be displayed or expressed in various forms like encyclopedias or papers. Second, the complexity of side information and knowledge is distinctive, for most of side information is only about the properties of users or items while knowledge is about things more versatile, almost everything in the real world like multi-modal information \cite{sun2020multi}. In addition, knowledge generally grows constantly and rapidly, making itself far more complex in semantics and multiplicity compared to side information. Third, knowledge is re-usable while side information seldom does the same because the properties of users and items usually change over time. All in all, the three distinctions between side information and knowledge make the techniques employing knowledge in recommendation more challenging than those for side information as a result of the large-scale, multiplicity and evolution characteristics of knowledge beyond side information (Sec.~\ref{LEMGE} gives details).

\subsubsection{Graph representations}
\label{graph_representation}

In the last section, observed information for recommendation is divided into three categories: user-item interactions, side information and knowledge, according to their respective distinguishable complexity. Before being employed in recommendation, the three kinds of information should be represented by graph representations including bipartite graphs, general graphs and knowledge graphs, correspondingly, which are model-readable forms. Tab.~\ref{graph_structures_for_recommendation} briefly compares the three kinds of graph representations. Without loss of generality, this article uses $\mathcal{G}=(E,R,\mathcal{E},\mathcal{R})$ to denote a graph representation, where $E$ denotes a node-set and $R$ denotes an edge-set. $\mathcal{E}$ denotes a type-set of different nodes and $\mathcal{R}$ denotes a type-set of different edges.

\begin{table}[!ht]
\footnotesize
\setlength{\abovecaptionskip}{0.3cm}
\setlength{\belowcaptionskip}{0.3cm}
\caption{\textbf{A brief comparison between bipartite graph, general graph and knowledge graph.} $|E|$ and $|R|$ represent the number of different node types and different edge types, respectively.}
\label{graph_structures_for_recommendation}
\begin{adjustbox}{center}
\begin{tabular}{@{}ccccc@{}}
\toprule
\textbf{Graph representation} & \textbf{Carried information} & \textbf{$|\mathcal{E}|$} & \textbf{$|\mathcal{R}|$} & \textbf{Relative complexity} \\ \midrule
bipartite graph           & user-item interactions       & $=2$           & $=1$            & low                 \\
general graph             & side information             & $\geq1$        & $\geq1$         & middle              \\
knowledge graph           & knowledge                    & $\to \infty$   & $\to \infty$    & high                \\ \bottomrule
\end{tabular}
\end{adjustbox}
\end{table}

\begin{figure}[!ht]
\centering
\includegraphics[scale = 0.9]{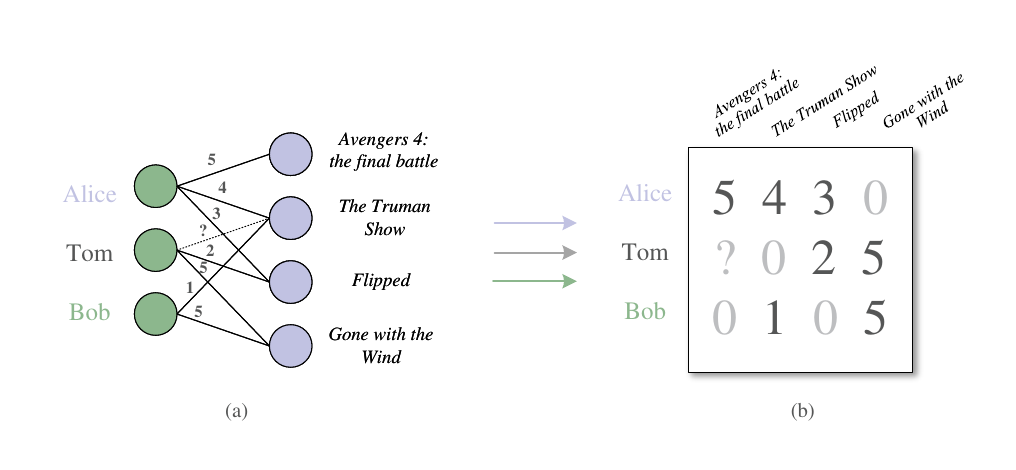}
\caption{\textbf{Represent user-item interactions by a bipartite graph and a matrix, respectively.} In (a), the weighted bipartite graph represents the explicit user-item interactions from a movie recommender system, where, for example, the edge weighted $5$ between Alice and \emph{Avengers 4: the final battle} represents an explicit interaction \{Alice, Avengers 4: the final battle, 5\}, meaning that Alice watched \emph{Avengers 4} and rated it five points. The bipartite graph in (a) can be directly converted into a weight matrix as shown in (b), where the value $0$ corresponds to the unobserved edges in (a), indicating the unobserved interactions between users and movies.}
\label{BG}
\end{figure}

As the most common-used graph representation for user-item interactions, a \textbf{bipartite graph} is formally defined as $\mathcal{G}_{bi}=(E,R,\mathcal{E},\mathcal{R})$ containing two types of nodes and one type of edges (\textit{i.e.}, $|\mathcal{E}|=2$ and $|\mathcal{R}|=1$), where edges only exist between nodes with different types. In practice, a bipartite graph can represent users and items in recommender systems with the two types of nodes. Based on that, it can represent implicit user-item interactions by adding edges between the corresponding node pairs and represent explicit user-item interactions by weighting the corresponding edges with ratings. Meanwhile, its use soon widened to algebra. By indexing user nodes and item nodes with rows and columns of a matrix, respectively, a bipartite graph can be directly converted into a matrix, where its elements indicate the existence (\textit{i.e.}, $0/1$) or weight (\textit{i.e.}, ratings) of edges in the bipartite graph, which enables the implementation of algebraic theory in recommendation (Secs.~\ref{sec:SVD} and \ref{MF-based_online_learning} give details). Fig.~\ref{BG} gives toy examples to illuminate how to represent explicit user-item interactions by a bipartite graph and meanwhile to convert it into a matrix.

A \textbf{homogeneous graph} \cite{taylor2007evolution} $\mathcal{G}_{\operatorname{homo}}=(E,R,\mathcal{E},\mathcal{R})$ where $|\mathcal{E}|=1$ and $|\mathcal{R}|=1$ or a \textbf{heterogeneous graph} \cite{shi2016survey,shi2017heterogeneous} $\mathcal{G}_{\operatorname{hete}}=(E,R,\mathcal{E},\mathcal{R})$ where $|\mathcal{E}|>1$ and/or $|\mathcal{R}|>1$ can be used to represent side information. Without loss of generality, this article terms the \textbf{general graph} to unify both of them. Under this definition, one could consider bipartite graphs as a subcategory of general graphs. However, a bipartite graph would prefer to represent side information but are constraind from doing so by that its edges can only exist between nodes with different types. Since side information could be both homogeneous and heterogeneous, edges should be allowed to exist between nodes of the same type. In contrast, a general graph is more flexible in representing side information, like representing user's social information containing one node type (\textit{i.e.}, user) and one edge type (\textit{i.e.}, friend relationship) by a homogeneous graph, or representing more enriched user's social information with attributes of users and items by a heterogeneous graph. To employ side information in recommendation, it's a common-used way to integrate a general graph with a bipartite graph through their jointly owned nodes as connections (because side information is generally about users and items, which are involved in user-item interactions).

\begin{figure}[!ht]
\centering
\includegraphics[scale=0.9]{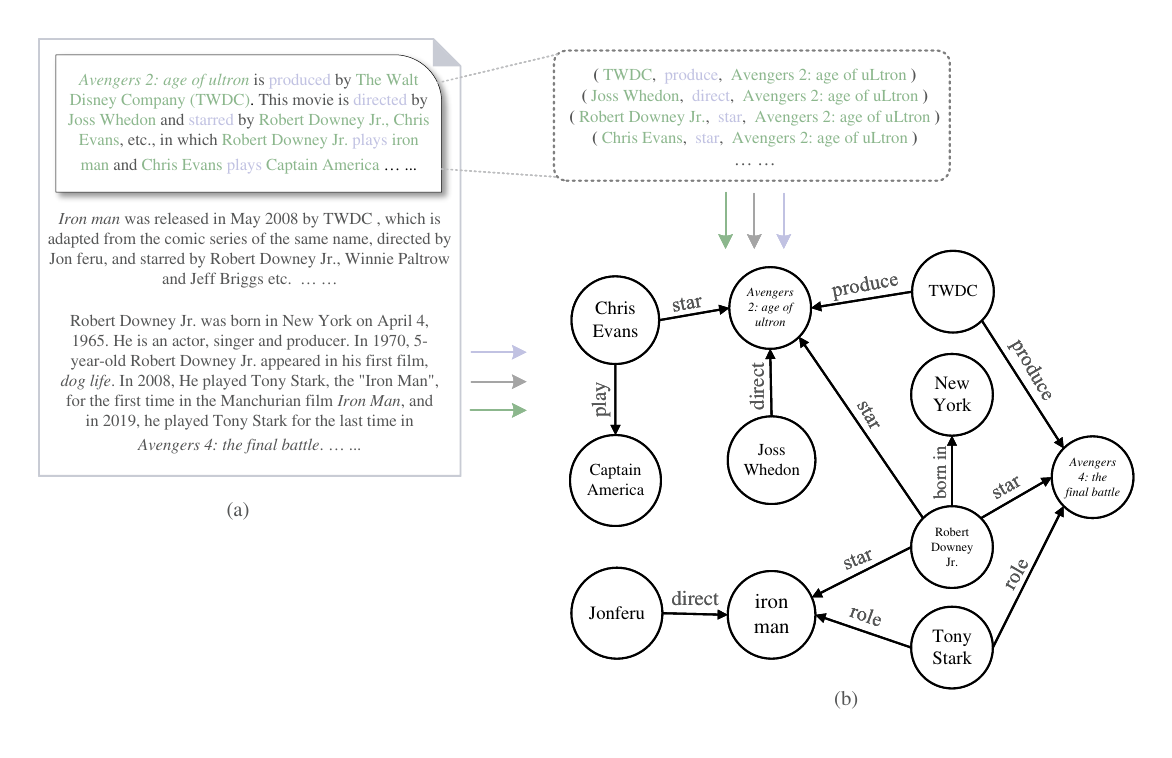}
\caption{\textbf{Represent knowledge by a knowledge graph.} In (a), the subjects, predicates and objects contained in the profile of a classic movie \textit{Avengers 2: age of ultron} can be split out and can be represented by multiple triples organized as (subject, predicate, object). Then, these triplets are used to construct a knowledge graph shown in (b).}
\label{KG}
\end{figure}

By using triplets (subject, predicate, object) to transform knowledge into a model-readable form, a \textbf{knowledge graph} \cite{weikum2020machine} can be used to represent the entities and their relations split out from knowledge. Formally, let $E$ denote the set of subjects and objects and let $R$ denote the set of predicates, knowledge can be represented by triplets $S=(h, r, t)$, where $h\in E$ is termed \textbf{head entity} (\textit{i.e.}, subject), $t \in E$ is termed \textbf{tail entity} (\textit{i.e.}, object), and $r \in R$ is a directed edge from $h$ to $t$. Based on these triplets, a knowledge graph can be constructed. Fig.~\ref{KG} gives toy examples to illuminate the construction process. Since the properties like boundless scale and complex semantics of knowledge, a knowledge graph $\mathcal{G}_{kg}=(E,R,\mathcal{E},\mathcal{R})$ generally contains extremely diverse node types and edge types such that $|\mathcal{E}| \to \infty$ and $|\mathcal{R}| \to \infty$, which can be considered as the most complex instance of general graphs. In order to sufficiently and efficiently represent these properties of knowledge graphs, research into \textbf{multi-viewed graphs} and \textbf{multi-layered graphs} has been a tendency recently (Sec.~\ref{LEMGE} gives details). In practice for commercial applications such as those in IBM \cite{ferrucci2012introduction}, Amazon \cite{dong2019building} or Alibaba \cite{luo2020alicoco}, representative knowledge graphs include YAGO \cite{suchanek2007yago}, WikiTaxonomy \cite{ponzetto2007deriving},  DBpedia \cite{auer2007dbpedia}, Wikidata \cite{vrandevcic2014wikidata} and WebOfConcepts \cite{dalvi2009web}, to name a few, constructed by automatic knowledge harvesting techniques \cite{suchanek2013knowledge, weikum2016ten, weikum2019knowledge}. These open knowledge graphs provide resources for recommendation. Similar in being employed in recommendation to side information dose, a knowledge graph can be integrated with a bipartite graph through their jointly owned nodes as connections. In the same way, a knowledge graph also can be integrated with a general graph.

\subsubsection{K-order proximity}
\label{similarity_and_proximity}

After representing observed information for recommendation by graph representations, it comes to measure the \textbf{proximity} \cite{yang2017fast, yang2018hop} between users and items (\textit{i.e.}, between the corresponding user nodes and item nodes in graph representations), which is a key step of recommendation's implementation, from the perspective of link (\textit{i.e.}, edge) prediction \cite{lu2011link, rossi2021knowledge}. Specifically, user-item proximity is used to record or predict the similarity between user-item node pairs both linked by edges or not, which can be used to quantify the likelihood of occurrence of unobserved user-item interactions in recommender systems. Intuitively, the higher similarity between a not-linked user-item node pair, the higher user's affection degree on the item, and then a higher likelihood of occurrence of the corresponding unobserved user-item interaction.

In general, the \textbf{first-order proximity} of a node pair $v_i-v_j$ is defined as their local pairwise proximity, which can be measured by the existence (\textit{i.e.}, $0/1$) or weight (\textit{i.e.}, rating if exists) of edge $(v_i,v_j)$. Take Fig.~\ref{BG} as an example, the first-order proximity of node pair Tom-\emph{Flipped} can be measured as $2$ by the edge's weight (\textit{i.e.}, rating), recording the similarity between them. Since the edge between Tom and \emph{Avengers 4: the final battle} did not exist, the similarity between Tom-\emph{Avengers 4: the final battle} need to be predicted based on the observed user-item proximity by methods (\textit{i.e.}, recommendation models). However, edges in a graph representation are usually in a small proportion as a result of the sparsity problem of recommendation, which casts the first-order proximity into doubt about its precision on recording or predicting user-item similarity. In fact, the proximity between two not-linked nodes does not always have to be zero measured by the first-order proximity, like in situations where the corresponding two users could still be intrinsically related. For example, the first-order proximity of Tom-Bob in Fig.~\ref{BG} is measured as zero while the two users might share common movie preferences, which can be intuitively inferred from the fact that they co-rated the movie \emph{Gone with the Wind} with the same points of five, and thus be intrinsically related. To make up for the flaw of the first-order proximity, the \textbf{second-order proximity} of a node pair $v_i-v_j$ is defined as the overall first-order proximity between the two nodes' respective neighborhoods. Formally, denote $S_i=\{s^{(1)}_{i,1}, s^{(1)}_{i,2},...,s^{(1)}_{i,|V|}\}$ as a collection of the first-order proximity $s^{(1)}_{i,j}$ between node $v_i$ and the other nodes $v_j$ in a graph representation, respectively. Following that, the second-order proximity between $v_i$ and $v_j$ is measured based on $S_i$ and $S_j$ by such as cosine index \cite{leydesdorff2008normalization}, Pearson coefficient (PC) \cite{lee1988thirteen} or Jaccard index \cite{hamers1989similarity}. Apparently, if two nodes $v_i$ and $v_j$ do not have any other co-lined node, the second-order proximity of $v_i-v_j$ is measured as zero, such as that of \emph{Avengers 4: the final battle}-\emph{Gone with the Wind} shown in Fig.~\ref{BG}. Iteratively, the \textbf{higher-order proximity} of a node pair can be analogously defined as above, which has been a popular research focus in recent years (Sec.~\ref{Meta-path-section} gives details).

Under these definitions of proximity with different orders, it is more obvious why is it feasible to alleviate the cold start and sparsity problems of recommendation by employing side information or knowledge. Take Fig.~\ref{BG} for example, when coming to a new movie \emph{Avengers 2: age of ultron} represented by a node, it is isolated in the bipartite. Consequently, it is impossible to measure the proximity between the new movie and any other user or movie, let alone to predict the similarity between them and estimate user's preferences on this new movie, which leads to the cold start problem as illustrated beforehand. However, after employing side information or knowledge in recommendation, it is another matter. For example, by integrating the knowledge graph shown in Fig.~\ref{KG} with the bipartite graph, more hidden (or indirect) relations between the new movie \emph{Avengers 2: age of ultron} and other movies can be uncovered, like that with \emph{Avengers 4: the final battle}. These new relations enable the implementation of measuring user-item proximity related to the new movie. Benefits are not limited to this, enriched relations uncovered by side information or knowledge can also make the original graph denser, with newly established edges between nodes. In this way, not only the sparse problem of recommendation could be alleviated but also user-item proximity could be put in more diverse orders, ensuring the precision of measured user-item similarity.

In different recommendation methods, the value of user-item proximity is generally entitled different practical meanings, like the ratings of items given by users by matrix factorization-based method (Sec.~\ref{sec:SVD} gives details), the Pearson coefficient (PC) \cite{lee1988thirteen} between nodes by k-nearest neighbors-based method (Sec.~\ref{conventional_models} gives details), the allocated resources on items diffused from users by diffusion-based method (Sec.~\ref{conventional_models} gives details) or the occurrence probability of user-item interactions by deep learning-based method (Sec.~\ref{DL-Bi} gives details). On the other hand, the value of user-item proximity could be meaningless, such as that in translation-based method (Sec.~\ref{Trans-section} gives details). Without loss of generality, this article generalizes the definition of proximity to be a metric, which can be used to quantify the relative magnitude of similarity between nodes in graph representations. Note that the term proximity also can be called similarity.

\subsubsection{Methods and conventional models}
\label{conventional_models}

As illustrated above, in order to quantify the likelihood of occurrence of unobserved user-item interactions, recommendation models aim to predict the similarity between unobserved user-item pairs based on observed user-item proximity, among which \textbf{collaborative filtering} \cite{sarwar2001item, herlocker2000explaining, linden2003amazon, su2009survey, koren2015advances}, \textbf{diffusion-based} \cite{zhang2007recommendation, yu2016network, wang2017diffusion, xu2020recommending} and \textbf{content-based} \cite{pazzani2007content, van2013deep, lops2019trends} are three prevalent ones. Respectively, based on the assumption that a user's preferences for items might be affected by those of his neighbors, collaborative filtering method predict the similarity between an unobserved user-item pair by analyzing the observed proximity between the item and the user's neighbors sharing interacted items with the user. Diffusion-based method pioneered a strategy for applying physic diffusion processes, such as heat spreading \cite{zhang2007heat} or mass diffusion \cite{zhang2007recommendation}, to recommendation. Content-based method runs by building user's profiles, which are used to match with item's attributes or descriptions. As for conventional recommendation models, user-based collaborative filtering (UBCF) \cite{sarwar2001item} and probabilistic spreading (ProbS) \cite{zhang2007recommendation} are two common-used ones of collaborative filtering method and diffusion-based method, respectively. The rest of this section illustrates the two models, used as instances to illuminate the rationale behind conventional recommendation implemented by directly analyzing graph topological characteristics. At the same time, the two models are used as experimental benchmarks in Sec.~\ref{evaluation}.

\begin{figure}[h]
\centering
\includegraphics[scale = 1.0]{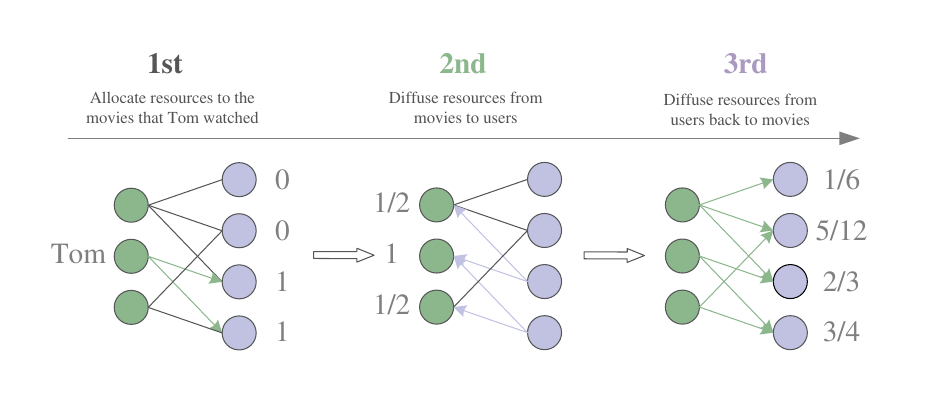}
\caption{\textbf{Schematics of ProbS.} The first step is to allocate resources. The second and third steps combine a two-step process of resource diffusion.}
\label{RS-HC}
\end{figure}

Based on the bipartite graph shown in Fig.~\ref{BG}, to predict Tom's similarity with the two movies not interacted with him, UBCF measures the second-order proximity between Tom and other users by Pearson coefficient (PC) \cite{lee1988thirteen} as
\begin{equation}
\displaystyle S_{i_1i_2}^{\operatorname{PC}} = \frac{\sum_{j \in \mathcal R(i_1)^{+} \cap \mathcal R(i_2)^{+} }(r_{i_1j}-\overline{r}_{i_1})(r_{i_2j}-\overline{r}_{i_2})}{\sqrt{\sum_{j \in \mathcal R(i_1)^{+} \cap \mathcal R(i_2)^{+}}(r_{i_1j}-\overline{r}_{i_1})^2} \sqrt{\sum_{j \in \mathcal R(i_1)^{+} \cap \mathcal R(i_2)^{+}}(r_{i_2j}-\overline{r}_{i_2})^2}},
\label{PC}
\end{equation}
where $\displaystyle \mathcal R(i_1)^{+} \cap \mathcal R(i_2)^{+}$ denotes the set of items rated by both users $i_1$ and $i_2$, and $\overline{r}_i$ denotes the average of ratings given by user $i$. Following that, the similarity between Tom and \textit{The Truman Show}, which represents the predicted rating $\hat{r}_{*j}$ of \textit{The Truman Show} (denoted by $j$) given by Tom (denoted by $*$), can be calculated by
\begin{equation}
\label{predicting_rating_by_UBCF}
\hat{r}_{*j}= \overline{r}_* + \alpha \sum_{i \in N(*)}S_{*i}(r_{ij}-\overline{r}_i),
\end{equation}
where $N(*)$ is a collection of neighbors within the second-order hops from Tom, and $\displaystyle \alpha=\frac{1}{\sum_{i \in N(*)}|S_{*i}|}$ is a normalization factor. For example, in Fig.~\ref{BG}, based on the observed proximity that Alice and Tom both rated \textit{Flipped} as well as Bob and Tom both rated \textit{Gone with the Wind}, and calculated by Pearson coefficient in Eq.~(\ref{PC}), the second-order proximity between Tom and Alice can be measured as $-0.9795$ and that between Tom and Bob can be measured as $0.8593$. Following that, based on the observed proximity that Alice rated \textit{The Truman Show} by $4$ points and Bob rated \textit{The Truman Show} by $1$ point, the similarity $\hat{r}_{*j}$ (\textit{i.e.}, predicted rating) can be calculated by $\displaystyle 3.5 + \frac{1}{1.8388}[-0.9795\times (4-4)+0.8593\times (1-3)]\approx 2.5654$, according to Eq.~(\ref{predicting_rating_by_UBCF}).

Different from UBCF resolving topological characteristics of users' co-interactions with common items, ProbS runs a mass diffusion process on a graph, using dynamically diffused and aggregated resources to represent the similarity between nodes. Take Tom in Fig.~\ref{RS-HC} as the user waiting to be recommended, in the first step, ProbS allocates a unit of resources for all the movies that interacted with Tom, respectively. Then, in the second step, the resources allocated to movies are equally distributed and diffused along edges from each movie to its interacted users, where the aggregated resources reached to Alice and Bob can be used to represent their respective second-order proximity with Tom. In the third step, these resources allocated to users are again equally distributed and diffused along edges from each user back to his interacted movies. Finally, the aggregated resources reached to movies can be used to represent their similarity with Tom, which indicates that Tom's preference for \textit{The Truman Show} could be beyond that for \textit{Avengers 4: the final battle}. Note that the second and third steps of ProbS combine a two-step diffusion process, which can be iterated to multiple rounds for a higher recommendation accuracy.

\subsection{Graph embedding}
\label{graph_embedding_and_RS}

In the last section, the rationale for conventional recommendation is illuminated by two models as instances, which are implemented by resolving graph's topological characteristics. However, this rationale prohibits conventional recommendation from efficiently implementing on different recommendation scenarios, especially with big data, because it has to repeat the resolution on graph representations mechanically, which lags behind graph embedding-based recommendation in scalability and migration. In contrast, graph embedding-based recommendation runs by directly reusing nodal embedding vectors, which represent users and items, once learned from graph representations. On top of that, after incorporating machine learning methodology, graph embedding-based recommendation can be equipped with abilities to pattern discovery, which contributes to a higher recommendation accuracy. This section presents basic concepts of graph embedding as well as its application in recommendation.

\subsubsection{Definitions and concepts}
\label{graph_embedding}

\textbf{Graph embedding} \cite{goyal2018graph, cai2018comprehensive, chen2020graph, barros2021survey} is a technique used to generate features of non-Euclidean data for machine learning-based downstream tasks, like node classification \cite{bhagat2011node, khattak2020tweets}, graph classification \cite{verma2017hunt, narayanan2017graph2vec, galland2019invariant, chen2019gl2vec, wang2021hyperparameter}, link prediction \cite{liben2007link, lu2011link, rossi2021knowledge, kumar2020link}, clustering \cite{wang2020commerce} and stuff. In general, in order to satisfy the input format of machine learning models \cite{shalev2014understanding}, features (usually represented by multidimensional vectors) of objects in data should be generated in the first place. For that purpose, until recently, researchers usually resorted to artificial generation methods \cite{cheng2016wide} implemented based on hand-engineering with expert knowledge and bag-of-words methods \cite{zhang2010understanding} as strategies for generating features of Euclidean data. However, as illustrated in Secs.~\ref{information} and \ref{graph_representation} that most of the information (or data) for recommendation is represented by graph representations with complex and diverse hidden relation (or connectivity) patterns, characterized by non-Euclidean. For that gap, in recent research, graph embedding techniques began, for the first time in large numbers, to be used to generate features of non-Euclidean data, which can project non-Euclidean graph representations into a low-dimensional Euclidean space consisting of embedding vectors (also called \textbf{embeddings}) of nodes, edges, subgraphs or whole graphs as their features. These embeddings preserve the intrinsic topological characteristics of graph representations, which can be used to reconstruct them.

Formally, in terms of generating features of nodes, graph embedding is defined as a mathematical process using a mapping $\Phi: \mathcal{G} \to \mathbb{R}^{n \times k}$ to project a graph representation $\mathcal{G}$ into a Euclidean space $\mathbb{R}^{n \times k}$, where $n$ is the size of $\mathcal{G}$ (\textit{i.e.}, $\mathcal{G}$ contains $n$ nodes) and $k$ ($k \ll n$) is the Euclidean space's dimension. Through the mapping, the embedding (a $k$-dimensional vector) of an arbitrary node $v_i$ in $\mathcal{G}$ can be generated as $\Phi(v_i) \in \mathbb{R}^k$. Based on these embeddings, the proximity between two nodes $v_i$ and $v_j$ can be measured as $\mathcal{F}(\Phi(v_i), \Phi(v_j))$, where $\mathcal{F}:(\cdot , \cdot)\to \mathbb{R}$ (like dot-product \cite{ogita2005accurate}) is a mapping that can project two embeddings to $\mathbb{R}$, where a greater value of $\mathcal{F}(\Phi(v_i), \Phi(v_j))$ is generally recognized as a higher possibility of the existence of edge $(v_i, v_j)$. Following that, $\mathcal{G}$ can be approximately reconstructed by adding these possible edges between node pairs. Intuitively, a well-performed graph embedding technique can determine a mapping $\Phi: \mathcal{G} \to \mathbb{R}^{n \times k}$ related to $\mathcal{F}:(\cdot , \cdot)\to \mathbb{R}$, whether it is directly set up or learned by machine learning methods, which is supposed to be used to approximately reconstruct $\mathcal{G}$'s topological characteristics as much as possible.

When it comes to machine learning-based graph embedding techniques, which are the most prevalent ones in recent years \cite{shalev2014understanding}, the process of determining the mapping $\Phi: \mathcal{G} \to \mathbb{R}^{n \times k}$ runs by optimization algorithms (Sec.~\ref{optimization_algorithm} gives details). Specifically, given a graph representation $\mathcal{G}$ with a node-set $V$, constructing \textbf{training samples} and \textbf{test samples} is the first step. With a set of node pairs $S^{\operatorname{train}}=\displaystyle \{(v_i^{\operatorname{train}},v_j^{\operatorname{train}})|v_i^{\operatorname{train}},v_j^{\operatorname{train}} \in V\}$ randomly sampled from $\mathcal{G}$ by a specified proportion and a set of their observed proximity $P^{\operatorname{train}}=\{p^{\operatorname{train}}_{ij} | p^{\operatorname{train}}_{ij} \in \mathbb{R}\}$, correspondingly, training samples can be constructed as  $(S^{\operatorname{train}}, P^{\operatorname{test}})$, and in the same way, so do test samples $(S^{\operatorname{test}}, P^{\operatorname{test}})$ that satisfy $S^{\operatorname{train}} \cap S^{\operatorname{test}} = \emptyset$. Then, in the second step, from a beforehand defined hypothesis space $\Phi$ \cite{shalev2014understanding} containing all possible mappings, machine learning-based graph embedding techniques aim to learn a candidate mapping $\Phi^{\operatorname{local}} \in \Phi$ that could minimize the average error between predicted proximity $\mathcal{F}(\Phi^{\operatorname{local}}(v_i^{\operatorname{train}}), \Phi^{\operatorname{local}}(v_j^{\operatorname{train}}))$ and observed proximity $p^{\operatorname{train}}_{ij}$ on training samples. Next, in order to assess their performance, the average error between $\mathcal{F}(\Phi^{\operatorname{local}}(v_i^{\operatorname{test}}), \Phi^{\operatorname{local}}(v_j^{\operatorname{test}}))$ and $p^{\operatorname{test}}_{ij}$ on test samples is calculated. Until the error is lower than a target precision, the learning process on training samples will iterate run by optimization algorithms, searching for the optimal $\Phi^{\operatorname{global}} \in \Phi$ that can minimize that average error on test samples. In fact, the rationale behind machine learning-based graph embedding lies in the (high-order) input-output data fitting, aiming to learn embeddings that can capture and preserve the complex patterns hidden in graph representations as much as possible, which contributes to reconstructing the original graphs. Unless otherwise specified, the graph embedding techniques retrospected in this article are all machine learning-based.

\subsubsection{Recommendation based on graph embedding}
\label{recommendation_based_on_graph_embedding}

When incorporating graph embedding techniques in recommendation, it comes to a sort of straightforward. By implementing graph embedding techniques on graph representations involving user nodes and item nodes, the embeddings of users and items can be learned to measure user-item proximity and predict their similarity for recommendation, which is called \textbf{graph embedding-based recommendation}. Methodologically, similar in the process of learning a mapping $\Phi:\mathcal{G} \to \mathbb{R}^{n \times k}$ to graph embedding techniques, in the first place graph embedding-based recommendation constructs two hypothesis spaces $U$ and $V$ for users and items, respectively, from which two mappings that project user nodes and item nodes of $\mathcal{G}$ into a common Euclidean space can be determined. In order to determine the optimal mappings, \textbf{objective functions} are set up to measure the average error between predicted proximity $\mathcal{F}(U(v_i), V(v_j))$ of user-item node pairs and observed ones $p_{ij}$ on both training samples and test samples, formally constructed as $E[L(\mathcal{F}(U(v_i), V(v_j)),p_{ij})]$ where $L$ is a \textbf{loss function} and $E$ is an objective (or expectation) function. Based on it, the embeddings of users and items can be learned by optimization algorithms. Using these embeddings, the probabilities of existence of each unobserved user-item interaction can be predicted by $\mathcal{F}(U(v_i), V(v_j))$, which is the ground for sorting candidate items in descending order and select the top-N ones as recommendations returned to users.

From a practical perspective, the embeddings of users and items learned from side information and knowledge related to users and items are reusable and are possibly optimal for preserving and reconstructing the original graph representations. In that case, they could intrinsically carry the properties of users and items as well as the hidden relations between users (since the embeddings of edges can be derived from the embeddings of their endpoints through some calculations). Resorting to combing these embeddings as a supplement with those learned from user-item interactions, enriched information can be employed in recommendation, which contributes to alleviating the cold start problem by building relations between ever non-interacted user-item pairs and the sparsity problem by uncovering more hidden relations between sparsely connected user-item pairs. As an overview,  categorized by different recommendation tasks, the graph embedding-based recommendation methods retrospected in this article are summarized in Tab.~\ref{a_comparison_of_graph_embedding_based_recommendation}, including their respective pros, cons and recent focus.

In practice, the applications of graph embedding techniques are not only throughout recommender systems but far outside it as well \cite{choudhary2021survey}, like knowledge graph completion, question answering and query expansion.

\begin{table}[!ht]
\footnotesize
\caption{\textbf{A comparison between different graph embedding-based recommendation methods.} Categorized by recommendation tasks, recent focuses concentrate on developing the pros and solving the cons of the respective methods.}
\label{a_comparison_of_graph_embedding_based_recommendation}
\begin{adjustbox}{center}
\begin{tabular}{@{}cccccc@{}}
\toprule
\textbf{Task} & \textbf{Method} & \textbf{Section} & \textbf{Pros} & \textbf{Cons} & \textbf{Recent focus} \\ \midrule
\multirow{4}{*}{\begin{tabular}[c]{@{}c@{}}Bipartite\\ graph\\ embedding \\ for recom-\\mendation\end{tabular}} & \begin{tabular}[c]{@{}c@{}}Matrix\\ factorization\\ (MF)\end{tabular} & \begin{tabular}[c]{@{}c@{}}\ref{sec:SVD}\\ \ref{Other_Models}\\ \ref{MF-based_online_learning}\end{tabular} & \begin{tabular}[c]{@{}c@{}}1. has well extensi-\\bility\\ 2. can achieve non-\\negative embedding\\ 3. can capture user's \\ long-term preferences\end{tabular} & \begin{tabular}[c]{@{}c@{}}1. faces non-convex \\ optimization problem\\ 2. is shallow learning\\ 3. could violate the  \\ triangle inequality\\ principle\end{tabular} & \begin{tabular}[c]{@{}c@{}}1. non-negative MF\\ 2. metric learning\\ 3. fast online learning\end{tabular} \\ \cmidrule(l){2-6}
 & \begin{tabular}[c]{@{}c@{}}Bayesian\\ analysis\end{tabular} & \ref{Bayesian_Sec} & \begin{tabular}[c]{@{}c@{}}1. can achieve automatic \\hyperparameter adjust-\\ment\\ 2. can achieve pair-\\wise ranking\end{tabular} & 1. is shallow learning & \begin{tabular}[c]{@{}c@{}}1. automatic machine \\ learning\\ 2. casual inference\end{tabular} \\ \cmidrule(l){2-6}
 & \begin{tabular}[c]{@{}c@{}}Markov \\ processes\end{tabular} & \ref{Markov_processes} & \begin{tabular}[c]{@{}c@{}}1. can capture user's \\ short-term preferences\end{tabular} & 1. is shallow learning & 1. combined with MF \\ \cmidrule(l){2-6}
 & \begin{tabular}[c]{@{}c@{}}Deep\\ learning\end{tabular} & \ref{DL-Bi} & \begin{tabular}[c]{@{}c@{}}1. can discover non-\\linear patterns \\ 2. can achieve fast \\parallel computing\\ 3. has well input \\compatibility\end{tabular} & \begin{tabular}[c]{@{}c@{}}1. lacks explainability\\ 2. faces higher hyper-\\ parameter adjustment\\ difficulty\end{tabular} & \begin{tabular}[c]{@{}c@{}}1. deep metric learning\\ 2. casual learning\\ 3. sequential recom-\\mendation\end{tabular} \\ \midrule
\multirow{3}{*}{\begin{tabular}[c]{@{}c@{}}General \\ graph \\ embedding \\ for recom-\\mendation\end{tabular}} & Translation & \begin{tabular}[c]{@{}c@{}}\ref{Trans-section}\\ \ref{RSwithSideInfo}\end{tabular} & \begin{tabular}[c]{@{}c@{}}1. can preserve local\\ topological features\\ 2. can flexibly distin-\\guish the multiplicity\\ of nodes and relations\end{tabular} & \begin{tabular}[c]{@{}c@{}}1. could lose global \\ topological features\end{tabular} & \begin{tabular}[c]{@{}c@{}}1. sequential recom-\\mendation\end{tabular} \\ \cmidrule(l){2-6}
 & \begin{tabular}[c]{@{}c@{}}Meta path\\ (Random walk)\end{tabular} & \begin{tabular}[c]{@{}c@{}}\ref{Meta-path-section}\\ \ref{RSwithSideInfo}\end{tabular} & \begin{tabular}[c]{@{}c@{}}1. can preserve global\\ topological features\end{tabular} & \begin{tabular}[c]{@{}c@{}}1. requires expert \\ knowledge for meta \\ path design\end{tabular} & \begin{tabular}[c]{@{}c@{}}1. random walk on \\ heterogeneous graphs\\ 2. combinations with MF\\ 3. automatic meta path \\ construction by using\\ graph topology\end{tabular} \\ \cmidrule(l){2-6}
 & \begin{tabular}[c]{@{}c@{}}Deep \\ learning\end{tabular} & \begin{tabular}[c]{@{}c@{}}\ref{DL-KG}\\ \ref{RSwithSideInfo}\end{tabular} & \begin{tabular}[c]{@{}c@{}}1. can preserve non-\\linear topological \\features\\ 2. can run (semi-)un-\\ supervised learning\end{tabular} & \begin{tabular}[c]{@{}c@{}}1. could lose infor-\\mation through the \\Encoder\end{tabular} & \begin{tabular}[c]{@{}c@{}}1. attention mechanism and\\ self-attention mechanism\end{tabular} \\ \midrule
\multirow{3}{*}{\begin{tabular}[c]{@{}c@{}}Knowledge\\ graph \\ embedding\\ for recom-\\mendation\end{tabular}} & \begin{tabular}[c]{@{}c@{}}Graph neural\\ network \\ (GNN)\end{tabular} &\begin{tabular}[c]{@{}c@{}}\ref{LEMGE} \\ \ref{KGERS} \end{tabular}  &\begin{tabular}[c]{@{}c@{}} 1. can achieve fast \\ parallel computing \\ 2. can capture the \\ multiplicity and dy-\\namics of knowledge \\graphs \end{tabular}  & \begin{tabular}[c]{@{}c@{}} 1. carries popularity  \\biases in negative \\sampling \end{tabular} & \begin{tabular}[c]{@{}c@{}} 1. propagation module \\ 2. sampling module \\ 3. pooling module \end{tabular} \\ \cmidrule(l){2-6}
 & \begin{tabular}[c]{@{}c@{}}Multi-viewed\\ graph\end{tabular} &  \begin{tabular}[c]{@{}c@{}}\ref{LEMGE} \\ \ref{KGERS} \end{tabular} & \begin{tabular}[c]{@{}c@{}} 1. can capture relati-\\onal multiplicity \end{tabular} & \begin{tabular}[c]{@{}c@{}} 1. could lose nodal \\ multiplicity  \end{tabular}  & \begin{tabular}[c]{@{}c@{}} 1. attention mechanism \\for weight distribution \end{tabular} \\ \cmidrule(l){2-6}
 & \begin{tabular}[c]{@{}c@{}}Multi-layered\\ graph\end{tabular} &\begin{tabular}[c]{@{}c@{}}\ref{LEMGE} \\ \ref{KGERS} \end{tabular}  & \begin{tabular}[c]{@{}c@{}} 1. can be used for \\cross-domain recom-\\mendation \end{tabular}  & \begin{tabular}[c]{@{}c@{}} 1. requires expert know-\\ledge for modeling \end{tabular}  &\begin{tabular}[c]{@{}c@{}} 1. translation method \\2. meta path method \\ 3. attention mechanism \end{tabular} \\ \bottomrule
\end{tabular}
\end{adjustbox}
\end{table}

\subsubsection{Optimization algorithms}
\label{optimization_algorithm}

Objective functions, once formulated, can be solved as optimization problems by numerical or analytical \textbf{optimization algorithms} \cite{sra2012optimization, parikh2014proximal, sun2019survey}, which are used to implement the learning process of searching for the optimal $U^{\operatorname{global}}$ and $V^{\operatorname{global}}$ from beforehand defined hypothesis spaces, in order to satisfy the extremum of objective functions. In this way, the effectiveness and efficiency of optimization algorithms determine the performance of graph embedding-based recommendation.

Briefly, common-used optimization algorithms for graph embedding-based recommendation include stochastic gradient descent (SGD) \cite{bottou1991stochastic} and its parallel version ASGD \cite{niu2011hogwild}, the two most popular ones out of their simplicity and efficiency, as well as other representative latest advances like Mini-bath Adagrad \cite{duchi2011adaptive}, nmAPG \cite{li2015accelerated}, Adam \cite{kingma2014adam} and ADMM \cite{boyd2011distributed}.

\subsection{A general design pipeline of graph embedding-based recommendation}
\label{pipeline}

\begin{figure}[!ht]
\centering
\includegraphics[scale=0.8]{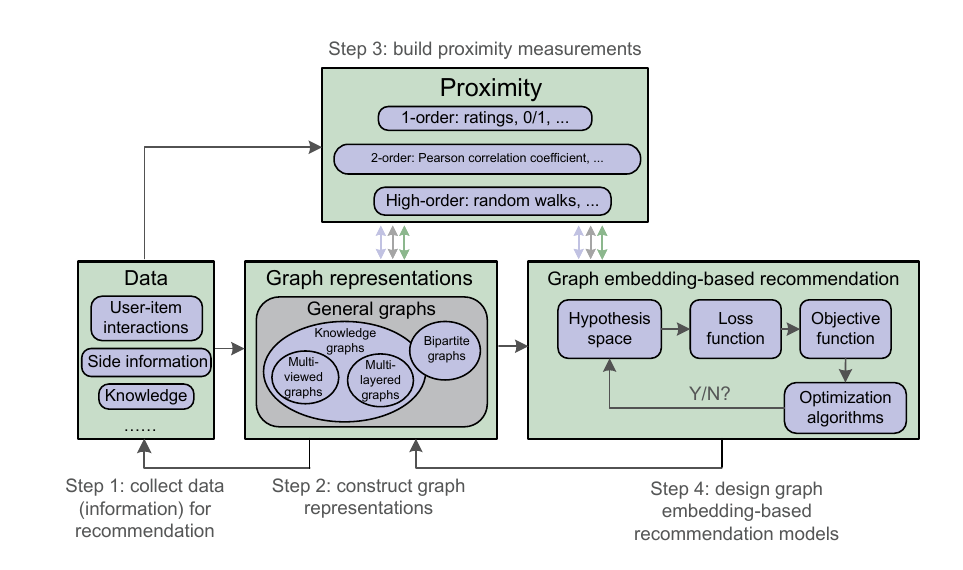}
\caption{\textbf{A general design pipeline of graph embedding-based recommendation.}}
\label{pipeline_fig}
\end{figure}

Overall, under the framework of machine learning methodology, a graph embedding-based recommendation model can be mathematically presented as
\begin{equation}
\label{pipline_eq}
\arg \min_{\Phi} E\bigg(L\big(\mathcal{F}(\Phi(v_i),\Phi(v_j)),p_{ij}\big), \Theta\bigg), \Phi \in H, \forall v_i,v_j \in \mathcal{G},
\end{equation}
where $\Phi:\mathcal{G} \to \mathbb{R}^{n \times k}$ denotes a mapping that projects a graph representation's $n$ nodes into an embedding matrix $\mathbb{R}^{n \times k}$ consisting of $k$-dimensional row vectors, $H$ denotes a hypothesis space designed beforehand, $\mathcal{F}:(\cdot, \cdot) \to \mathbb{R}$ denotes a proximity measurement, $L$ denotes a loss function which is used to measure the error between predicted values and observed (or true) values on training samples or test samples, $E$ denotes an objective function measuring the expectation of overall loss, $p_{ij}$ is the observed proximity between nodes $i$ and $j$, and $\Theta$ denotes a hyper-parameter set. From the perspective of Eq.~(\ref{pipline_eq}), this section decomposes the designing process of graph embedding-based recommendation into several steps and proposes a general design pipeline of that, as shown in Fig.~\ref{pipeline_fig}.

Specifically, in Fig.~\ref{pipeline_fig}, the first step is to collect information for recommendation from such as public data sets, practical recommender systems, Internet of things or other commercial data products, among which public data sets are recognized as the priority of designing and evaluating a recommendation model because of their cost-effective and widespread access. Then, representing the information by graph representations is the second step. In this step, selecting an appropriate graph representation $\mathcal{G}$ that can capture and preserve the complexity (or multiplicity) involved in original information as much as possible is crucial, which directly determines the design of recommendation models and the accuracy of recommendation results. After that, the third step is to build proximity measurements $\mathcal{F}:(\cdot, \cdot) \to \mathbb{R}$, used to measure the proximity between node pairs in graph representations. Methodologically, as for a specific recommendation scenario, once identified, the first-order proximity could give clue to the possible forms of proximity measurements fitting this scenario, which helps researchers build higher-order ones.

When it comes to designing a graph embedding-based recommendation model (the fourth step in Fig.~\ref{pipeline_fig}), a hypothesis space $H$ should be constructed in the first place, from which the optimal mapping $\Phi$ in Eq.~(\ref{pipline_eq}) can be searched out. On training samples, after determining an initial mapping $\Phi$ from a constructed hypothesis space, nodal embeddings can be generated through it. As the input of $\mathcal{F}$, these embeddings can be used to measure the observed proximity between node pairs. In order to assess the precision, loss function $L$ is designed to calculate the error between predicted proximity and the corresponding observed (\textit{i.e.}, true) proximity of a node pair. After being implemented on all node pairs of training samples, the expected loss can be calculated by an objective function $E$, an expectation function. Searching out (or training) the optimal mapping $\Phi$ on training samples from the hypothesis space runs by optimization algorithms, which will be used to predict unobserved proximity on test samples. Methodologically, designing an appropriate loss function, objective function or optimization algorithm is generally not hectic because a lot of related theories and experiences have matured as a recipe, which can be easily accessed from previous research. In fact, what really matters in step $4$ is to construct a hypothesis space fitting to a specific recommendation task. Completely pioneering a novel model is usually not easy. In this regard, as references, Tabs.~\ref{SVD}, \ref{LMF-KNN}, \ref{Bayes}, \ref{SVD dynamics}, \ref{Markov}, \ref{Trans} and \ref{MetaPath} summarize several common-used architectures of different graph embedding techniques and different recommendation methods, including their respective hypothesis spaces, loss functions or objective functions.

As shown in Fig.~\ref{pipeline_fig}, the four steps are recurrent as an iteratively revising and refining process. For example, when designing a recommendation model, if its performance has hit its ceiling while still not being able to reach one's expectation, it could be helpful to modify or even reconstruct a more appropriate hypothesis space or proximity measurement. In this process, as said before, constructing appropriate graph representations which can capture and preserve the complexity (or multiplicity) of original information is crucial. For that purpose, combined with analyzing the topology characteristics of a graph representation constructed in step $2$, going back to step $1$ to check if it can well fit the original information is a strategy, which gives clues to refine the structures of graph representations.

On the other hand, the design pipeline on its face is data-oriented (or task-oriented), which might be preferred by computer science researchers, who generally design recommendation models starting from specific tasks related to collected data (\textit{i.e.}, information). To be sure, by means of data mining techniques, this data-oriented designing strategy could quickly dig out the hidden patterns of data and incorporate them in modeling, which can achieve a higher recommendation accuracy on a specific task. However, these models designed in this way face fundamental limits on their generalization to other tasks, because they are data-oriented while different tasks generally carry distinctive data patterns, scales or sparsity. On top of that, by starting from step $4$, the designing process based on this pipeline can also be run by generalization-oriented, aiming to design versatile recommendation models fitting to diverse tasks with different data properties, in which case physicists and mathematicians might prefer. For all they are two different perspectives for designing models, there is no priority between data-oriented and generalization-oriented strategies. In fact, by combining their respective advantages, it could be more beneficial for researchers to design recommendation models, where multi-task learning \cite{crawshaw2020multi, zhang2021survey} seems to be a promising direction.

At the end of this section, notations used in this article are presented in Tab.~\ref{Notation}. The following Secs.~\ref{BiGE}, \ref{GGE} and \ref{KGBE} retrospect embedding techniques for bipartite graphs, general graphs and knowledge graphs, respectively, as well as their corresponding applications in recommendation.

\begin{table}[!ht]
\small
\setlength{\abovecaptionskip}{0.3cm}
\setlength{\belowcaptionskip}{0.3cm}
\caption{\textbf{Notations used in this article.}}
\label{Notation}
\begin{adjustbox}{center}
\footnotesize
\begin{tabular}{cl}
\toprule
\textbf{Notations} & \multicolumn{1}{c}{\textbf{Meaning}} \\
\hline \addlinespace $R$ & \tabincell{l}{The observed user-item rating matrix, where its element $r_{ij}$ represents the rating of item $j$ given \\by user $i$.}\\  \addlinespace
 $\hat{R}$ & The predicted user-item rating matrix. \\ \addlinespace
 $\mathcal{R}(i)^+, \mathcal{R}(i)^-$ & The set of items rated (unrated) by user $i$. \\ \addlinespace
 $N(i)^+, N(i)^-$ & The set of items implicitly interacted (non-interacted) by user $i$.\\ \addlinespace
 $U$ & The user embedding matrix consisting of row vectors $U_i$ of each user $i$.\\ \addlinespace
 $V$ & The item embedding matrix consisting of row vectors $V_j$ of each item $j$.\\ \addlinespace
 $S$ & \tabincell{l}{The item-item (user-user) proximity (or similarity) matrix, where $S_{j_1 j_2}$ ($S_{i_1 i_2}$) is the proximity \\ between items $j_1$ and $j_2$ (users $i_1$ and $i_2$).}\\ \addlinespace
 $b_{ij}$ & \tabincell{l}{The user-item interaction bias involved in rating $r_{ij}$, consisting of user bias $b_i$ and item bias $b_j$.} \\ \addlinespace
 $\mathcal {G}_{kg}=(E, R,\mathcal{E}, \mathcal{R})$ & \tabincell{l}{A knowledge graph, where $E$ is the set of entities and $R$ is the set of relations, $\mathcal{E}$ and $\mathcal{R}$ represent \\ the set of node types and the set of relation types, respectively.} \\ \addlinespace
 $(h, r, t)$ & \tabincell{l}{A knowledge triplet, where $h, t \in E$ represent a head entity and a tail entity, respectively; $r \in R$ \\ represents the relation between entities.}\\ \addlinespace
 $\bm{h}, \bm{t}, \bm{r}$ & The embedding vectors of $h,t$ and $r$.\\ \addlinespace
 $s_{ij}^{(1)}, s_{ij}^{(2)}$ & The first-order proximity and second-order proximity between nodes $v_i$ and $v_j$. \\ \addlinespace
\bottomrule
\end{tabular}
\end{adjustbox}
\end{table}

\section{Bipartite graph embedding for recommendation}
\label{BiGE}

To reveal user's preferences for items, recommendation models are generally run by analyzing user-item relations which are directly recorded in observed user-item interactions and also can be uncovered with side information or knowledge. As the bedrock, recommendation models based on bipartite graphs are of top priority in research, which can be generalized to recommendation models based on general graphs or knowledge graphs. According to the taxonomy of user-item interactions (illustrated in Secs.\ref{information}), Secs.~\ref{without:dynamics} and \ref{with:dynamics} retrospect recommendation models based on bipartite graph embedding techniques for static user-item interactions and temporal user-item interactions, respectively.

\subsection{Recommendation with static user-item interactions}
\label{without:dynamics}

In general, recommendation models based on bipartite graph embedding techniques for static user-item interactions can be divided into three categories: those based on methods of matrix factorization, Bayesian analysis and deep learning. From an overview, as the pioneer of bipartite graph embedding techniques, the matrix factorization method has a virtue of extensibility dear to researchers. As a probabilistic version of the matrix factorization method, the Bayesian analysis method can alleviate the non-convex optimization issue out of data sparsity problem suffered by the matrix factorization method, in a manner that setting model's regularization terms with prior knowledge, like the fact that the error follows a Gaussian distribution. As for learning and preserving the non-linear patterns involved in data, the deep learning method has significant advantages over the above two methods.

\subsubsection{Models based on matrix factorization}
\label{sec:SVD}

The rationale behind matrix factorization-based recommendation models basically lies in the singular value decomposition (SVD) \cite{stewart1993early}, which can decompose a matrix $A_{M \times N}$ into $A=U \Sigma V^T$, where $U$ and $V$ are two orthonormal Eigen matrices and $\Sigma$ is a diagonal matrix composed of $A$'s singular values. In turn, implementing matrix product on $U,V$ and $\Sigma$ can approximately reconstruct $A$. Within this framework, through SVD, a user-item rating matrix $R$ is supposed to be decomposed into such elements as the embedding matrices of users and items, based on which $R$ can be approximately reconstructed by implementing matrix product on the embedding matrices as well.

For that purpose, latent semantic analysis (LSA) \cite{deerwester1990indexing} is recognized as one pioneer of SVD's application in textual information retrieval. Based on documents and terms appearing in at least two documents, LSA firstly constructs a term-document matrix $\displaystyle A$ where its element $a_{ij}$ denotes the frequency of term $i$'s appearance in document $j$. Then, through truncated SVD \cite{hansen1990truncated} (an accelerated version of SVD), $A$ can be decomposed by $\displaystyle A \approx \hat{A}=U_k \Sigma_k V_k^T$, based on which the embedding of term $i$ can be represented by the $i$-th row of matrix $U_k \Sigma_k$ and that of document $j$ can be represented by the $j$-th row of matrix $V_k \Sigma_k$, which are both in a common $k$-dimensional vector space. To complete a user's information retrieval with a query $q$ (a set of query words), LSA can generate the embedding of $q$ as $\displaystyle \hat{q}=q^TU_k\Sigma_k^{-1}$, which will be used to measure the query $q$'s proximity with each of the documents, by doing operations (such as dot product) on their corresponding embeddings.

Feasible as LSA in theory, when coming to recommender systems where the number of users and items are generally hundreds of millions, LSA becomes unfeasible in decomposing such an extremely huge user-item interaction matrix $R$ as a result of the high complexity of SVD and the sparsity of $U$ and $V$ which brings the NP-hard problem \cite{davis1997adaptive}. To break those limitations, on his blog Simon Funk proposed FunkSVD inherited from LSA's idea, which resorts to optimization algorithms as a strategy for efficiently running on a large-scale matrix for recommendation (Tab.~\ref{SVD} gives details). Slightly different from LSA, FunkSVD does not directly decompose $R$ by $R=U\Sigma V^T$ but rather hypothesizes that $R$ can be represented by the dot product of two matrices $U$ and $V$, the embedding matrices of users and items. After initializing their element values, FunkSVD will search the optimal $U$ and $V$ by optimization algorithms, satisfying $\displaystyle UV^T=\hat{R} \approx R$ as approximately as possible.

\begin{table}[!ht]
\tiny
\setlength{\abovecaptionskip}{0.3cm}
\setlength{\belowcaptionskip}{0.3cm}
\caption{\textbf{Examples of modeling matrix factorization-based recommendation.} (1) In FunkSVD, by penalizing the magnitudes of parameters, $\lambda(\|U_i\|_2^2+\|V_j\|_2^2)$ is used as a regularization term for preventing from overfitting \cite{hawkins2004problem}. (2) In BiasSVD, $\mu$ is the average of the values in $R$ and $b_i$ ($b_j$) is the user (item) bias. (3) In SVD++, $|N(i)^+|^{-\frac{1}{2}}\sum_{k \in N(i)^+}y_k$ represents user $i$'s preferences for his implicitly interacted items, where $y_k$ is item $k$'s embedding vector. (4) In SRui,  $\displaystyle \mathcal{F}^+(i)$ and $\displaystyle \mathcal{Q}^+(j)$ represent the top-N social neighbors of user i and item j, respectively, whose proximity $s_{**}$ is measured by Pearson correlation coefficient \cite{breese2013empirical}. (5) In NCRPD-MF, $v_n$ and $v_c$ represent the features of item's geography information and category, respectively; and the received review words $v_w$ of an item can be used to represent its intrinsic characteristics with $V_j$. $\displaystyle \alpha_1,\alpha_2 \in [0,1]$ correspondingly control the influence of geographical neighborhood and category; and $z$ represents both popularity and geographical distance. (6) In FM, in order to predict the rating $\hat{r}({\bm{x}})$ of an item given by a user, a feature vector $\bm{x}$ is constructed, consisting of the features of the user and the item both represented by one-hot encoding and the ratings of other items rated by the user and stuff. $w_0, w_i \in \mathbb{R}$ represent the global bias and the strength of the $i$-th variable, respectively; and $\bm{v}_i$ represents the embedding of the $i$-th feature in all $n$ features.}
\label{SVD}
\begin{adjustbox}{center}
\begin{tabular}{ccc}
\toprule \addlinespace \textbf{Model} & \textbf{Factorization (hypothesis space)} & \textbf{Objective function}\\ \addlinespace
\midrule \addlinespace FunkSVD & $\displaystyle \hat{r}_{ij}=U_i V_j^T$ & $\displaystyle \operatorname*{arg\,min}_{U,V}\sum_{(i,j) \in \mathcal{R}(i)^+}(r_{ij}-U_iV_j^T)^2+\lambda(\|U_i\|_2^2+\|V_j\|_2^2)$ \\ \addlinespace

\addlinespace BiasSVD & $\displaystyle \hat{r}_{ij}=\mu+b_i+b_j+U_i V_j^T$ & \tabincell{c}{$\displaystyle \operatorname*{arg\,min}_{U,V,b_*} \sum_{(i,j) \in \mathcal{R}(i)^+}(r_{ij}-\mu-b_i-b_j-U_i V_j^T)^2$\\$\displaystyle +\lambda (\|U_i\|_2^2+\|V_j\|_2^2+b_i^2+b_j^2) $} \\ \addlinespace

\addlinespace SVD++ & \tabincell{c}{$\displaystyle \hat{r}_{ij}=b_{ij}+\bigg (U_i$\\ $ \displaystyle +|N(i)^+|^{-\frac{1}{2}}\sum_{k \in N(i)^+}y_k \bigg) V_j^T$} & \tabincell{c}{$\displaystyle \operatorname*{arg\,min}_{U,V,b_*,y_*} \sum_{(i,j) \in \mathcal{R}(i)^+}\bigg(r_{ij}-b_{ij}- \bigg(U_i+|N(i)^+|^{-\frac{1}{2}} \sum_{k \in N(i)^+} y_k \bigg)V_j^T\bigg)^2 $\\ $ \displaystyle  +\lambda (\|U_i\|_2^2+\|V_j\|_2^2+b_i^2+b_j^2+\sum_{k \in N(i)^+}y_k) $}  \\ \addlinespace

\addlinespace  SRui& see \cite{ma2013experimental} & \tabincell{c}{$\displaystyle \operatorname*{arg\,min}_{U,V} \frac{1}{2}\sum_{(i,j) \in \mathcal{R}(i)^+} (r_{ij}-U_iV_j^T)^2+\frac{\alpha}{2}\sum_{i=1}^m \sum_{f \in \mathcal{F}^+(i)}s_{if}\|U_i-U_f\|_F^2 $\\$\displaystyle +\frac{\beta}{2}\sum_{j=1}^n \sum_{q \in \mathcal{Q}^+(j)}s_{jq}\|V_j-V_q\|_F^2+\frac{\lambda_1}{2}\|U\|_F^2+\frac{\lambda_2}{2}\|V\|_F^2$} \\ \addlinespace

\addlinespace  NCRPD-MF &  \tabincell{c}{$\displaystyle \hat{r}_{ij}=\mu+b_i+b_j+z$\\ $\displaystyle +U_i\bigg( \frac{1}{|W_j|}\sum_{w \in W_j}v_{w}+\frac{\alpha_1}{|N_j|}\sum_{n \in N_j} v_{n}$\\ $\displaystyle +\frac{\alpha_2}{|C_j|}\sum_{c \in C_j} v_{c}\bigg)^T$} & \tabincell{c}{$\displaystyle \operatorname*{arg\,min}_{U,V,b_*,v_*,\beta_*} \sum_{(i,j) \in \mathcal{R}(i)^+} (r_{ij}-\hat{r}_{ij})^2+\lambda_1 \bigg( \|U_i\|^2+\sum_{w \in W_j}\|v_w\|^2\bigg)$\\ $\displaystyle +\lambda_2(b_i^2+b_j^2+\beta_i^2+\beta_j^2)+\lambda_3 \bigg( \sum_{n\in N_j}\|v_n\|^2+\sum_{c \in C_j}\|v_c\|^2 \bigg)$}  \\ \addlinespace

\addlinespace  FM &  \tabincell{c}{$\displaystyle \hat{r}({\bf{x}})=w_0+\sum_{i=1}^n w_i x_i$ \\ $\displaystyle +\sum_{i=1}^n \sum_{j=i+1}^n \langle \bm{v}_i, \bm{v}_j \rangle x_i x_j$} & see \cite{rendle2010factorization}  \\ \addlinespace

\bottomrule
\end{tabular}
\end{adjustbox}
\end{table}

FunkSVD has several virtues dear to researchers. One remarkable aspect of those is its salient extensibility, which makes it compatible with auxiliary information (such as user biases or item biases) contributing to a higher recommendation accuracy. In view of that, FunkSVD's variants soon widened in subsequent research. For instance, by defining the biases in ratings as a term $\displaystyle b_{ij}=\mu+b_i+b_j$ linearly appended to $\displaystyle UV^T$, BiasSVD \cite{koren2009matrix} can incorporate user biases $b_i$ and item biases $b_j$ (Tab.~\ref{SVD} gives details) into FunkSVD. By defining user's preferences as a term (Tab.~\ref{SVD} gives details) appended to $U_i$, SVD++ \cite{koren2009matrix} can further incorporate user's implicit interactions into BiasSVD, which decreases the deviations between $b_{ij}$ and $U_i V_j^T$. Auxiliary information that can be incorporated into recommendation is not limited to these forms. For instance, on the advice of pattern mining and data analysis, Hu et al. \cite{hu2014your} discovered that a positive correlation could hide out between an individual business' ratings given by customers and those of its geographical neighbors (regardless of their business type), revealing that the market environment might play an influential role in an individual business' popularity. After quantifying this correlation with terms, Hu et al. proposed NCRPD-MF, which can incorporate the discovered auxiliary information into BiasSVD (Tab.~\ref{SVD} gives details).

In addition, the strong extensibility of FunkSVD and its variants can also enable them to be integrated with k-nearest neighborhood-based (KNN) recommendation models. For instance, Koren et al. \cite{koren2008factorization} proposed a 3-tier SVD++ model, which can integrate the item-item proximity calculated by KNN models with SVD++. From another perspective, instead of calculating the item-item proximity matrix by similarity metrics (like Pearson correlation coefficient) as KNN models do, Slim \cite{ning2011slim} learns this matrix by means of optimization algorithms under the framework of FunkSVD (Tab.~\ref{LMF-KNN} gives details). Moreover, by applying matrix factorization to learn two embedding matrices $P$ and $Q$ preserving the patterns between items, FISM \cite{kabbur2013fism} can estimate the item-item proximity matrix $S$ with $P$ and $Q$, which is used to be integrated with KNN models. All in all, these integrated methods help out of the dependence on users' co-interactions with items in terms of calculating the item-item proximity matrix $S$, which has been a fundamental limitation on the accuracy of KNN models as a result of the sparsity problem of recommendation.

However, like any model, FunkSVD and its variants have their critics. Yet much of the criticism is based on the following two flaws. The first one is that the embeddings of users and items learned by the FunkSVD framework could involve negative values, which have difficulties in being well interpreted in practice due to the general meaninglessness of negative values in reality. One way out of this dilemma is to develop methods of non-negative matrix factorization \cite{lee1999learning,zhang2006learning,xu2012alternating,hernando2016non,luo2015nonnegative, gouvert2020ordinal, liu2021factor}. The other flaw is that the implementation of the FunkSVD framework generally violates the triangle inequality principle \cite{tversky1982similarity,ram2012maximum} because it is put in Hilbert space to measure the user-item proximity by dot-product, which could hinder the preservation of find-grained user preference. To tackle this issue, methods based on metric learning \cite{kulis2012metric, suarez2021tutorial} of measuring the user-item proximity can satisfy the triangle inequality principle since they are put in a metric (or Banach) space. In detail, these methods run by constructing a transformed user-item rating matrix $R$ (like by converting a method to convert $R$ into a distance matrix \cite{zhang2018metric}) in the metric space and factorizing it as the FunkSVD framework does \cite{hsieh2017collaborative, zhang2018metric, ma2020probabilistic, kraeva2021application, li2021dual}. Note that the metric space is unnecessary to be Euclidean. For instance, constructing and factorizing the matrix $R$ in hyperbolic space can also work well \cite{vinh2020hyperml}.

In mathematics, most of the aforementioned recommendation models are built on a global low-rank assumption of matrix factorization. Differently, Lee et al. \cite{lee2013local} built an assumption that the user-item matrix $R$ is partially observed, which is characterized by a low-rank matrix restricted in the vicinity of certain row-column combinations. Aharon et al. \cite{aharon2006k} overturned the conventional assumption that a transform matrix should always be observed and fixed. Halko et al. \cite{halko2011finding} built a randomization assumption, which contributes to a fast matrix factorization on large-scale data. Establishing novel factorization frameworks based on other assumptions from a perspective of mathematics appears to be a challenging, intriguing and promising direction of research into matrix factorization-based recommendation models.

\begin{table}[h]
\linespread{1.5}
\tiny
\setlength{\abovecaptionskip}{0.3cm}
\setlength{\belowcaptionskip}{0.3cm}
\caption{\textbf{Examples of integrating matrix factorization-based models with neighborhood-based collaborative filtering methods.} (1) In FISM, $\hat{r}_{ij}=b_i+b_j+(n_i^+-1)^{-\alpha} \sum_{k \in \mathcal R(i)^+ \setminus \{j\}} P_k Q_j^T$, where $n_i^+$ is used to control the agreement between the items rated by user $i$ with respect to their respective similarity to item $k$. (2) In SVD with prior, $E$ is set up to measure the squared loss, absolute loss or generalized Kullback-Liebler divergence. $R(U,V)$ is a regularization term and $\alpha$ is a coefficient used to balance the effects of unobserved ratings.}
\label{LMF-KNN}
\begin{adjustbox}{center}
\begin{tabular}{cc}
\toprule \addlinespace \textbf{Model} & \textbf{Objective function} \\ \addlinespace
\midrule \addlinespace Slim &  $\displaystyle \operatorname*{arg\,min}_S \frac{1}{2}\|R-RS\|_F^2+\frac{\beta}{2}\|S\|_F^2+\lambda \|S\|_1$ \\ \addlinespace
 \addlinespace  FISMrmse & $\displaystyle \operatorname*{arg\,min}_{P, Q} \frac{1}{2} \sum_{i, j}\|r_{ij}-\hat{r}_{ij}\|_F^2+\frac{\beta}{2}(\|P\|_F^2+\|Q\|_F^2)+\frac{\lambda}{2}\|b_i\|_2^2+\frac{\gamma}{2}\|b_j\|_2^2 $  \\ \addlinespace
 \addlinespace FISMauc &  $\displaystyle \operatorname*{arg\,min}_{P,Q} \frac{1}{2} \sum_{i} \sum_{j_1 \in \mathcal R(i)^+, j_2 \in \mathcal R(i)^-} \|(r_{ij_1}-r_{ij_2})-(\hat{r}_{ij_1}-\hat{r}_{ij_2})\|_F^2+\frac{\beta}{2}(\|P\|_F^2+\|Q\|_F^2)+\frac{\gamma}{2}\|b_{j_1}\|_2^2$ \\ \addlinespace
 \addlinespace SVD with prior & $\displaystyle \operatorname*{argmin}_{U,V} \sum_{(i,j)\in \mathcal{R}(i)^+} E(r_{ij},U_iV_j^T)+\alpha \sum_{(i,j)\in \mathcal{R}(i)^-}E(\hat{r}_0, U_i V_j^T) +R(U,V)$\\ \addlinespace
\bottomrule
\end{tabular}
\end{adjustbox}
\end{table}

\subsubsection{Models based on Bayesian analysis}
\label{Bayesian_Sec}

In practice, since the giant amount of users and items while generally very sparse interactions between them in recommender systems, the matrix factorization method could face the non-convex optimization problem \cite{srebro2003weighted} when factorizing such a huge and sparse user-item rating matrix, in which case, at best, its recommendation accuracy could largely fluctuate flowing from different model hyper-parameters settings and at worst, its convergence in training (or learning) by optimization algorithms could even be damaged as a result of setting inappropriate model hyper-parameters. As a tool of automatic hyper-parameter adjustment \cite{he2021automl,waring2020automated}, Bayesian analysis method, to some extent, can be used to guide the proper settings of hyper-parameters in matrix factorization-based recommendation models, like by defining regularization terms involved with prior knowledge.

The rationale behind Bayesian analysis-based recommendation can be illuminated by probabilistic matrix factorization (PMF) \cite{mnih2007probabilistic}, a probabilistic version of FunkSVD (Tab.~\ref{Bayes} gives details). In detail, by hypothesizing that the error $r_{ij}-\hat{r}_{ij}$ obeys the Gaussian distribution as $\mathcal N(r_{ij}-U_iV_j^T|0,\sigma^2)$, where user embedding $U_i$ and item embedding $V_i$ obey the zero-mean spherical Gaussian priors \cite{tipping1999probabilistic}, respectively, PMF can maximize the log of the posterior distribution by
\begin{equation}
\label{PMF}
\arg \min_{U,V} \frac{1}{2}\sum_{i=1}^M\sum_{j=1}^N I_{ij}(r_{ij}-U_iV_j^T)^2+\frac{\lambda_U}{2}\sum_{i=1}^M\|U_i\|_2^2+\frac{\lambda_V}{2}\sum_{j=1}^N\|V_j\|_2^2,
\end{equation}
where $\displaystyle \lambda_U=\frac{\sigma^2}{\sigma_U^2}, \lambda_V=\frac{\sigma^2}{\sigma_V^2}$. Similar in form of the objective function to FunkSVD, Eq.~(\ref{PMF}) is further involved with prior knowledge (i.e., the value ranges of $\lambda_U$ and $\lambda_V$), which can definitely contribute to the settings of hyper-parameters in FunkSVD. However, when inappropriately setting the $\sigma$ and $\sigma^2_*$ in Eq.~(\ref{PMF}), which are still hyper-parameters, PMF could be over-fit in training samples. Faced with this situation, instead of following the hypothesis of PMF that $U$ and $V$ are independent, Bayesian PMF (BPMF) \cite{salakhutdinov2008bayesian} argues that the distributions of $U$ and $V$ are supposed to be non-Gaussian and that $\lambda_U$ and $\lambda_V$ can both obey the Gaussian distribution (Tab.~\ref{Bayes} gives details). Moreover, in light of the Markov random field \cite{li1994markov}, Mrf-MF \cite{peng2016n} hypothesizes that the prior distributions of $U$ and $V$ should be relevant to user's neighborhood (Tab.~\ref{Bayes} gives details). The contributions of Bayesian analysis method are not only throughout the FunkSVD framework but far outside it as well, like those in recommendation based on ordinal data \cite{johnson2006ordinal} by means of Poisson factorization \cite{gopalan2014content}, Bernoulli-Poisson factorization \cite{acharya2015nonparametric} or OrdNMF \cite{gouvert2020ordinal}. Since a widespread tool for hyper-parameter adjustment, Bayesian analysis method can exert its great value in automatic machine learning \cite{yao2018taking, waring2020automated, he2021automl}, which has been the focus of recent research into recommendation and other machine learning-based fields.

In addition, Bayesian analysis method can be used to design new strategies for ranking items. Until recently, most recommendation models adopt a point-wise strategy for comparing user's preferences for different items by ranking one's ratings on items, according to their relative size. In recent research, BPR-OPT \cite{rendle2012bpr} (based on Bayesian analysis method) pioneered a pair-wise ranking strategy, comparing user's preferences for each pair of different items, in a more fine-grained way. In other words, it hypothesizes that one generally prefers his interacted items more than non-interacted ones (Tab.~\ref{Bayes} gives details). Methodologically, for each user $i$, BPR-OPT builds a training sample $D_S:=\{(i,j_m,j_n)|j _m\in \mathcal{R}(i)^+ \wedge  j_n \in \mathcal{R}(i)^- \}$, abbreviated as $>_i$, where the tuple $(i, j_m, j_n)$ represents that the user prefers item $j_m$ more than item $j_n$. Based on the built training samples for all users, maximizing the log of the posterior distribution runs by
\begin{equation}
\label{BPR}
\arg \min _{\Theta} \sum_{(i,j_m,j_n)\in D_S} -\ln \frac{1}{1+e^{-\hat{x}_{ij_mj_n}}}+\lambda_{\Theta}\|\Theta\|^2,
\end{equation}
where $\lambda_{\Theta}$ represents regularization parameters. Besides, the pair-wise ranking strategy can also be integrated with the matrix factorization-based method by, for instance, defining $\displaystyle \hat{x}_{i j}=U_i V_j$ to represent $\hat{x}_{ij_mj_n}$ with $\hat{x}_{ij_m}-\hat{x}_{ij_n}$ \cite{rendle2012bpr} under the BPR-OPT framework. So this way, the matrix factorization method, usually oriented to recommendation based on explicit user-item interactions, can be implemented on that based on implicit ones.

\begin{table}[h]
\tiny
\setlength{\abovecaptionskip}{0.3cm}
\setlength{\belowcaptionskip}{0.3cm}
\caption{\textbf{Examples of modeling Bayesian analysis-based recommendation.} (1) In PMF, $I_{ij}$ is an indicator function, equaling to $1$ if user $i$ rated item $j$ and $0$ otherwise. (2) In BPMF, $\mathcal{W}$ is the Wishart distribution related to the freedom degree $\nu_0$ and a $D \times D$ identity matrix $W_0$, where $\nu_0=D$ and $\mu_0=0$. (3) In mrf-MF, $U_{-i}$ is a user set where user $i$ is removed. $\tilde{U}_i=\frac{\sum_{i'}K_k^U(i,i') U_{i'}}{K_U}$ where $K_U$ is the number of neighbors, $K_k^U(i,i')$ contains user $i$'s k-nearest neighbors, and $\Sigma_U=I \times \sigma_U^2$. The same is true of items. (4) In BPR-OPT, $\hat{x}_{ij_mj_n}$ is a function of vector $\Theta$, used to represent the relationship among $i,j_m$ and $j_n$. (5) In RBM, $\tilde{R}_i \in \mathbb{R}^{m \times L}$ is an observed binary indicator matrix of user $i$, with values $\tilde{R}_{ij}^r=1$ if the user rated movie $j$ by score $r$ and $0$ otherwise.}
\label{Bayes}
\begin{adjustbox}{center}
\begin{tabular}{cccc}
\toprule \addlinespace \textbf{Model} & \textbf{Prior distribution} & \textbf{Posterior distribution (hypothesis space)} & \textbf{Objective function}\\ \addlinespace
\midrule \addlinespace  PMF & \tabincell{c}{$\displaystyle P(R|U,V,\sigma^2)=\prod_{i=1}^M\prod_{j=1}^N[\mathcal N (r_{ij}|U_i V_j^T,\sigma^2)]^{I_{ij}}$, \\$ \displaystyle P(U|\sigma_U^2)=\prod_{i=1}^M \mathcal N(U_i|0,\sigma_U^2 \bm{I}), P(V|\sigma_V^2)=\prod_{j=1}^N \mathcal N (V_j|0,\sigma_V^2 \bm{I})$} & \tabincell{c}{$\displaystyle P(U,V|R,\sigma^2,\sigma_U^2,\sigma_V^2) \propto$ \\$\displaystyle P(R|U,V,\sigma^2)P(U|\sigma_U^2)P(V|\sigma_V^2)$}  & see Eq.~(\ref{PMF}) \\ \addlinespace

\addlinespace BPMF & \tabincell{c}{BPMF gives the prior distributions of $\mu$ and $\sigma$ for PMF:  \\ $\displaystyle P(\mu_U,\sigma_U^2I|\mu_0,\sigma_0^2I)=\mathcal{N}(\mu_U|\mu_0,(\beta_0 \sigma_U^2I)^{-1}) \mathcal{W}(\sigma_U^2I|W_0,\nu_0)$\\ $\displaystyle P(\mu_V,\sigma_V^2I|\mu_0,\sigma_0^2I)=\mathcal{N}(\mu_V|\mu_0,(\beta_0 \sigma_V^2I)^{-1}) \mathcal{W}(\sigma_V^2I|W_0,\nu_0)$ } & see \cite{salakhutdinov2008bayesian} & \tabincell{c}{see \cite{salakhutdinov2008bayesian}, used \\ Markov Chain Monte \\ Carlo approximation} \\ \addlinespace

\addlinespace  mrf-MF & \tabincell{c}{$\displaystyle P(U_i|U_{-i})=\frac{1}{\sqrt{(2 \pi)^{d_U} \Sigma_U}}\exp (-\frac{1}{2}(U_i-\tilde{U}_i)^T\Sigma_U^{-1}(U_i-\tilde{U}_i))$\\ $\displaystyle P(V_j|V_{-j})=\frac{1}{\sqrt{(2 \pi)^{d_V} \Sigma_V}}\exp (-\frac{1}{2}(V_j-\tilde{V}_j)^T\Sigma_V^{-1}(V_j-\tilde{V}_j))$} & $\displaystyle P(U,V| \Omega, \Theta)=\frac{P(R^+,U,V|\Omega,\Theta)}{\iint P(R^+,U,V|\Omega,\Theta) d_U d_V } $  & see \cite{peng2016n}  \\ \addlinespace

\addlinespace  BPR-OPT & \tabincell{c}{$\displaystyle P(>_i|\Theta)= \prod_{(i,j_m,j_n) \in D_S} P(j_m>_i j_n |\Theta)$\\$\displaystyle =\prod_{(i,j_m,j_n) \in D_S}\frac{1}{1+e^{\hat{x}_{ij_mj_n}(\Theta)}}, \,P(\Theta) \sim N(0, \Sigma_{\Theta})$} & \tabincell{c}{$\displaystyle P(\Theta|>_i) \propto$\\$\displaystyle P(>_i|\Theta)P(\Theta)$}  & see Eq.~(\ref{BPR}) \\ \addlinespace

\addlinespace RBM & \tabincell{c}{$\displaystyle P(\tilde{R}_{ij}^r=1|U_i)=\frac{\exp(b_j^r+\sum_{k=1}^K U_{ik}W_{jk}^r)}{\sum_{l=1}^{L}\exp(b_j^l+\sum_{k=1}^K U_{ik}W_{jk}^l)}$, \\ $\displaystyle P(U_{ik}=1|\tilde{R}_i)=\sigma(b_k+\sum_{j=1}^m \sum_{l=1}^LR_{ij}^l W_{jk}^l)$}   & $\displaystyle P(\tilde{R}_i)=\sum_{U_i}\frac{\exp(-E(\tilde{R}_i,U_i))}{\sum_{\tilde{R}'_i,U'_i}\exp(-E(\tilde{R}'_i,U'_i))}$  & see \cite{salakhutdinov2007restricted}\\

\bottomrule
\end{tabular}
\end{adjustbox}
\end{table}

As a promising research direction, by means of the prior knowledge of Bayesian analysis method, causal inference \cite{liang2016causal, wang2020causal} aims to understand user's behaviors in recommendation, which contributes to the explainability of recommendation results or even models.

\subsubsection{Models based on deep learning}
\label{DL-Bi}

Despite general acknowledgement of the feasibility and effectiveness of matrix factorization and Bayesian analysis methods, they are generally based on shallow learning, only being able to capture and preserve the linear patterns involved in user-item interactions in recommendation. From a mathematical point of view, by representing the features of user $i$ and item $j$ with vectors $v_i \in \mathbb{R}^n$ and $v_j \in \mathbb{R}^n$, respectively, the implementation of recommendation can be described as the process of learning a mapping $f: \mathbb{R}^n \times \mathbb{R}^n \to \mathbb{R}$ such that $f(v_i, v_j)=\hat{r_{ij}} \approx r_{ij}$. However, if leaned by matrix factorization method or Bayesian analysis method, the mapping $f$ has to be linear, which is insufficient in fitting any non-linear relation between $(v_i, v_j)$ and $r_{ij}$ (in fact, these relations in practical recommender systems are generally non-linear). Given these inadequacies, recent years have witnessed a boom in applying deep learning methods \cite{wang2003artificial} into recommendation, in order to build deep learning-based recommendation models with non-linear mappings.

\begin{figure}[!ht]
\centering
\includegraphics[scale=0.245]{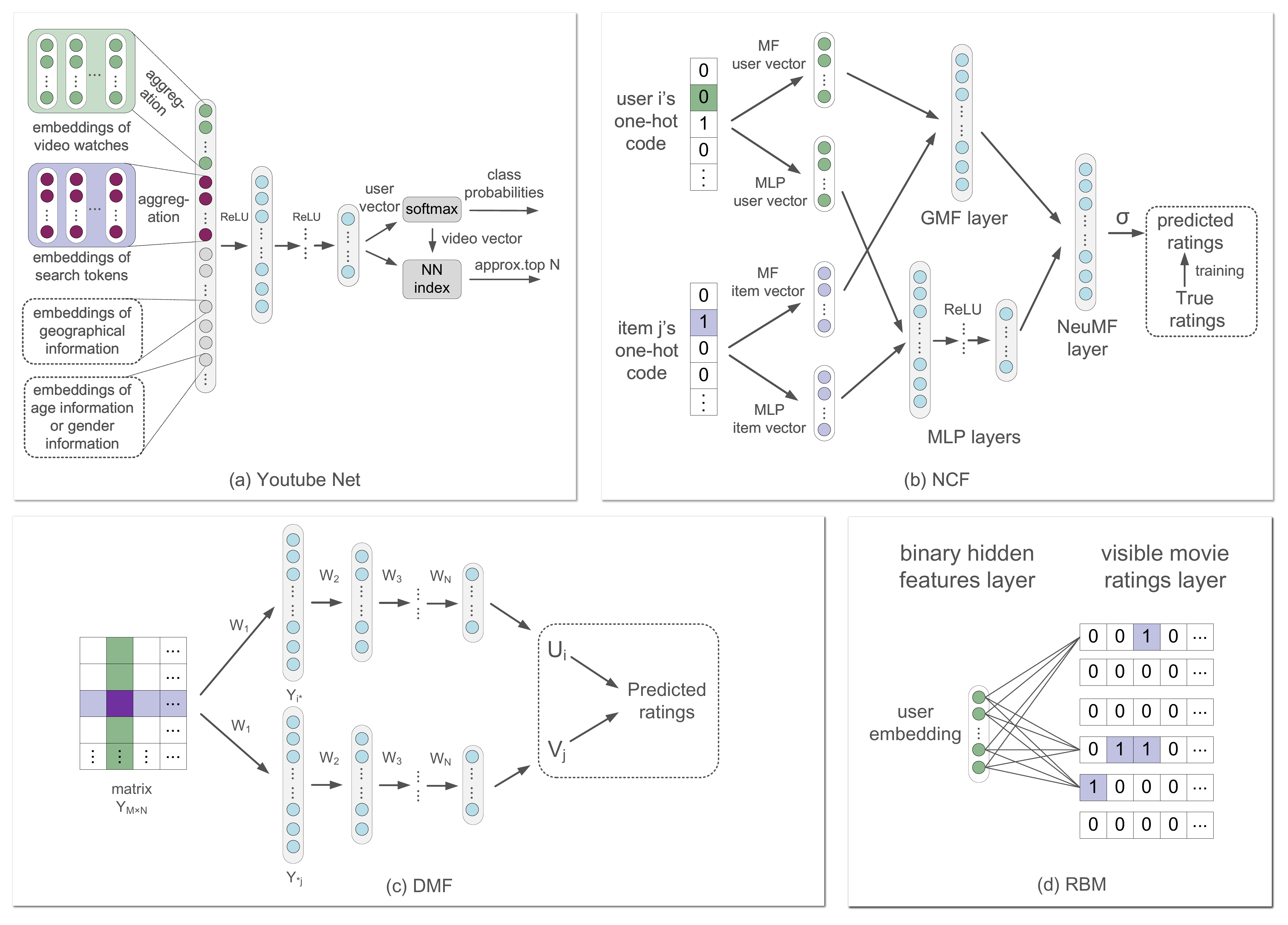}
\hspace{0.1in}
\caption{\textbf{Schematics of Youtube Net, NCF, DMF and RBM.} In (a), the generation component for recommendations of Youtube Net is presented, which runs by minimizing the cross-entropy loss with a descent on the output through sampled softmax. In (b), NCF firstly learns four mappings, used to project user i's one-hot code to $U_i^{\operatorname{MLP}}$ and $U_i^{\operatorname{MF}}$ and to project item j's one-hot code to $V_i^{\operatorname{MLP}}$ and $V_i^{\operatorname{MF}}$, respectively. Then, cross-combinations on them are implemented by $\displaystyle \phi^{\operatorname{GMF}}=U_i^{\operatorname{MF}} \odot V_j^{\operatorname{MF}}$ and $\displaystyle \phi^{\operatorname{MLP}}=a_L(W_L^T(a_{L-1}(...a_2(W_2^T(U_i^{\operatorname{MLP}}\,V_j^{\operatorname{MLP}})^T +b_2))...))+b_L )$, respectively, used as the input of neural networks to learn user i's embedding and item j's embedding, based on which the final prediction can be outputted as $\hat{y}_{ij}=\sigma(h^T (\phi^{\operatorname{GMF}} \, \phi^{\operatorname{MLP}})^T)$. In (c), the corresponding row in matrix $Y$ indexed by user $i$ is mapped to $Y_{i*}$ as user i's embedding $U_i$, and the same is true of item j's embedding $V_j$. Based on $U_i$ and $V_j$, predicting their interaction $S_{ij}=\operatorname{cosine}(U_i,V_j)$ can be performed by minimizing the objective function $L=-\sum_{(i,j)\in \mathcal{R}(i)^+ \cup \mathcal{R}(i)^-} (\frac{R_{ij}}{\max(R)} \log \hat{R}_{ij}+(1-\frac{R_{ij}}{\max(R)}) \log (1- \hat{R}_{ij}))$. In (d), RBM has a two-layer architecture based on a neural network framework, where the left layer with $K$ units (\textit{i.e.}, the $K$ elements of user's feature vector) represents user's binary hidden features, and the right layer represents the one-hot encoding of all items. Edges are built between the user and his interacted items, which can be weighted by scores from $1$ to $L$.}
\label{graph_set_one}
\end{figure}

Among these models, Youtube Net \cite{covington2016deep} is a pioneer whose schematics are shown in Fig.~\ref{graph_set_one}. Through pre-training, the embeddings of all Youtube videos are learned, based on which the feature vector of each user can be constructed, according to one's video watches. Besides, a user's feature vector is also involved with one's side information like search tokens, geographical information of gender information. After that, the feature vector will be taken as the input of a deep neural network with multiple layers, used to learn user's embedding. Finally, unobserved implicit user-item interactions can be predicted by using all of the user's embeddings. Researchers have come to see the merits of Youtube Net, including its fast parallel computing \cite{takefuji2012neural} and non-linear mapping learning (since it adopts a deep learning framework). Later, its use soon widened to various practical applications, which benefit from its high compatibility of diverse data forms as input.

Under the Youtube Net framework, specific implementations have been proposed. For instance, in order to prepare the embeddings for constructing user's feature vector, neural collaborative filtering (NCF) \cite{he2017neural} bases the pre-training on user's implicit interactions with items combined with user's characteristics. The same is true of constructing item's feature vector. After that, as shown in Fig.~\ref{graph_set_one}, when coming to predicting an implicit user-item interaction, NCF concatenates the feature vectors of the corresponding user and item, used as the input of a generalized matrix factorization (GMF) layer and multiple MLP layers, respectively, which are both jointed to a NeuMF layer outputting the final prediction. As for being compatible with side information as input, ConvMF \cite{kim2016convolutional} proposes an enhanced PMF \cite{mnih2007probabilistic} based on convolution neural networks (CNNs) \cite{li2016survey}, which can be used to learn the representations of documents. However, these deep learning models could encounter the over-fitting problem. To alleviate that, Cheng et al. \cite{cheng2016wide} proposed to combine deep learning with wide learning \cite{pandey2014go} as a strategy.

In addition, as for measuring user-item proximity in recommendation, compared with linear operators (like dot product) pervasively adopted by models based on shallow learning, the deep learning framework can be generalized as a non-linear operator for proximity measurement, which is more robust since it could uncover the complex non-linear relations between user-item pairs. For instance, by means of a deep neural network, NCF can learn the non-linear relations between an implicit user-item interaction and the two embeddings of the corresponding user and item. Its implementation was soon generalized to based on explicit user-item interactions by deep matrix factorization (DMF) \cite{xue2017deep}, which can learn the non-linear relations between an explicit user-item interaction and the corresponding values in a user-item matrix $Y$ constructed from both implicit and explicit user-item interactions, as shown in Fig.~\ref{graph_set_one}.

One difficulty, however, was that deep learning-based recommendation models generally lack well explainability, since the bedrock of these models is the fitting and optimization theory, long been concerned as almost a black-box. Faced with this situation, causal learning (or casual inference) \cite{gopnik2007causal, bonner2018causal, yao2020survey, xu2021causal, xu2021learning} seems to be a potential solution, among which restricted Boltzmann machine (RBM) \cite{salakhutdinov2007restricted} pioneered the application of causal learning in recommendation. As shown in Fig.~\ref{graph_set_one}, for a user RBM takes each element of the user's feature vector as an independent unit. Then, it builds the so-called causal relations from these units to the user's interacted items encoded by one-hot, used to model the causality between the user's features and his behaviors (\textit{i.e.}, represented by his interactions with items). As a result, by learning the weights of relations, RBM could unravel how user's features influence his behaviors correspondingly and give them practical meanings oriented to different scenarios, which are definitely helpful to explain the formation of user-item interactions. On the other hand, factorization machine (FM) \cite{rendle2010factorization} can be recognized as another pioneer, resorting to casual learning as a strategy for alleviating the sparsity problem in recommendation. In detail, from a fine-grained perspective, FM builds the causal relations between each pair of elements in user's feature vector, named feature interactions, by integrating support vector machine (SVM) with SVD (Tab.~\ref{SVD} gives details). In that case, as a supplement, these discovered causality hidden in user's feature vector contributes to enriching recommendation information. Although RBM and FM are models based on shallow learning, their rationales soon widened to based on deep learning and motivated a variety of deep causal learning-based recommendation models, including DeepFM \cite{guo2017deepfm}, xDeepFM \cite{lian2018xdeepfm}, deep Boltzmann machine \cite{salakhutdinov2009deep, hu2018deep} and stuff.

\subsubsection{Other models}
\label{Other_Models}

Until recently, in the popular conception it was usually claimed that the user exposure assumption \cite{liang2016modeling} should be the bedrock of recommendation models, which hypothesized that the reason for the existence of unobserved user-item interactions lies in user's limited view of items in recommender systems. In other words, the non-interacted items for a user were those that haven't ever been exposed to the user, thereby being considered valueless as information for recommendation (because no positive or negative preference of the user had any chance for these items). However, this assumption is not invariably true. In recent research, there are conditions under which as Devoogth et al. \cite{devooght2015dynamic} argued that non-interacted items might not always be beyond a user's view but could be eschewed by the user just as a result of dislikes (the so-called not missing at random assumption \cite{marlin2012collaborative,steck2010training}) can contradict it. In that case, the user exposure assumption will neglect user's negative preferences for items, which in fact are valuable to be used to construct negative samples \cite{chen2017sampling} for improving training precision. As Devoogth et al. \cite{devooght2015dynamic} put it, by defining a term $\alpha \sum_{(i,j)\in \mathcal{R}(i)^-}E(\hat{r}_0,U_i V_j^T)$ ($\hat{r}_0$ is a prior estimation on predicted ratings) to measure the probability that an item could be eschewed by a user, the accuracy of the SVD model can be promoted built on the not missing at random assumption (Tab.~\ref{LMF-KNN} gives details). In addition, by using a matrix constructed based on the Bernoulli condition, Liang et al. \cite{liang2016modeling} proposed to represent the probability of an item's exposure to a user, as a supplement to recommendation models.

In practice, there are still two flaws in the implementation of the not missing at random assumption. First, this assumption is not fit to implicit user-item interactions because implicit ones only record user's interactions with items without directly carrying user's preferences, which is the so-called positive-unlabeled problem \cite{hu2008collaborative, elkan2008learning, jannach2018recommending}. Second, the general strategy adopted by recommendation models built on the not missing at random assumption is to represent user's negative preferences by equally allocated weights, which could eliminate the deviations in user's negative preferences for different items in reality. In respect to the two flaws, for instance, Saito et al. \cite{saito2020unbiased} used a bias to distinguish user's affection degrees toward different items. In addition, by replacing the second term in the objective function of SVD with prior shown in Tab.~\ref{LMF-KNN} with $\sum_{(i,j) \in \mathcal{R}(i)^-} c_jE(\hat{r}_0, U_i V_j^T)$, where $c_j$ is the confidence of item $j$ to be non-interacted by users caused by a true negative preference, He et al. \cite{he2016fast} proposed a negative weighting allocation strategy considering the popularity of user's non-interacted items.

By bridging two or more recommender systems together and sharing information for recommendation as supplements to either side, cross-domain recommendation \cite{man2017cross, zang2021survey, zhu2021cross} seems to be a promising direction, which contributes to alleviating the cold start and sparsity problems. In detail, the cross-domain \cite{man2017cross} refers to two types of domains: one is target domain in which the final recommendation is directly implemented, and the other is source domain consisting of other recommender systems that could provide observed user-item interactions as supplements corresponding to the unobserved ones in target domain. By utilizing these supplemental user-item interactions from source domain, hidden user-item relations in target domain can be uncovered as enriched observed ones for recommendation. This is sort of similar to the utilization of side information and knowledge into recommendation but is fundamentally different in rationales. For illustration, suppose a user has zero or very few interactions with items in target domain, leading to the cold start and sparsity problems in recommendation. Meanwhile, in source domain if the user ever interacted with sufficient items that also existed in target domain, his preferences for items can still be analyzed and uncovered in source domain, which can be transferred and utilized in target domain to complete recommendation. Methodologically, transfer learning \cite{pan2009survey,liu2013multi, long2013transfer, wang2019multi, zhuang2020comprehensive} is a prevalent technique to realize that mechanism. For instance, in order to transfer user embeddings from source domain to those in target domain, EMCDR \cite{man2017cross} learns a one-to-one mapping to bridge the users and items commonly existed in the two domains, in which case the embedding of a cold-start user in target domain can be transferred from that learned in source domain. Besides, based on the assumption that user's preferences for items are almost consistent in different recommender systems, DDTCDR \cite{li2020ddtcdr} can further support the information exchange between the two domains back and forth by means of dual transfer learning \cite{long2012dual,xia2016dual, xia2017dual}.

In the past, the theories in optimization algorithms used in recommendation models have escaped widespread investigation. It is only recently when data scale increased dramatically that research into the efficiency of optimization algorithms has attracted any substantial scholar attention, aiming to realize the optimal performance of recommendation models as quickly as possible while consuming possibly less computing resources. For that purpose, speeding up the learning (\textit{i.e.}, convergence) process of recommendation models plays a large role. Methodologically, for instance, in order to reduce the time complexity of ALS to linearity, He et al. \cite{he2016fast} proposed an element-wise ALS (eALS). By switching the constraints and regulation terms in objective functions, Boyd et al. \cite{steck2020admm} applied ADMM \cite{boyd2011distributed} to speed up the optimization process of SLIM. GFNLF \cite{yuan2020generalized} achieves a faster convergence process of non-negative matrix factorization (NMF) via adopting $\alpha-\beta-$divergence in objective functions and incorporating a generalized momentum method.

\subsection{Recommendation with temporal user-item interactions}
\label{with:dynamics}

Temporal factors of user-item interactions primarily flow from the following situations: (1) New user-item interactions are constantly occurring by existed or new coming user's new interactions with other items unobserved before or newly entered. (2) User's long-term and short-term preferences for items could be changed. As a feasible tool, real-time recommendation \cite{abbar2013real, albalawi2019toward, ma2020temporal} relies for capturing these temporal factors on dynamically updating the embeddings of users and items, which can learn and preserve user's behavioral changes and contribute to the recommendation accuracy.

The common-used methods of real-time recommendation include the online learning (or online recommendation) method \cite{khan2017collaborative, hoi2018online}, which enables the matrix factorization-based recommendation models to absorb temporal (or newly occurring) user-item interactions in a low computing complexity, and the Markov processes method, which can represent the changes in user's short-term preferences for items.

\subsubsection{Models based on matrix factorization}
\label{MF-based_online_learning}

A simple strategy for accommodating newly occurring user-item interactions into recommendation is to reload recommendation models and to repeatedly learn the embeddings of users and items based on the whole interactions combined with the existed and newly occurring ones. Apparently, this strategy is so extremely high in computing resources that could be infeasible in large-scale data. Alternatively, by using only the newly occurring interactions, online learning (or online recommendation) method \cite{khan2017collaborative, hoi2018online} can directly update the embeddings of users and items based on those previously learned ones. In general, online learning methods implement based on the matrix factorization framework, whose objective functions or optimization frameworks can be perfectly compatible with appended terms representing newly occurring interactions. For instance, as shown in Tab.~\ref{SVD dynamics}, the objective function of SL with prior can be separated into $n$ blocks, each of which can be used to measure the changes in user's embeddings. Based on the optimization framework of FunkSVD, SGD-PMF and DA-PMF \cite{ling2012online} define appended terms to accommodate a newly occurring interaction $(u_i,v_j,r_{ij})$, incorporated into the updating process of embeddings $U_i$ and $V_j$ (Tab.~\ref{SVD dynamics} gives details). So this way, the changes of user's short-term preferences for items can be captured dynamically.

In terms of capturing the changes of user's long-term preferences for items, the matrix factorization framework is also compatible with temporal factors, by setting independent variables of objective functions or optimization algorithms to represent the changes in embeddings of users and items. For instance, as an extension of SVD++, TimeSVD++ \cite{koren2009collaborative} extends the terms $b_i$, $b_j$ and $U$ of SVD++ to $b_i(t)$, $b_j(t)$ and $U_i(t)$ related to time variables. Other methods include moving the time window \cite{vidmer2016essential} or setting instance-decay \cite{yong2015distance}.

\begin{table}[!ht]
\tiny
\setlength{\abovecaptionskip}{0.3cm}
\setlength{\belowcaptionskip}{0.3cm}
\caption{\textbf{Examples of modeling temporal matrix factorization-based recommendation.} (1) In SL with prior, $S^V=\sum_j V_j^T V_j$ is a $k \times k$ matrix, which is independent from $i$. (2) In SGD-PMF, $\eta$ is the step size controlling the convergence rate during updating iterations. (3) In DA-PMF, $Y_{U_i}$ is the approximation of $(\sum(\hat{r}_{ij}-r_{ij})\hat{r}'_{ij}V_j)/|\mathcal{R}(i)^+|$. }
\label{SVD dynamics}
\begin{adjustbox}{center}
\begin{tabular}{ccc}
\toprule \addlinespace \textbf{Model} & \textbf{Objective function} & \textbf{Optimization}\\ \addlinespace

\midrule \addlinespace SL with prior & \tabincell{c}{$\displaystyle \operatorname*{argmin}_{U_i,V_j} \sum_{i \in \mathcal{R}(i)^+} \sum_{j \in \mathcal{R}(i)^+} [(r_{ij}-U_i V_j^T)^2-\alpha  (U_iV_j^T)^2]$ \\$\displaystyle +\alpha U_i S^V U_i^T $} &  \tabincell{c}{using randomized block coordinate descant \cite{richtarik2014iteration} \\ and line search \cite{boyd2004convex}}  \\ \addlinespace

\addlinespace  SGD-PMF& \tabincell{c}{$\displaystyle \operatorname*{argmin}_{U_i,V_j} \quad (r_{ij}-\hat{r}_{ij})^2+\frac{\lambda_U}{2}\|U_i\|^2_2$ \\ $\displaystyle +\frac{\lambda_V}{2}\|V_j\|^2_2$}& \tabincell{c}{$\displaystyle U_i \leftarrow U_i-\eta((\hat{r}_{ij}-r_{ij})\hat{r}'_{ij}V_j+\lambda_U U_i)$ \\ $\displaystyle V_j \leftarrow V_j-\eta((\hat{r}_{ij}-r_{ij})\hat{r}'_{ij}U_i+\lambda_V V_j)$}\\\addlinespace

\addlinespace DA-PMF & \tabincell{c}{$\displaystyle \operatorname*{argmin}_{U_i,V_j} \quad (r_{ij}-\hat{r}_{ij})^2+\frac{\lambda_U}{2}\|U_i\|^2_2$ \\ $\displaystyle +\frac{\lambda_V}{2}\|V_j\|^2_2$} & \tabincell{c}{$\displaystyle Y_{U_i} \leftarrow \frac{|\mathcal{R}(i)^+|-1}{|\mathcal{R}(i)^+|}Y_{U_i}+\frac{1}{|\mathcal{R}(i)^+|}(\hat{r}_{ij}-r_{ij}) \hat{r}'_{ij}V_j$ \\$\displaystyle Y_{V_j} \leftarrow \frac{|\mathcal{N}(j)^+|-1}{|\mathcal{N}(j)^+|}Y_{V_j}+\frac{1}{|\mathcal{N}(j)^+|}(\hat{r}_{ij}-r_{ij}) \hat{r}'_{ij}U_i$ \\ $\displaystyle U_i=\operatorname*{argmin}_{\omega} \{ Y_{U_i}^T \omega +\lambda_U \|\omega\|^2_2\}$ \\$\displaystyle V_j=\operatorname*{argmin}_{\omega} \{ Y_{V_j}^T \omega +\lambda_V \|\omega\|^2_2\}$} \\ \addlinespace

\bottomrule
\end{tabular}
\end{adjustbox}
\end{table}

\subsubsection{Models based on Markov processes}
\label{Markov_processes}

Markov processes is another method used to capture the changes of user's short-term preferences, implemented based on analyzing user's sequential activities \cite{xu2019survey}. Its idea lies in learning an overall-shared transition matrix \cite{rendle2010factorizing} which can capture and represent the latest user-item interactions (\textit{i.e.}, could reveal user's latest preferences for items), where its elements are transition probabilities between item pairs, that is, the probabilities that a user will interact with each of his unobserved items after having interacted with previously observed ones (\textit{i.e.}, could reveal the probabilities that user's preferences would transition from his interacted items to each of the unobserved ones). Through the learned transition matrix, unobserved interactions can be predicted by ranking the items that mostly meet the transition probability from one's interacted items to unobserved ones. During the whole process, learning an accurate transition matrix is crucial for Markov processes-based recommendation, in which case considering environmental factors \cite{zhao2017sequential} in modeling the transition probability seems to contribute a lot.

\begin{table}[!htbp]
\tiny
\setlength{\abovecaptionskip}{0.3cm}
\setlength{\belowcaptionskip}{0.3cm}
\caption{\textbf{Examples of modeling Markov processes-based recommendation.} (1) In Fossil, the model can be reduced to the first-order when $L=1$. (2) In MFMP, $X_i(t) \sim \mathcal N(X_i(t)|0,\sigma_U^2 \bm{I})$ and $Y_j(t) \sim \mathcal N(Y_j(t)|0,\sigma_V^2 \bm{I})$. The parameters in Eq.~(\ref{MPMF}) are defined as  $\displaystyle \rho_U :=\sigma^2/ \Sigma_U^2$, $\rho_V :=\sigma^2/ \Sigma_V^2$, $\lambda_U := \lambda^2 / \sigma_U^2$ and $\lambda_V := \sigma^2/ \sigma_V^2$.}
\label{Markov}
\begin{adjustbox}{center}
\begin{tabular}{cccc}
\toprule \addlinespace \textbf{Model} & \textbf{Prior distribution} & \textbf{Posterior distribution (hypothesis space)}& \textbf{Loss function}\\ \addlinespace

\midrule  Fossil & see \cite{he2016fusing} & \tabincell{c}{$\displaystyle P_i(j|\mathcal R(i)^+_{t-1},\mathcal R(i)^+_{t-2},...,\mathcal R(i)^+_{t-L}) $ \\ $\displaystyle \propto  \beta_j+\langle \frac{1}{|\mathcal R(i)^+ \setminus \{j\}|^{\alpha}}  \sum_{k \in \mathcal R(i)^+ \setminus \{j\} }P_k $ \\ $\displaystyle +\sum_{l=1}^L(\eta_l+\eta_l^i)\cdot P_{\mathcal R(i)^+_{t-l}}, Q_j\rangle$} & S-BPR \cite{he2016fusing}  \\ \addlinespace

\addlinespace  MFMP & \tabincell{c}{$\displaystyle P(R(t)|U(t),V(t))=\prod_{t=0}^T \prod_{i,j \in \mathcal{R}(i)^+_t} \mathcal N(r_{ij}(t)|U_i(t)V_j(t)^T,\sigma^2)$\\ $\displaystyle U_i(t+1)=U_i(t)+X_i(t),\quad V_j(t+1)=V_j(t)+Y_j(t)$ \\ $\displaystyle P(U_i(0))=\mathcal N(U_i(0)|0,\Sigma_U^2 \bm{I}), \quad P(V_j(0))=\mathcal N(V_j(0)|0,\Sigma_V^2 \bm{I})$} & $P(U(t), V(t)| R(t))$  & see Eq.~(\ref{MPMF})  \\ \addlinespace

\bottomrule
\end{tabular}
\end{adjustbox}
\end{table}

However, Markov processes-based recommendation models cannot capture the changes of user's long-term preferences. To overcome the flaw, combining the Markov processes framework with the matrix factorization framework is a promising strategy since the latter generally performs well in capturing user's long-term preferences. Methodologically, for instance, by means of the Markov processes framework, FPMC \cite{rendle2010factorizing} builds several transition matrices corresponding to each user used to aggregate into a tensor, where the missing values correspond to unobserved user-item interactions. After that, through Tucker decomposition (TD) \cite{kim2007nonnegative}, FPMC factorizes the tensor into the embedding matrices of users and items, used to approximately reconstruct the original tensor in order to predict the missing values, as the matrix factorization framework does. Besides, by fusing user's long-term preferences learned by the matrix factorization framework and user's short-term preferences captured by a high-order Markov chain, Fossil \cite{he2016fusing} can represent user's hybrid preferences for items. By applying embedding techniques to build the transition matrices of the Markov processes framework, Wu et al. \cite{wu2013personalized} enabled the transition probabilities to involve user's long-term preferences. Since most of these works fuse user's long-term and short-term preferences linearly, they inevitably lose the higher-order patterns hidden in user-item interactions. In this regard, Wang et al. \cite{wang2015learning} proposed a two-layer structure constructed with different aggregation operations.

In addition, based automatic hyper-parameters adjustment on Bayesian analysis method, the accuracy of Markov processes-based recommendation models can be enhanced to some extent. Take MFMP \cite{zhang2019movie} as an instance, the probabilistic version of TimeSVD++ \cite{koren2009collaborative}. By hypothesizing that the changes of $U_i(t)$ and $V_j(t)$ over time follow the Gaussian Hidden Markov processes rule, maximizing the posterior distribution can run by
\begin{equation}
\begin{aligned}
\label{MPMF}
&\arg \min_{U(t), V(t)}\sum_{t=0}^T \sum_{(i,j) \in \mathcal K(t)}\bigg(R_{ij}(t)-U_i(t)V_j(t)^T \bigg)^2+\rho_U\|U(0)\|_2^2+\rho_V\|V(0)\|_2^2\\
&+\lambda_U \sum_{t=1}^T\|U(t)-U(t-1)\|_2^2+\lambda_V \sum_{t=1}^T\|V(t)-V(t-1)\|_2^2.
\end{aligned}
\end{equation}




\subsection{Summary}
\label{summary_for_chapter_three}

\begin{figure}[!ht]
\centering
\includegraphics[scale=0.65]{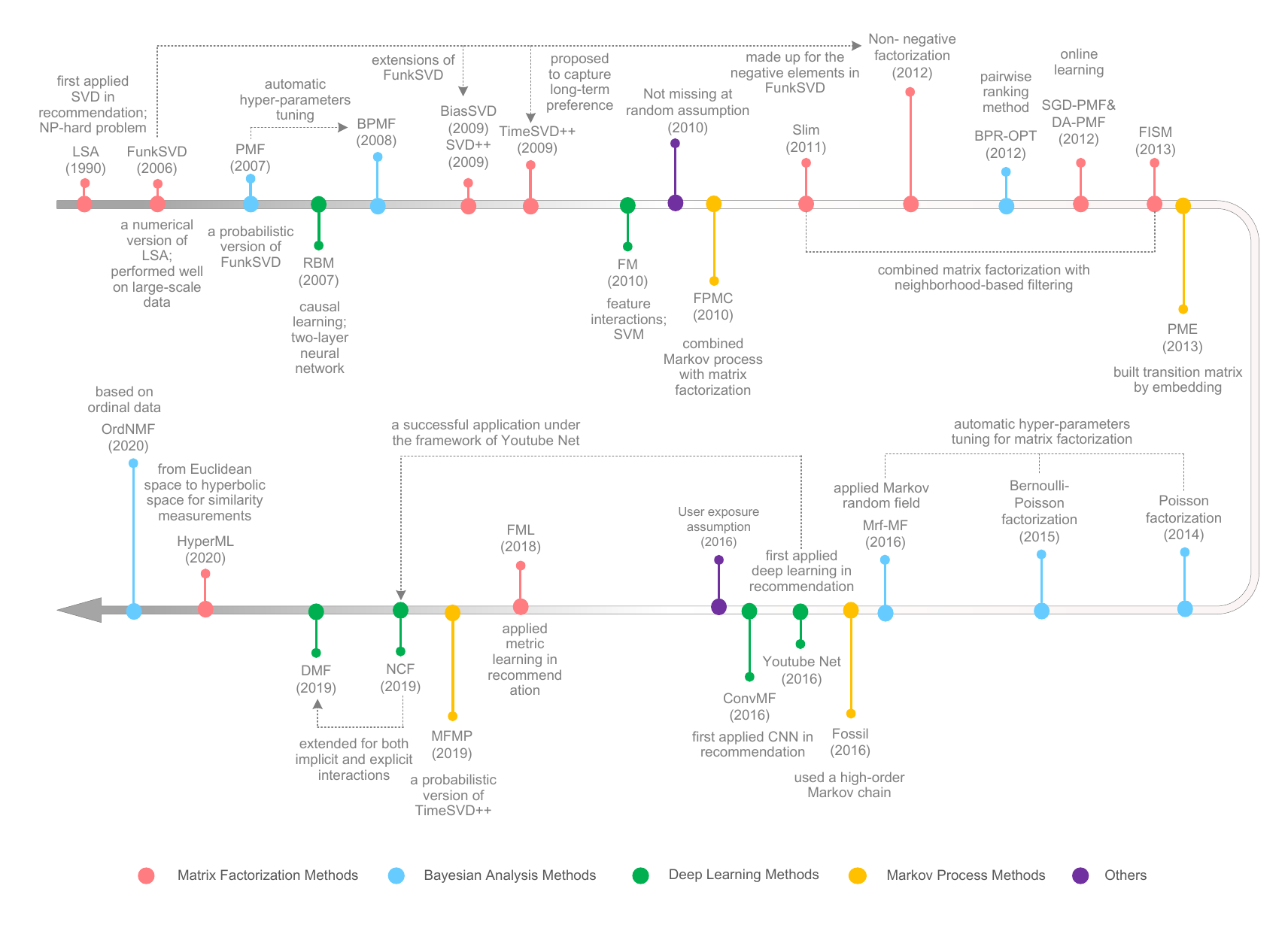}
\hspace{0.1in}
\caption{\textbf{Timeline of key developments in bipartite graph embedding recommendation.}}
\label{Ch4_Summary}
\end{figure}

The key developments of bipartite graph embedding-based recommendation models retrospected in this section are concisely summarized in Fig.~\ref{Ch4_Summary}, among which it is overt that the matrix factorization framework has long been concerned as the most extensible one succeeding various subsequent variants. However, the current focus has substituted the deep learning framework for the matrix factorization framework, for the advantages in non-linear pattern discovery and parallel computing of the former one is meeting the technical requirements of increasing complexity and scale of data for recommendation. As for the Bayesian analysis framework, its breakthrough runs through the entire timeline, indicating its indispensable contributes to the other methods for automatic hyper-parameter adjustment. On the other hand, the Markov method does not seem to be common in use.

\section{General graph embedding for recommendation}
\label{GGE}

As illustrated in Sec.~\ref{S2}, employing side information (like the properties of users and items) in uncovering hidden (indirect) user-item relations can alleviate the cold start and sparsity problems, which contributes to the recommendation accuracy. However, since side information is usually represented by general graphs which are far beyond bipartite graphs in complexity, the matrix factorization method as well as other methods designed for bipartite graph embedding face fundamental limits on the practice of the recommendation involving side information. To make up for the flaw, until recently, researchers have tried to extend the matrix factorization method for general graph embedding, like methods of collective matrix factorization \cite{nickel2011three,singh2008relational} and spectral \cite{belkin2001laplacian,tenenbaum2000global,cox2008multidimensional,jenatton2012latent}. Although being feasible, these new methods still cannot run efficiently enough on large-scale data attributed to their high computing complexity.

In recent research, in order to uplift the model scalability on general graphs with large-scale, general graph embedding techniques have been developed. In the first place, Sec.~\ref{Fundamental-Technologies} divides these techniques into three categories by those based on methods of translation, meta path and deep learning. Based on that, Sec.~\ref{RSwithSideInfo} retrospects their corresponding application in recommendation from two different perspectives: technique-oriented and scenario-oriented.

\subsection{Three categories of techniques for general graph embedding}
\label{Fundamental-Technologies}

Techniques of general graph embedding can be divided into three categories: those based on methods of translation, meta path and deep learning. From an overview, built on algebraic theory, the translation-based techniques have come to see the merit of being able to sufficiently preserve local topological features in a graph. To further capture a graph's global topological features, the Meta path-based techniques are run by random walking across nodes. Being similar in mechanism to the deep learning-based models for bipartite graph embedding, the deep learning techniques can also capture and preserve the non-linear topological features hidden in general graphs.

\subsubsection{Techniques based on translation}
\label{Trans-section}

The triplets $(h,r,t)$ illustrated in Sec.~\ref{graph_representation} describes a relation $r$ from the head node (or entity) $h$ to the tail node $t$, whose embeddings are denoted by $\bm{h},\bm{r}$ and $\bm{t}$, respectively.

From the perspective of algebraic theory, TransE \cite{bordes2013translating} takes $\bm{r}$ as a translation from $\bm{h}$ to $\bm{t}$ in a metric space, which satisfies that when $(h,r,t)$ existed $\bm{h}+\bm{r} \approx \bm{t}$ indicating that $\bm{t}$ is supposed to be one of the nearest neighbors of $\bm{h}+\bm{r}$ while $\bm{h}+\bm{r}$ should be far away from $\bm{t}$ otherwise. However, in practice, a node shared by multiple communities in a general graph could play different roles like, for example, that a head node $h$ could have positive impacts on other tail nodes in a community (\textit{e.g.}, a boss could bring interests to the staffs of his own incorporation) but may have negative impacts on another community (\textit{e.g.}, the boss could earn some profits from the staffs of competitive incorporation), in which case $h$ should be denoted by $h_0$ and $h_1$ oriented to its different roles (\textit{i.e.}, impacts), respectively. Consequently, representing $h_0$ and $h_1$ by a common embedding $\bm{h}$ as TransE does would not be able to distinguish the different practical meanings. TransH \cite{wang2014knowledge} gives clues to out of this dilemma. Based on the assumption that a node's distinctive roles of different communities in a general graph could be revealed (or represented) by its diverse relations with other nodes (\textit{i.e.}, the positive relations of the node with other nodes could reveal its positive role in the related community and vice versa), TransH maps each pair of $(h, t)$ to multiple relation-specific hyper-planes $\bm{w_r}$, which are used to represent the diverse relations between nodes. Analogously, in a general graph (especially a heterogeneous graph), a relation between two nodes could also play multiple roles. For example, a triplet (location, contains, location) can be interpreted by multiple semantics such as country-contains-city, country-contains-university or something. Consequently, representing the relation by one embedding $\bm{r}$ is insufficient in distinguishing such abundant relational semantics. In this regard, TransR \cite{lin2015learning} separately maps nodes and relations into a node space and different relation spaces corresponding to the diverse relational semantics between head-tail pairs, respectively. Moreover, on the advice of the diverse semantics of a specific relation $r$, CTransR \cite{lin2015learning} clusters $r$'s linked node pairs $(h, t)$ into multiple groups, corresponding to the different semantics of $r$. In these ways, TransR and CTransR can preserve multiplex semantics of nodes and relations in a general graph. Tab.~\ref{Trans} gives details of these techniques.

Besides, in a general graph, the multiplicity (\textit{i.e.}, diverse types or appended attributes) of nodes and relations is far beyond their multiple semantics, which plays a large role in representing a general graph as precisely as possible. In view of that, TransD \cite{ji2015knowledge} takes the multiple types of nodes and relations into account. Ji et al. \cite{ji2016knowledge} proposed that relations could belong to different graph patterns like that some of them are linked with a large number of node pairs while others may not. In addition, Ji et al. also discovered that relations could have uneven balances that the quantity of node pairs linked with relations could differ a lot. Moreover, by means of the Chinese restaurant process (CRP) \cite{blei2010nested}, TransG \cite{xiao2015transg} can cluster the semantic components $\pi_{r,m}$ in $(h,r,t)$ of nodes and relations. Tab.~\ref{Trans} gives details of these techniques.

An efficient objective function for the above translation-based techniques is
\begin{equation}
L=\sum_{(h,r,t)\in S}\sum_{(h',r',t')\in S'}[f_r(\bm{h},\bm{t})+\gamma-f_{r}(\bm{h'},\bm{t'})]_{+},
\end{equation}
which can be run by minimizing a margin-based ranking criterion in optimization, where $[x]_{+}$ denotes the positive part of $x$, and $\gamma>0$ is a margin hyper-parameter. $f_r(\bm{h}, \bm{t})$ is the loss function such as those shown in Tab.~\ref{Trans}. $S=\{(h,r,t)|h,t \in E\}$ is the golden triplets set or named training samples, and $\displaystyle S'=\{(h',r,t)|h' \in E\}\cup \{(h,r,t')|t' \in E\}$ is the negative triplets set or named negative samples, whose efficiency could largely influence the training process, which is generally constructed by replacing either the head node or the tail node with a negative one selected by random (but usually not both at the same time).

Nevertheless, since built on the mechanism of algebraic transformation on local topology, the above translation-based techniques could lose the global topological features in a general graph. In this case, researchers resorted to deepening the proximity order of transformation as a strategy for capturing and preserving global topology under the translation-based framework. For instance, by building a path space used to preserve relational semantics, RPE \cite{lin2019relation} can measure the higher-order proximity of non-adjacent node pairs connected by paths in the space. However, a limitation of extending the translation-based framework to higher-order ones is clear: high computing complexity. In order to reach higher efficiency in representing the global topological features in a general graph, meta path-based techniques are hitherto prevalent, as illustrated in the next section.

\begin{table}[h]
\tiny
\setlength{\abovecaptionskip}{0.3cm}
\setlength{\belowcaptionskip}{0.3cm}
\caption{\textbf{Examples of modeling translation-based techniques} (1) In TransH, after being mapped to a hyper-plane $\bm{w_r}$, the embeddings of nodes $h$ and $t$ in a pair $(h,t)$ are denoted by $\bm{h_{\bot}}$ and $\bm{t_{\bot}}$, respectively. Correspondingly, the embedding of the relation between them on $\bm{w_r}$ with a specific meaning is denoted by $\bm{d_r}$. (2) In TransR, $\bm{M_r}$ is a transformation matrix used to map a head-tail pair with a relation $r$ into different relation-specific spaces corresponding to distinctive relationsl semantics. (3) In TransD, $\bm{h_p}, \bm{t_p} \in \mathbb{R}^n$ represent the semantics of head node and tail node, respectively, and $\bm{r_p} \in \mathbb{R}^m$ represents the semantics of the relation between them. $\bm{M_{rh}}, \bm{M_{rt}} \in \mathbb{R}^{m \times n}$ are two transformation matrices. (4) In TranSparse, ${\theta}$ measures the sparse degree of a matrix, recording the fraction of zero elements over the total number of elements. For each relation $\bm{r}$, a sparse transfer matrix $\bm{M_r}(\theta) \in \mathbb{R}^{m \times n}$ is constructed. $0 \leq \theta_{\min} \leq 1$ is a hyper-parameter, which denotes the minimum sparse degree, $N_r$ denotes the number of node pairs linked by the relation $r$, and $N_{r^*}$ denotes the maximum number in $N_r$.}
\label{Trans}
\begin{adjustbox}{center}
\begin{tabular}{ccc}
\toprule \addlinespace \textbf{Technique} & \textbf{Mapping} & \textbf{Loss function}\\ \addlinespace

\midrule \addlinespace TransE & none  &  $f_r(\bm{h},\bm{t})=\|\bm{h}+\bm{r}-\bm{t}\|_2^2$ \\ \addlinespace

\addlinespace TransH & $\displaystyle \bm{h}_{\bot}=\bm{h}-\bm{w}_r^T\bm{h}\bm{w}_r, \bm{t}_{\bot}=\bm{t}-\bm{w}_r^T\bm{t}\bm{w}_r$ & $\displaystyle f_r(\bm{h},\bm{t})=\|\bm{h}_{\bot}+\bm{r}-\bm{t}_{\bot}\|_2^2$ \\ \addlinespace

\addlinespace TransR & $\bm{h}_r=\bm{h} \bm{M}_r, \bm{t}_r=\bm{t} \bm{M}_r$ & $f_r(\bm{h},\bm{t})=\|\bm{h}_r+\bm{r}-\bm{t}_r\|_2^2$ \\ \addlinespace

\addlinespace CTransR & $\bm{h}_{r,c}=\bm{h} \bm{M}_r, \bm{t}_{r,c}=\bm{t} \bm{M}_r$ & $f_r(\bm{h},\bm{t})=\|\bm{h}_{r,c}+\bm{r_c}-\bm{t}_{r,c}\|_2^2+\alpha \|\bm{r}_c-\bm{r}\|_2^2$ \\ \addlinespace

\addlinespace TransD & \tabincell{c}{$ \displaystyle \bm{M}_{rh}=\bm{r}_p \bm{h}_p^T + \bm{I} ^{m \times n}, \bm{M}_{rt}=\bm{r}_p \bm{t}_p^T + \bm{I} ^{m \times n}$\\ $\displaystyle \bm{h}_{\bot}=\bm{M}_{rh} \bm{h}, \bm{t}_{\bot}=\bm{M}_{rt} \bm{t}$} & $\displaystyle f_r(\bm{h},\bm{t})=-\|\bm{h}_{\bot}+\bm{r}-\bm{t}_{\bot}\|_2^2$ \\  \addlinespace

\addlinespace TranSparse(share) & \tabincell{c}{$ \displaystyle \theta_r=1-(1-\theta_{\min})N_r/N_{r^*}$\\  $\bm{h}_p=\bm{M}_r(\theta_r)\bm{h}, \bm{t}_p=\bm{M}_r(\theta_r) \bm{t}$} & $f_r(\bm{h},\bm{t})=\|\bm{h}_p+\bm{r}-\bm{t}_p\|_2^2$ \\ \addlinespace

\addlinespace TranSparse(separate) & \tabincell{c}{$\displaystyle \theta_r^l=1-(1-\theta_{\min}N_r^l/N_{r^*})^{l^*} (l=h,t)$\\ $ \bm{h}_p=\bm{M}_r^h(\theta_r^h)\bm{h}, \bm{t}_p=\bm{M}_r^t(\theta_r^t)\bm{t}$} &$f_r(\bm{h},\bm{t})=\|\bm{h}_p+\bm{r}-\bm{t}_p\|_2^2$ \\ \addlinespace

\addlinespace TransG & see \cite{xiao2015transg} & \tabincell{c}{ $\displaystyle P\{(h,r,t)\} \propto \sum_{m=1}^{M_r} \pi_{r,m} P(\bm{u_{r,m}}|h,t)$ \\ $\displaystyle =\sum_{m=1}^{M_r}\pi_{r,m} e^{-\frac{\|\bm{u_h}+\bm{u_{r,m}}-\bm{u_t}\|_2^2}{\sigma_h^2+\sigma_t^2}}$} \\ \addlinespace

\bottomrule
\end{tabular}
\end{adjustbox}
\end{table}

\subsubsection{Techniques based on meta path}
\label{Meta-path-section}

\begin{figure}[!ht]
\centering
\includegraphics[scale=0.33]{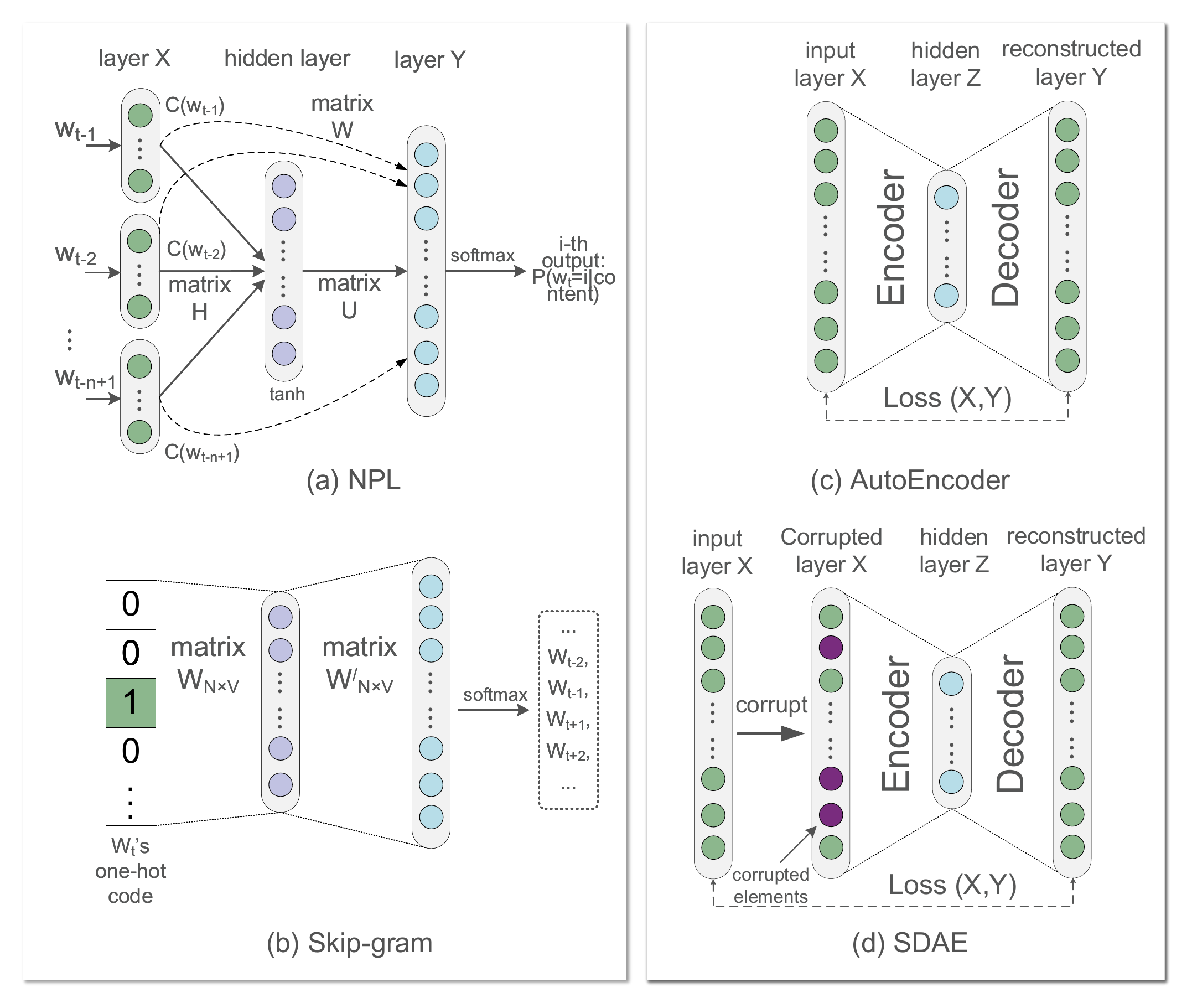}
\hspace{0.1in}
\caption{\textbf{Schematics of NPL, Skip-gram, AutoEncoder and SDAE.} In (a), a legal sentence $\displaystyle w_t,w_{t-1},...,w_{t-n+1}$ is made up with $n$ sequential words, where $w_*$ records a word's index in a corpus. NPL takes the sentence as a training sample and transforms it into $\displaystyle x=(C(w_{t-1}),C(w_{t-2}),...,C(w_{t-n+1}))$, where $C(w_*)$ is the one-hot code of $w_*$. The training process can be run by maximizing $\displaystyle P(w_t=\alpha|w_{t-1},...,w_{t-n+1})=\operatorname{softmax}(b+Wx+U \tanh (d+Hx))$, which represents the probability of $w_t$'s index to be $\alpha$. In (b), from a reverse perspective, by taking $w_t$ as the input, Skip-gram can reconstruct $w_t$'s surrounding context as the output. In (c), according to different tasks, the Encoder can be flexibly constructed, such as by an LSTM \cite{hochreiter1997long} in a task of machine translation (MT) or by a CNN \cite{rakhlin2016convolutional} in a task of computer vision (CV). Since it could largely determine the precision of learned embedding in $Z$ as well as the lower bound of reconstruction error between $X$ and $Y$, appropriately constructing the Encoder is an important step. In (d), diverse strategies can be adopted in corrupting $X$ to $\tilde{X}$, like Additive isotropic Gaussian noise (GS), Masking noise (MN) and Salt-and-pepper noise (SP), among which SP is only for binary data.}
\label{graph_set_two}
\end{figure}

Meta path-based techniques were inspired from the basic thought of word representation \cite{naseem2021comprehensive} in natural language processing (NLP). Given $N$ words chosen from a corpus containing hundreds of millions of words, a legal sentence like $w_1,w_2,...,w_N$ ($w_*$ represents a word) with effective meanings can be made up with them. Since words appearing in the same context of a sentence are hypothesized to be similar in meanings, the representations (i.e., embeddings) of these words are supposed to be close in proximity. Methodologically, by taking legal sentences as training samples, NPL \cite{bengio2003neural} can learn the representation of a given word, which could involve the word's context (Fig.~\ref{graph_set_two} gives details). From a reverse perspective, Skip-gram \cite{mikolov2013efficient} aims to learn the representations of words in a given context around the context's central word, which could capture and preserve the linguistic patterns among these words (Fig.~\ref{graph_set_two} gives details). Compared with NPL, Skip-gram appears to be more available for large-scale data. Furthermore, by using hierarchical softmax \cite{morin2005hierarchical} to identify phrases, Word2vec \cite{mikolov2013distributed} can extend Skip-gram from a word-based model to a phrase-based one. Tab.~\ref{MetaPath} gives details of these models.

In the same way, if taking a graph as the analogy of a corpus, the thought of word representation can be applied in graph embedding. Given $N$ nodes picked up from a graph by random walking which starts or ends at an arbitrary node, a nodal sequence can be made up with them. After repeating random walking multiple times, a set of nodal sequences can be generated. Much as words in sentences, the nodes appearing in a nodal sequence with higher frequency are hypothesized to be represented by embeddings with closer proximity. The feasibility of this transferred thought was firstly proven by DeepWalk \cite{perozzi2014deepwalk,yang2015comprehend}. In detail, Deepwalk adopts the depth-first searching strategy for generating random walks $\mathcal{W}_{v_r}=(\mathcal{W}_{v_r}^1, \mathcal{W}_{v_r}^2,...,\mathcal{W}_{v_r}^k)$ rooted at an arbitrary node $v_{r}$, where $\mathcal{W}_{v_r}^{k+1}$ is a node randomly picked up from $\mathcal{W}_{v_r}^k$'s neighbors. In reality, path $\mathcal{W}_{v_r}$ can be recognized as a particular ``sentence''. After denoting the embedding of each node $v_i \in \mathcal{W}_{v_r}$ by $\bm{v_i} \in \mathbb{R}^d$ and denoting its surrounding context by $v_j \in \mathcal{W}_{v_r}[i-w:i+w]$ (where the window size is $2 \omega+1$), the training process can be run by maximizing $P(v_j|v_i)$ with hierarchical softmax \cite{morin2005hierarchical} (Tab.~\ref{MetaPath} gives details). So this way, DeepWalk can well capture and preserve the high-order proximity between nodes. On the other hand, by adopting the breadth-first searching strategy for generating random walks, LINE \cite{tang2015line} can concentrate on the first-order and second-order proximity between nodes, which is generally called ``WideWalk'' (Tab.~\ref{MetaPath} gives details).

However, being in sharp contrast to word representation whose legitimacy of constructed sentences generally can be guaranteed by human linguistic knowledge, the legitimacy of random walks still lacks a recognized examination standard, which could hurt the accuracy of learned embeddings since inaccurate or even incorrect random walks are insufficient to capture a graph's global topology. Under this condition, based on the frameworks of DeepWalk or LINE, designing more intelligent random walking rules has been the focus of subsequent research into meta path-based techniques. For instance, by defining a more flexible notion of node's neighbors, Node2vec \cite{grover2016node2vec} equips random walking with abilities in identifying nodes of a common community or of similar roles in a general graph. In detail, Node2vec designs a biased random walking rule guided by a return parameter $p$ and an in-out parameter $q$ as
\begin{equation}
\label{Node2vec}
P(v_i|v_{i-1})=\frac{1}{Z} \alpha_{pq}(i-2,i) \cdot \omega_{i-1,i}, \quad \alpha_{pq}(i-2,i)=
\left \{ \begin{aligned}
&1/p & \operatorname{if} d_{i-2,i}=0, \\
&1 & \operatorname{if}  d_{i-2,i}=1, \\
&1/q  &\operatorname{if}  d_{i-2,i}=2,
\end{aligned} \right.
\end{equation}
where $v_{i-1}$ walks from $v_{i-2}$, $v_i$ is a neighbor of $v_{i-1}$, $d_{i-2,i}$ is the shortest path between $v_{i-2}$ and $v_i$, $Z$ is a normalization constant, and $\omega_{i-1,i}$ is the weight of edge $(i-1,i)$ (Tab.~\ref{MetaPath} gives details). Moreover, as for designing new random walking rules from a deeper perspective in algebra, NetMF \cite{qiu2018network} unifies DeepWalk, LINE and Node2vec into a matrix factorization framework.

\begin{table}[!ht]
\tiny
\setlength{\abovecaptionskip}{0.3cm}
\setlength{\belowcaptionskip}{0.3cm}
\caption{\textbf{Examples of modeling meta path-based techniques.} (1) In Skip-gram, $w_O$ and $w_I$ denote a word $w$'s input and output embedding, respectively. $N$ is the number of words in a corpus, $L$ is the number of words in a sentence (\textit{i.e.}, a training sample) and $c$ is the size of a context. (2) In Word2vec, $N'$ is the number of negative samples, and $P_n(w)$ is a noise distribution. (3) In LINE, the embedding of $v$ is denoted by $\bm{v}$ when treating $v$ as a vertex while by $\bm{v}'$ when a ``context''. (4) In PTE, nodes in a bipartite graph are divided into two sets $V_A$ and $V_B$. $P(v_j|v_i)$ denotes the conditional probability that node $v_j$ in set $V_B$ is walked from node $v_i$ in set $V_A$.}
\label{MetaPath}
\begin{adjustbox}{center}
\begin{tabular}{ccc}
\toprule  \addlinespace \textbf{Technique} & \textbf{Random walk strategy} & \textbf{Loss function}  \\ \addlinespace

\midrule \addlinespace Skip-gram & depth-first search & $\displaystyle P(w_O|w_I)=\frac{\exp(\bm{w_O}^T \bm{w_I})}{\sum_{k=1}^{N} \exp(\bm{w_{kO}}^T \bm{w_{kI})}}$ \\ \addlinespace

\addlinespace Word2vec &depth-first search  & $\displaystyle P(w_O|w_I)=\log \sigma(\bm{w_O}^T \bm{w_I})+\sum_{k=1}^{N'} \mathbb{E}_{w_k~ \sim P_n(w)}[\log \sigma(-\bm{w_{kO}}^T \bm{w_I)}]$ \\ \addlinespace

\addlinespace DeepWalk &depth-first search & $\displaystyle P(u_j|v_i)=\prod_{l=1}^{\lceil \log |V| \rceil} P(b_l|v_i)$ \\\addlinespace

\addlinespace LINE & breadth-first search &  \tabincell{c}{$\displaystyle P_1(v_i,v_j)=\frac{1}{1+\exp(-\bm{v_i} \cdot \bm{v_j})},$ \\$ P_2(v_j|v_i)=\log \sigma(\bm{v_j}^{'T} \cdot \bm{v_i})+\sum_{k=1}^{N'}\mathbb{E}_{v_k \sim P_n(v)}[\log \sigma(-\bm{v_k}^{'T} \cdot \bm{v_i})] $ }\\ \addlinespace

\addlinespace PTE &  breadth-first search & $\displaystyle P(v_j|v_i)=\frac{\exp(\bm{v_j}^T \cdot \bm{v_i})}{\sum_{k \in V_B}\exp(\bm{v_k}^T \cdot \bm{v_i})}$    \\ \addlinespace

\addlinespace Node2vec & see Eq.~(\ref{Node2vec}) & $\displaystyle P(v_j|v_i)=\frac{\exp(\bm{v_j}^T \cdot \bm{v_i})}{\sum_{k \in V}\exp (\bm{v_k}^T\cdot \bm{v_i})}$ \\ \addlinespace

\addlinespace Metapath2vec & see Eq.~(\ref{Metapath2vec}) &  $\displaystyle P(v_j|v_i)=\frac{\exp{(\bm{v_j}^T \bm{v_i}})}{\sum_{k \in V}\exp{(\bm{v_k}^T \bm{v_i}})}$  \\ \addlinespace

\addlinespace Metapath2vec++ & see Eq.~(\ref{Metapath2vec}) &   $\displaystyle P(v_{j_t}|v_i)=\frac{\exp{(\bm{v_{j_t}}^T \bm{v_i}})}{\sum_{k_t \in V_t}\exp{(\bm{v_{k_t}}^T \bm{v_i}})}$  \\ \addlinespace
\bottomrule
\end{tabular}
\end{adjustbox}
\end{table}

In practice, as illustrated in Sec.~\ref{Trans-section}, nodes and relations in general graphs carry diverse semantics, types and attributes. However, the above techniques are basically oriented to homogeneous graphs. In order to capture and preserve the multiplicity of heterogeneous graphs (most general graphs are heterogeneous ones), for instance, PTE \cite{tang2015pte} extends LINE to be available on bipartite graphs (Tab.~\ref{MetaPath} gives details). In addition, by decomposing a bipartite graph into two homogeneous graphs, BiNE \cite{gao2018bine} designs a random walking rule of ``rich nodes are getting richer'', in order to satisfy the power-law distribution phenomenon. Recently, by means of manually built meta paths \cite{hussein2018meta, wang2020survey} based on expert knowledge, designing random walking rules on heterogeneous graphs has been a promising solution, which aims to cover the multiplicity by meta paths as sufficient as possible. HIN2Vec \cite{fu2017hin2vec} is a case in point. By capturing multiple relational types between nodes, HIN2Vec can jointly learn the embeddings of nodes based on training samples $\langle u,v,r,L(u,v,r) \rangle$, which indicate that a relation $r$ exists between $u$ and $v$ when $\langle u,v,r \rangle=1$ and vice versa. Moreover, during random walking, Metapath2vec/ Metapath2vec++ \cite{dong2017metapath2vec} can capture the diverse nodal and relational types in meta paths by defining a translation probability between nodes as
\begin{equation}
\label{Metapath2vec}
P(v_{i}|v_{i-1}^{t-1})=
\left\{ \begin{aligned}
&\frac{1}{|N_t(v_{i-1}^{t-1})|} & (v_{i},v_{i-1}^{t-1})\in E, \,&\phi(v_i)=t,\\
&0 & (v_i,v_{i-1}^{t-1})\in E, \, &\phi(v_i) \neq t,\\
&0 & (v_i, v_{i-1}^{t-1})\notin E.
\end{aligned} \right.
\end{equation}
Where $v_{i-1}^{t-1} \in V_{t-1}$, $N_t(v_{i-1}^{t-1})$ denotes the $V_t$ type of $v_{i-1}^{t-1}$'s neighborhood and $v_i \in V_t$. Besides, there are also techniques \cite{chang2015heterogeneous, gui2016large, shang2016meta} used to represent the multiplicity of relations in general graphs.

\subsubsection{Techniques based on deep learning}
\label{DL-KG}

As illustrated in Sec.~\ref{DL-Bi} that deep learning methods are equipped with abilities in capturing the non-linear features in a bipartite graph, by the same token it is also true of those in a general graph. Among them, AutoEncoder \cite{rumelhart1985learning} is a representative one, an unsupervised deep learning framework, which is different from those based on supervised deep learning frameworks in Sec.~\ref{DL-Bi}. In detail, as shown in Fig.~\ref{graph_set_two}, AutoEncoder consists of two components: the Encoder and the Decoder. The Encoder can learn the input $X$'s embedding, which will be stored in the hidden layer $Z$. Then, taking the learned embedding as input, the Decoder is used to reconstruct $X$ by $Y$, which is supposed to be approximate to $X$ as much as possible, by minimizing the reconstruction error between $X$ and $Y$. Subsequently, this framework has matured as different variants. For instance, by corrupting the input $X$ to $\tilde{X}$ and minimizing the reconstruction error between $\tilde{X}$ and $Y$, Vincent et al. \cite{vincent2010stacked} can enhance the robustness of the learned embedding (\textit{i.e.}, $Z$) (Fig.~\ref{graph_set_two} gives details). Salakhutdinov et al. \cite{salakhutdinov2009semantic} proposed a multi-layer version of AutoEncoder named Deep AutoEncoder. More variants of AutoEncoder include SCAE \cite{masci2011stacked}, generalized AutoEncoder (GAE) \cite{wang2014generalized}, variational AutoEncoders (VAEs) \cite{rezende2014stochastic} and deep hierarchical variational AutoEncoder (Nvae) \cite{vahdat2020nvae}, to name a few. When it comes to generalizing the AutoEncoder framework and its variants to general graph embedding, SDNE \cite{wang2016structural} provides a semi-supervised deep model, preserving the second-order proximity between node pairs by reconstructing their common neighborhoods with two deep AutoEncoders sharing common parameters. Meanwhile, SDNE can also preserve the first-order proximity between nodes by using a  Laplacian Eigenmaps-based supervised component.

\begin{figure}[!ht]
\centering
\includegraphics[scale=0.3]{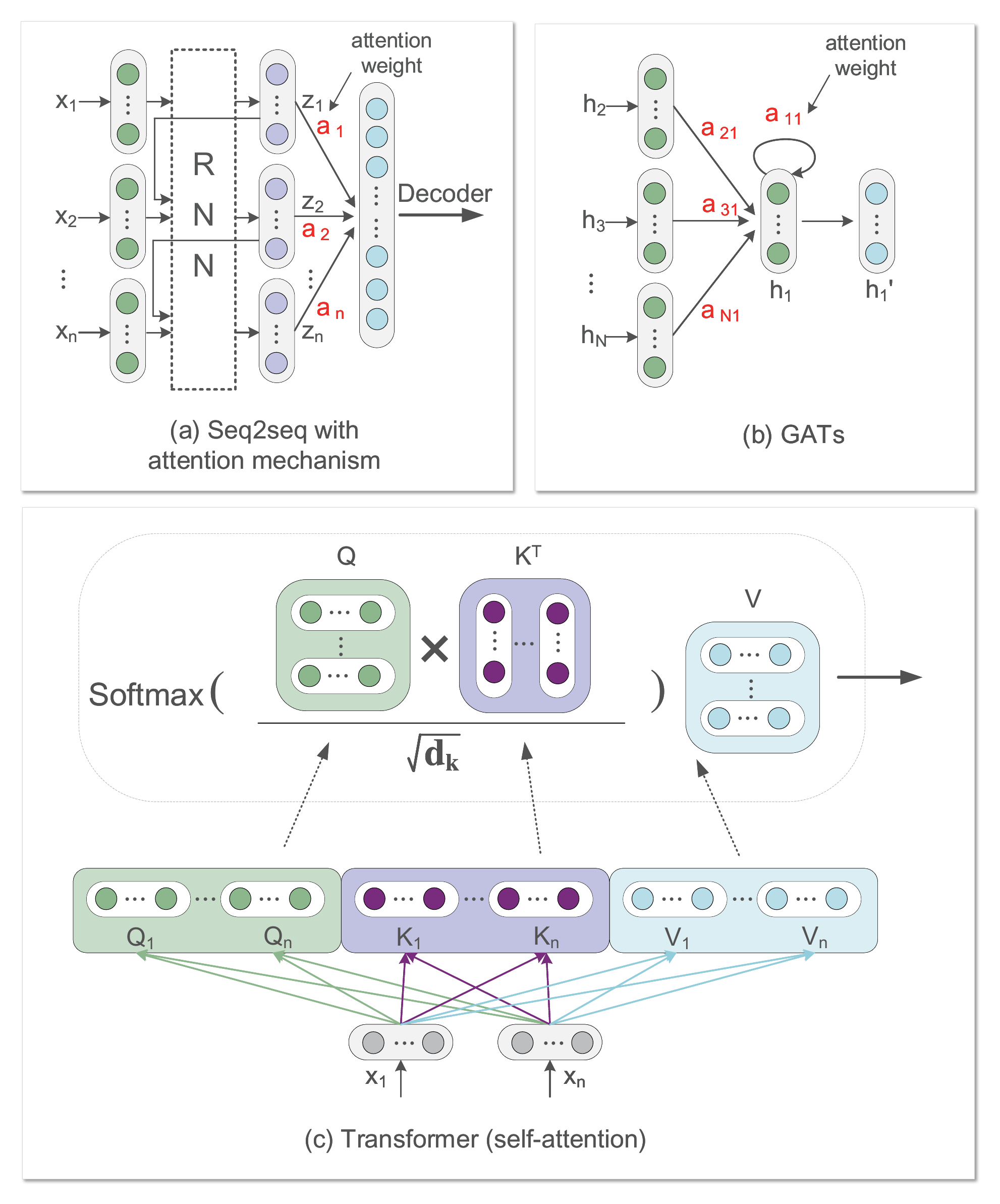}
\hspace{0.1in}
\caption{\textbf{Schematics of Seq2seq with attention mechanism, Transformer(self-attention) and GATs.} In (a), as an Encoder, Seq2seq can learn the embedding $Z$ with a fixed dimension compressed from $z_1,...,z_n$ (generated from the input $x_1,...,x_n$) by minimizing $p(y_1,...,y_{T'}|x_1,...,x_T)= \prod_{t=1}^{T'}p(y_t|Z,y_1,...,y_{t-1})$, where $X=(x_1,..,x_T)$ is the input, $y_1,...,y_{T'}$ is the output, and $T$ is generally not equal to $T'$. After incorporating the attention mechanism into Seq2seq, attention weights can be learned by $a\sim p(a|K,Q)$, where $K$ represents the input $X$ and $Q$ is the result of the last iteration in the Decoder. Different from the pure Seq2seq where the output $Z$ is directly compressed from $z_1,...,z_n$, in Seq2seq with attention mechanism the importance of $z_*$ can be assigned by attention weights $a_*$. In (b), as for a node $i$, GATs takes the node's neighborhood $h_* \in \mathbb{R}^F$ as the input, based on which the attention weights between node $i$ and its neighbors can be learned by $a_{ij}=\frac{\exp (\operatorname{LeakyReLU(a^T[\emph{W} h_i || \emph{W} h_j])})}{\sum_{k \in N_i}\exp (\operatorname{LeakyReLU(a^T[\emph{W} h_i || \emph{W} h_k])})}$, where $W \in \mathbb{R}^{F' \times F}$ is a shared weight matrix, $a \in \mathbb{R}^{2F'}$ is a weight vector, $N_i$ is some neighborhood of node $i$, and $||$ is the concatenation operator. After that, the embedding of the input $h_i' \in \mathbb{R}^{F'}$ can be learned by $h_i=\sigma(\sum_{j\in N_i} a_{ij}W h_j)$. In (c), the input $X=(x_1,...,x_n)$ is firstly mapped into $Q_*,K_*$ and $V_*$ through transformation matrices $W^Q, W^K$ and $W^V$, respectively, as model parameters. Based on them, $z_1,...,z_n$ can be generated by $Z=\operatorname{softmax} (\frac{Q K^T}{\sqrt{d_k}})V$, where $d_k$ is the dimension of $Q_*$ and $K_*$. $Q,K$ and $V$ are all vectors, among which $K$ is equal to $V$.}
\label{graph_set_three}
\end{figure}

However, one fundamental limitation of the AutoEncoder framework is that the dimension of hidden layer $Z$ is fixed, and so are its variants. In truth, if the input $X$'s dimension is far higher than that of $Z$, from $X$ to $Z$ essential information could be lost through the Encoder, which is insufficient to represent the original $X$. For instance, as shown in Fig.~\ref{graph_set_three}, if the input sequence $(x_1,x_2,...,x_n)$ of Seq2seq \cite{sutskever2014sequence} is high in dimension, compressing $z_1,z_2,...,z_n$ (generated from $(x_1,x_2,...,x_n)$ correspondingly) into an embedding with fixed dimension could lose essential patterns of correlation among $X$'s elements. To deal with the issue, by means of the attention mechanism \cite{bahdanau2014neural, cho2015describing, lee2019attention, hu2019introductory}, attention weights $a_*$ for $z_*$ of Seq2seq can be learned, used to quantify $z_*$'s importance. In other words, Seq2seq with attention mechanism aims to distinguish the (generally different) importance of $z_*$ by learning their respective attention weights $a_*$, in order to pick up the most representative ones which can carry most of $X$'s information and to represent them with an embedding as output. This way enables the learned embedding to maximally represent the primary information of $X$ with a limited dimension. Methodologically, attention weights can be learned by an attention function mapping a query $Q$ and a set of key-value pairs $K, V$ to an output (Fig.~\ref{graph_set_three} gives details). Subsequently, Kim et al. \cite{kim2017structured} proposed a structured attention network (SAN), which can take the structured dependency of the attention layer into account. By substituting a self-attention component used to learn the attention weights for the RNN structure in the Encoder, Transformer \cite{vaswani2017attention} can support parallel computing (Fig.~\ref{graph_set_three} gives details). The attention and self-attention mechanisms can also be applied to the AutoEncoder framework-based general graph embedding. For instance, GATs \cite{velivckovic2017graph} can construct a masked self-attention block layer for graph convolution (Fig.~\ref{graph_set_three} gives details).

\subsection{Recommendation involving side information}
\label{RSwithSideInfo}

This section retrospects recommendation models involving side information from two different perspectives: technique-oriented and scenario-oriented.

Firstly, in technique, extracting properties of users and items from general graphs and incorporating them into recommendation can be implemented by the three categories of techniques retrospected in Sec.~\ref{Fundamental-Technologies}. Based on the translation-based general graph embedding framework, for instance, TransRec \cite{he2017translation} can generalize TransE to sequential recommendation, in a manner that regarding a user as the translation between the pairs of items he sequentially interacted during a certain period (of course these item pairs are adjacent in the timeline in terms of this user's interactions), which means that as for a user embedding $\stackrel{\longrightarrow}{\mathbf{\operatorname{user}}}$ the relation $\displaystyle \stackrel{\longrightarrow}{\mathbf{\operatorname{prev.item}}}+\stackrel{\longrightarrow}{\mathbf{\operatorname{user}}} \,\approx \, \stackrel{\longrightarrow}{\mathbf{\operatorname{next.item}}}$ should always hold between the embedding of the user's previous interacted item $\stackrel{\longrightarrow}{\mathbf{\operatorname{prev.item}}}$ and that of his next (or sequential) interacted items $\stackrel{\longrightarrow}{\mathbf{\operatorname{next.item}}}$. However, TransRec inherits the defect of TransE that only be suited for $1$-to-$1$ relations (\textit{i.e.}, take different relational types as the same one). In practice, since user's next interacted items are usually diverse in types, distinguishing the diverse relational types between these item pairs is supposed to be a necessity. By building multiple semantic-specific matrices for the transition between distinctive item types, CTransRec \cite{li2019translation} can further represent the different relational types between item pairs, extending TransRec to the $1$-to-$n$ relational translation.

In order to employ the meta path-based general graph embedding framework into recommendation, a common-used strategy is to base a user-item proximity matrix on random walks, which will be used as the input of the matrix factorization framework to implement recommendation. Take HIN \cite{yu2014personalized} for instance. In a general graph, HIN hypothesizes that a user's preferences for items could be visualized by random walking along designed meta paths from this user to items within a boundary centered on this user. Under this assumption, the accumulated frequency of passes through an item on the user's random walks could be recognized as the item's proximity with the user, which could further be used to quantify the user's preference for this item. Intuitively, the higher the accumulated frequency is, the more indirect relations (\textit{i.e.}, paths) between the user-item pair, so the more possible that an unobserved interaction exists between them. After implementing the diffusion mechanism on each user, a global user-item proximity matrix can be built, based on which the matrix factorization will run for recommendation. Similarly, by generating multiple local user-item proximity matrices corresponding to different meta paths, FMG \cite{zhao2017meta} can learn multiple distributed embeddings for each user-item pair. These embeddings will be used as the input of FM \cite{rendle2010factorization}, which enables FMG to represent the feature interactions between inter-meta paths. In the practice of various recommendation scenarios, under the meta path-based framework, designing efficient meta paths which could maximally capture the diverse characteristics of a general graph to rule random walking directly determines a model's recommendation accuracy. At this point, for instance, by remaining only the nodes of users and items in meta paths, Shi et al. \cite{shi2018heterogeneous} designed a graph schema used to distinguish the two nodal types during random walking. However, manually designing meta paths could require much expert knowledge. In order to automatically generate meta paths by random walking without hand-craft design, by triggering multiple ``ripples'' from user's historical interacted items as centers, RippleNet \cite{wang2018ripplenet} can stimulate user's preference diffusion in a general graph along hierarchical propagation traces as multiple meta paths. Besides, by using an LSTM layer \cite{hochreiter1997long} to identify the holistic semantic of meta paths, KPRN \cite{wang2019explainable} can realize reasoning on meta paths. Chen et al. \cite{chen2021temporal} further took into account the temporal factors in general graphs and proposed a temporal meta path guided explainable recommendation (TMER).

When it comes to applying the deep learning-based general graph embedding framework into recommendation, the AutoEncoder framework motivated a variety of recommendation models. For instance, according to explicit user-item interactions (\textit{i.e.}, ratings, where the unobserved ones are represented by value $0$), AutoRec \cite{sedhain2015autorec} firstly builds a rating vector with a dimension equaling the number of all items. Then, this rating vector will be used as the input of AutoEncoder, which can complete the missing elements (\textit{i.e.}, ratings $0$) in the input as predicted ratings by reconstructing it. In terms of recommendation based on implicit user-item interactions, by assuming that user's preferences for items should be represented with the corrupted layer of SDAE because of the possible incompleteness of one's observed implicit interactions, Wu et al. \cite{wu2016collaborative} applied SDAE to recommendation. Besides, Liang et al. \cite{liang2018variational} firstly applied variational AutoEncoders (VAEs) \cite{rezende2014stochastic} to recommendation. As a Bayesian version of VAEs, CVAE \cite{li2017collaborative} can simultaneously employ explicit user-item interactions as well as item's profiles into recommendation. Similar in being equipped with the attention and self-attention mechanisms to the AutoEncoder framework, recommendation models can also be incorporated with them, For instance, by means of the attention mechanism, AFM \cite{xiao2017attentional} can weigh the feature interactions in FM \cite{rendle2010factorization}. Xu et al. \cite{xu2021long} proposed a multi-layered long- and short-term self-attention network (LSSA) for sequential recommendation.

It is positive that the above thoughts of employing general graph embedding techniques into recommendation have also made successful attempts in the industry. For instance, by taking the records about user's session-based online activities in Taobao as side information, Alibaba \cite{wang2018billion} constructed a weighted and directed network that is comprised of user-item interactions in a continuous period \cite{wang2019survey} and that can be used to extract hidden consumption habits of users to promote recommendation accuracy. By taking the semantic networks, social networks and profile networks of users as side information simultaneously, SHINE \cite{wang2018shine} can predict the sign of a sentiment link (\textit{i.e.}, each of user's attitudes towards items) without analyzing textual information like user's comments on items.

On the other hand, from a scenario-oriented perspective, the rest of this section retrospects the utilization of user's social information \cite{he2010social, zhang2018social, chen2019integrating} and location information \cite{ding2018objectives} as side information in recommendation, in which case recent years have pioneered the location-based social recommendation \cite{werneck2020survey} thriving for a long period. In general, user's online information can be collected from online social platforms where users are allowed to express their needs, desires or attitudes towards items and events by such as tweeting or posting. This online information usually carries user's preferences for items, which is more timely and explicit compared with user-item interactions \cite{zhao2014we}. Collecting user's online information across multiple social platforms could link these different platforms together by their shared users. In other words, diverse side information could be therefore integrated in this way, which contributes to improving recommendation accuracy by collaboratively revealing more about user's preferences. In this regard, it is not to deny that users are encouraged to use their social platform accounts (like Twitter or Weibo) to log on to other online platforms like Amazon or Taobao, which enables user's comments on products, a type of social information revealing one's hobbies or demands shared on social platforms, to be employed in e-commerce recommender systems.

For the purpose of incorporating user's online social information in recommendation, multitudinous recommendation models were proposed. For instance, by collecting user's properties from microblogging, METIS \cite{zhao2014we} can learn item's demographics which could be depicted with the properties of its interacted users. Methodologically, by means of classification algorithms METIS can detect user's purchase intents and at the same time can extract item's demographic attributes from user's microblogging information in real-time, based on which recommendation can be run by matching users and items. Take MART \cite{zhao2015connecting} for another instance. Built on user's online social information, MART firstly constructs several social networks (\textit{i.e.}, a type of general graphs) related to recommender systems by users as bridges. When coming to a cold-start user, MART can retrieve the user's social information from these social networks, which would provide valuable information for recommendation such as his attributes like gender, age or career, aiming to analyze his preferences for items. The rationale behind MART lies in learning a transformation matrix used to map user's information of properties from social networks to his corresponding embeddings in recommender systems. Apart from user's properties, there is more valuable information from social networks that could be employed in recommendation. Another instance is user's social ties \cite{tilly2015identities}, a concept used to distinguish the strong and weak ties of users' friendship. With respect to it, Granovetter et al. \cite{granovetter1973strength} exemplified the more crucial role of weak ties in social networks compared with strong ones, because the weak ties can largely determine the connection patterns of clustered components in social networks. Wang et al. \cite{wang2016social} discovered that by distinguishing user's social ties in a social network the recommendation accuracy can be improved. Based on the discovery, by measuring the tie strength between node pairs with Jaccard's coefficient to classify them into strong or weak ones, Wang et al. proposed TBPR, a recommendation model that can distinguish each user's interacted items into five types in order to quantify user's different preferences for them. Finally, these distinguished preferences will be the input of the BPR framework \cite{rendle2012bpr}, used to perform the pair-wise ranking strategy for training.

Besides, methods of employing user's location information in recommendation are also a recent research focus. For instance, Wang et al. \cite{wang2013location} discovered that more than 80\% of a user's visited new places are located within a 10km vicinity of the user's latest visited places, unraveling the influence of one's location information on his decisions of future visits. The resources of user's location information are diverse, among which user's spatial trajectory information is a general one collected from user's mobile techniques, like GPS \cite{ladle2018measuring}, WiFi \cite{traunmueller2018digital} or ad-hoc networks \cite{wang2018spatial} with permission. Besides, user's check-in information of visits is also a source. In contrast with spatial trajectory information, it requires lower privacy rights because it can be directly collected from user's visit records. However, it cannot be updated in real-time as the spatial trajectory information does. In fact, a user's location information is usually related to his online social information, for one would like to share profiles of his visited places on social platforms, such as comments, experiences, photos or moods. To some extent these profiles can be used to express the motivation of one's visits to some places \cite{zhao2020personalized} and further be used to uncover his preferences for them. In this regard, recent research has explored enhancing the recommendation accuracy by employing user's location information as a supplement to online social information, called the location-based social recommendation.

For the purpose of incorporating user's location information in recommendation, models of location-based social recommendation are proposed. For instance, by means of user's check-in records to link users and locations, LFBCA \cite{wang2013location} constructs a bipartite graph, based on which the proximity between user-location pairs can be measured by analyzing users' co-visits under the UBCF framework. In addition, based on PageRank with BCA \cite{berkhin2006bookmark}, LFBCA can also utilize user's social relationships in uncovering (hidden) relations between users in the bipartite graph, in order to enhance the precision of proximity measurement. However, two flaws of LFBCA were clear. In the first place, ignoring the timeliness of user's location information could cause embarrassment, like recommending a place that is far away from one's current location. To deal with it, users can be subdivided into two categories: in-town and out-town \cite{ference2013location}. Second, user's location information is usually sparse because of one's limited visits to new places due to the constraints of his economic or time income. For that, digging out more side information of places, like their profiles given by users or local people's preferences for them \cite{yin2013lcars}, as a supplement is helpful. Among them, place profiles are particularly valuable because they can reveal sentimental attributes, like user's positive or negative attitudes towards these places, which can be used to infer whether a place meets a user's sentimental preferences and needs. For instance, sentimental attributes can be used to measure the proximity between user-location or location-location pairs \cite{zhao2020personalized}, in which case the techniques of textual sentiment analysis \cite{shi2019survey,pereira2021survey} in NLP could wield its predominance. In recent years, taking into account temporal factors in the location-based social recommendation is setting off a new tendency \cite{yao2018exploiting, jiang2015author}.

\subsection{Summary}

\begin{figure}[!ht]
\centering
\includegraphics[scale = 0.65]{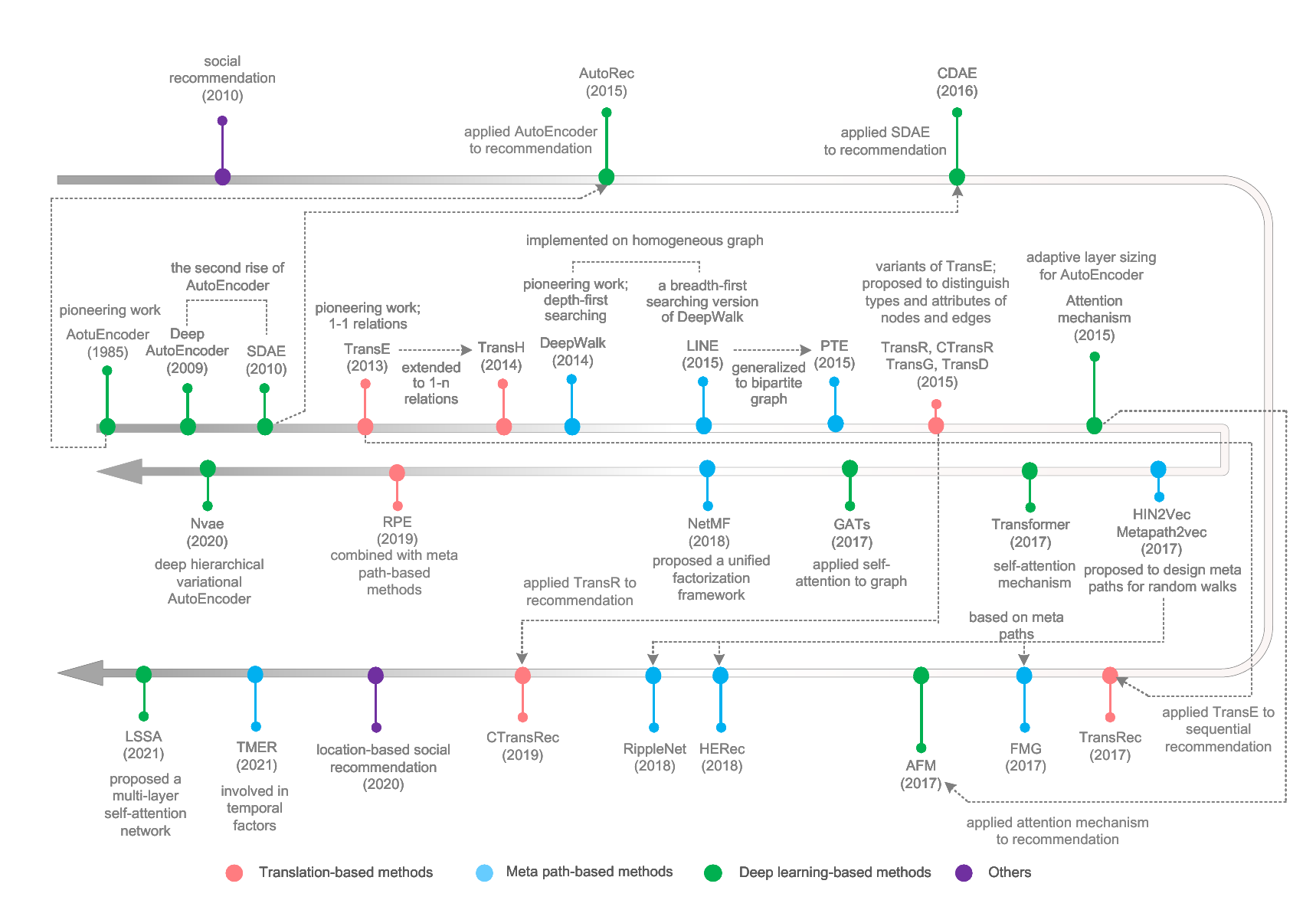}
\hspace{0.1in}
\caption{\textbf{Timeline of key developments in general graph embedding for recommendation.}}
\label{Ch5_Summary}
\end{figure}

Fig.~\ref{Ch5_Summary} concisely summarizes the key developments of general graph embedding techniques (\textit{i.e.}, by the inner timeline in the figure) and recommendation models based on these techniques (\textit{i.e.}, by the outer timeline of the figure), respectively. Apparently, all the three categories of general graph embedding techniques can be applied in recommendation, in which case those based on meta path and deep learning are the broadest ones. Relevant research in recent years has concentrated more on taking into account the temporal factors involved in a general graph and also more on representing the multiplicity of heterogeneous graphs.

Integrating general graphs with user-item bipartite graphs by linking their co-existed nodes is a direct strategy for providing the bedrock of employing general graph embedding techniques in recommendation. However, this strategy could dilute the significance of users and items as a result of the diversity of nodal types in a general graph. It seems that resorting to the attention and self-attention mechanisms as means to highlight the two most salient nodal types (\textit{i.e.}, users and items) in recommendation is a promising focus.

\section{Knowledge graph embedding for recommendation}
\label{KGBE}

As illustrated in Sec.~\ref{graph_representation}, knowledge graphs are usually recognized as the most complex case of general graphs. The complexity of knowledge graphs is characterized by an extreme abundance of node types and edge types, making knowledge graph embedding confront several challenges, like the large-scale, multiplicity and evolution of knowledge graphs. On the other hand, benefit from their complexity property of far more diverse carried information compared to general graphs, knowledge graphs have a virtue of uncovering abundant hidden user-item relations dear to researchers, in order to enhance the recommendation accuracy.

Feasible as general graph embedding techniques are in employing knowledge graphs in recommendation, the scalability of them is constrained by the large-scale of knowledge graphs. Defects are not limited to this, general graph embedding techniques are usually insufficient in capturing and preserving the features in a high multiplicity of knowledge graphs. Faced with this situation, one way out of the dilemma was to develop more efficient techniques for knowledge graph embedding. For that, in the first place, Sec.~\ref{LEMGE} illustrates the three challenges of knowledge graph embedding and retrospects the corresponding solutions (\textit{i.e.}, techniques for knowledge graph embedding). Based on that, Sec.~\ref{KGERS} retrospects the application of these techniques in recommendation from two perspectives: embedding-based and path-based. Finally, Sec.~\ref{KGERS} highlights two promising research directions.

\subsection{Three challenges of knowledge graph embedding}
\label{LEMGE}

In general, knowledge graph embedding meets three challenges \cite{auer2018towards}: the large-scale, multiplicity and evolution of knowledge graphs. The large-scale challenge means that knowledge graphs are usually too enormous in scale to be completely observed, making general graph embedding techniques have difficulty in running with an acceptable computing complexity. Besides, the multiplicity of knowledge graphs is generally characterized by the following three cases: graphs with multi-typed nodes and single-typed edges, graphs with single-typed nodes and multi-typed edges or graphs with both multi-typed nodes and edges. In view of these cases, sufficiently capturing and preserving the multiplicity is firmly anchored in precisely representing a knowledge graph. For that purpose, general graph embedding techniques would prefer to be used to represent knowledge graphs but are constrained from doing so by the lack of versatile expert knowledge in designing meta paths for such complex knowledge graphs and the general inability of translation-based techniques in capturing higher-order topological features. Moreover, since knowledge is constantly and rapidly growing in scale and multiplicity, the evolution of knowledge graphs largely challenges the computing complexity in updating the embeddings once learned from knowledge graphs.

In the face of the large-scale challenge, graph neural networks (GNNs) \cite{wu2020comprehensive, gao2020deep, zhou2020graph} can provide an efficient framework supporting parallel computing and fast response. The rationale behind GNNs lies in label propagation (or message passing) \cite{wang2007label, wang2019knowledge}, which is built on the assumption that each node with its features can be reconstructed by means of its connection with neighbors as well as its neighbors' features in a graph. Methodologically, in GNNs each node consists of two components: aggregator and updater. By collecting and aggregating the features from a node's neighbors, the aggregator can build a node's context embedding. After that, the node's context embedding combined with other input information (like side information) will be used to generate its embedding by the updater. Furthermore, by stacking each node's $K$ different adjacent neighbors or repeating the propagation process $K$ times on each node, the receptive field of GNNs can be expanded to a $K$-hop graph neighborhood. Subsequently, under the GNNs framework, more variants were proposed, among them transductive learning \cite{joachims1999transductive} and inductive learning \cite{belkin2005manifold, hamilton2017inductive} are two prevalent strategies for implementing the label propagation, where the former aims to infer the labels of unlabelled nodes based on labeled ones and the latter aims to learn a global function for labeling. In detail, for example, the inductive learning strategy takes each node in a graph as both the receiver and disseminator of information (like features) from and to its neighbors within specified hops, in which case the features of nodes can be updated during such a continuous information exchanging process, a parallel computing process of label propagation used to refine the embeddings of all nodes until which reaching a global convergence. Moreover, by extracting a graph's critical components based on the spectral sparsification theory \cite{cheng2015efficient}, spectral methods \cite{qiu2019netsmf, qiu2018network} can achieve a fast response. By generating new relationships between hidden semantic properties (such as analogical properties \cite{liu2017analogical} or circular correlation \cite{nickel2016holographic}) to equip a knowledge graph with enriched topological properties, generative graph methods \cite{bahdanau2014neural, liu2017analogical, nickel2016holographic} can speed up the learning process on large-scale graphs.

\begin{figure}[!ht]
\centering
\includegraphics[scale = 0.48]{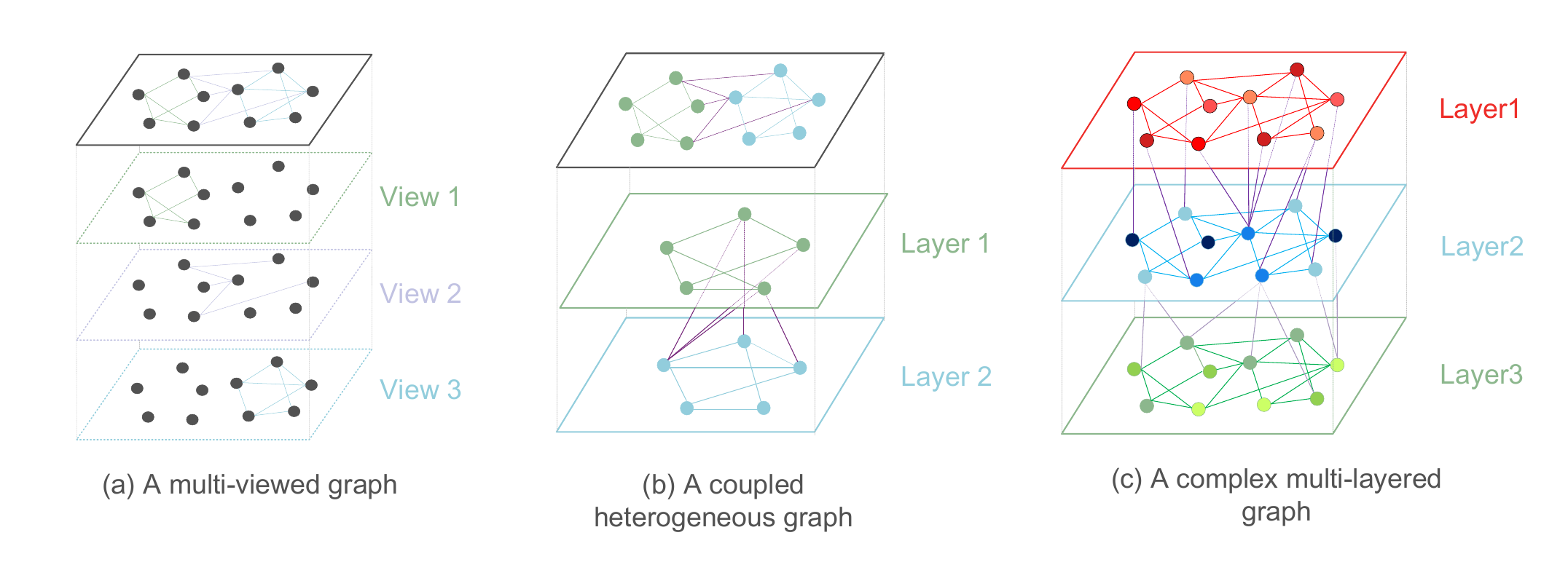}
\caption{\textbf{Examples of a multi-viewed graph, a coupled heterogeneous graph and a complex multi-layered graphs.} In (a), the multi-viewed graph represents a heterogeneous graph containing single-typed nodes colored in grey and three types of edges colored in green, purple and blue, respectively, where each view represents a homogeneous component connected by edges of the same type. In (b), the multi-layered graph represents a heterogeneous graph containing two types of nodes and three types of edges by clustering the nodes into two layers colored in green and blue, respectively, where each layer contains nodes of the same type and the interconnections between layers are colored in purple. In (c), it is a complex multi-layered heterogeneous graph with cross-domains, where the intra-connections in each layer are one-typed that belong to a particular domain while the nodes are heterogeneous.}
\label{graph_set_four}
\end{figure}

When it comes to capturing and preserving the multiplicity of knowledge graphs, GNNs have become adept in representing both the features and topological structures of knowledge graphs, in a manner of the label propagation between nodes. Because GNNs have come to see the merits of constructing highly expressive embeddings by composing extracted multi-scale localized spatial features like convolutional neural network (CNN) \cite{li2021survey} dose while being suitable for non-Euclidean data like graphs. In addition, in terms of building other effective graph representations which could carry the multiplicity of a graph as much as possible, on which embedding techniques run, multi-viewed graphs \cite{xu2013survey} can provide a clearer view of depicting a heterogeneous graph containing single-typed nodes and multi-typed edges (Fig.~\ref{graph_set_four} gives examples). Based on them, multi-viewed graph embedding aims to integrate the information carried by the different views of a multi-view graph together (termed as collaboration \cite{shi2018mvn2vec}) and at the same time to preserve their distinctive properties carried by each view (termed as preservation \cite{shi2018mvn2vec}). In this way, the learned embeddings can represent both the local features of nodes in each homogeneous community (\textit{i.e.}, views) as well as their global features across different communities. Methodologically, multi-viewed clustering \cite{chaudhuri2009multi, kumar2011co, xia2010multiview} and multi-viewed matrix factorization \cite{greene2009matrix, liu2013multi} were two attempts while still lacking sufficient collaboration as a consequence of their simple and independent concatenations of the learned embeddings from different views. In fact, completing both the collaboration and preservation is inextricably tied up with the good performance of multi-viewed graph embedding. For that purpose, more efficient methods have been proposed recently. For instance, by building random walk pairs across different views, Mvn2Vec \cite{shi2018mvn2vec} can capture the global structure of a graph. In addition, for each node $i$, by learning an overall-shared embedding $\displaystyle \bm{h}_i^c  \in \mathbb{R}^d$ of it as well as its distinctive embeddings $\displaystyle \bm{h}_i^v \in \mathbb{R}^s$ in a specific view $v$, MNE \cite{zhang2018scalable} can combine $\bm{h_i^c}$ and $\bm{h_i^v}$ by $\displaystyle \bm{h}_i^v=\bm{h}_i^c+\omega^v \bm{X}^{vT} \bm{h}_i^v$, where $\bm{X}^{vT}\in \mathbb{R}^{s \times d}$ is a transformation matrix and $\omega^v$ indicates the global importance of view $v$, in order to complete the preservation and collaboration for node $i$. Among these works, apparently, distributing proper importance weights $\omega^v$ to different views is essential \cite{qu2017attention}, in which case the attention mechanism seems to be promising assistance.

As for the other case of multiplicity that graphs with both multi-typed nodes and edges, multi-layered graphs \cite{benson2016higher, de2015structural, hu2014conditions} provide another category of intuitive graph representations, which have diverse applications ranging from multi-scale graph embedding \cite{perozzi2017don, ribeiro2017struc2vec} to cross-domain scenarios like critical infrastructure systems \cite{ouyang2014review} or collaboration platforms \cite{nambisan2009platforms}. Among them, the simplest form of multi-layered graphs is coupled heterogeneous graph, which consists of two types of nodes and three types of edges (Fig.~\ref{graph_set_four} gives examples). When it comes to a knowledge graph about actors and movies, the two layers in Fig.~\ref{graph_set_four} can be constructed by an actor community representing the cooperation among actors and a movie community representing the category relations between movies, respectively. The edges between the two layers can be used to represent the participation relationships between actors and movies. Methodologically, a variety of translation-based methods have been proposed recently for coupled heterogeneous graph embedding. For instance, as a node-oriented model, EOE \cite{xu2017embedding} can map nodal embeddings from different layers to each other through an overall-shared harmonious matrix. However, such a node-oriented strategy could neglect the influences of inter-connection edges between different layers. To deal with that, an edge-oriented strategy \cite{chang2015heterogeneous,li2018multi} was proposed to take into account the inter-connection edges. When coming to a multi-layered graph with enormous layers, applying meta path-based methods can achieve better computing efficiency. For instance, by building ruled random walks across different layers, PMNE \cite{liu2017principled} can capture the global topological features of a coupled heterogeneous graph. Furthermore, based on the previous example of an actor-movie knowledge graph, if the attributes of actors and movies need to be represented with nodes in multi-types in layers, a more complex multi-layered graph that can distinguish different types of nodes and edges is required, as shown in Fig.~\ref{graph_set_four} for example. For that, promising models include GATNE-T/GATNE-I \cite{cen2019representation} and DMNE \cite{ni2018co}, to name a few. One common idea of these models is to apply fast learning algorithms to represent the heterogeneous information in each layer efficiently and use an attention mechanism to learn their different importance.

Capturing and preserving the temporal factors involved in dynamical knowledge graphs in the face of the evolution challenge has become a research direction in recent years. Benefitting from its structure and label propagation (or message passing) mechanism, GNNs can be used to represent dynamic graphs \cite{zhou2020graph}. Other attempts include ctRBM \cite{li2014deep}, M$^2$DNE \cite{lu2019temporal}, HTNE \cite{zuo2018embedding} and TGAT \cite{xu2020inductive}, to name a few. See \cite{kazemi2020representation} for an in-depth review.

\subsection{Recommendation involving knowledge}
\label{KGERS}

Embedding-based and path-based are two typical strategies for equipping the knowledge graph embedding techniques retrospected in Sec.~\ref{LEMGE} in recommender systems, in order to efficiently employ knowledge in recommendation for higher accuracy and better interpretability.

By learning the supplemental features as embeddings of users and items from knowledge graphs, the embedding-based strategy can employ knowledge in recommendation by methods of cross-domain recommendation or transfer learning retrospected in Sec.~\ref{Other_Models}. Built on this strategy, various models were proposed. For instance, by means of a textual knowledge graph, kaHFM \cite{anelli2019make} can learn item's embeddings of semantic features as a supplement to factorization machine (FM) \cite{rendle2010factorization}. After extracting the structural, textual and visual features of items from three different knowledge graphs, CKE \cite{zhang2016collaborative} can construct a synthetic embedding for each item as a supplement. In general, since the efficiency in representing the features of entities and relations in knowledge graphs, it is a common-used method to apply GNNs to the embedding-based strategy \cite{wang2019knowledge, wang2019kgat}. Among them, as for a user, how to accurately measure the high-order proximity between his interacted items and those far away from him (named ``remote'' items) in a knowledge graph is an open and important issue because the ``remote'' items can be recognized as candidates of diverse recommendations to the user. For such purposes, for instance, by deepening the range of neighborhood from one hop to multiple hops away, KGCN \cite{wang2019knowledge} can measure the proximity between nodes with a longer distance. Meanwhile, discriminating the unequal influences on a node from its different neighbors is another focus. Based on GATs \cite{velivckovic2017graph}, KGAT \cite{wang2019kgat} builds an attention-based aggregator that can learn the different contribution weights from a node's neighbors. In the same way, in the hyperbolic space Hyper-Know \cite{ma2021knowledge} builds an aggregator that can learn hyperbolic attention with Einstein midpoint. In the training process of these models, unobserved edges were used to be commonly considered as negative samples \cite{pan2008one, he2016fast}. However, Togashi et al. \cite{togashi2020alleviating} uncovered that such a negative sampling strategy could aggravate the cold start problem as a consequence of taking new items as negative samples, leading to biased and sub-optimal recommendation results toward already popularized items. To make up for the flaw, by applying pseudo-labeling \cite{lee2013pseudo} to distinguish possible weak-positive pairs in unobserved edges, Togashi et al. proposed KGPL. Besides, more subsequent works \cite{wang2020reinforced,chen2020jointly} have been devoted to this focus in recent years.

In contrast to the embedding-based strategy which aims to learn supplementary features as embeddings of users and items from knowledge graphs, the rationale for path-based strategy lies in extracting user-item relations from knowledge graphs through random walking across multiple types of nodes and edges guided by designed meta paths, in order to achieve explainable recommendation \cite{zhang2018explainable, gao2019explainable, palmonari2020knowledge, xie2021explainable}. By following the extracted user-item relations along multi-hop meta paths, user's preferences for items could be better understood \cite{wang2019explainable}. Intuitively, take Fig.~\ref{KG} for example, by following the relations (or paths) \textit{(Alice, Watched, Avengers4)$\wedge$(Avengers4, ProduceBy, TWDC)$\wedge$(TWDC, Produce, Avengers2)} and \textit{(Alice, Watched, Avengers4)$\wedge$(Avengers4, StarBy, Robert)$\wedge$(Robert, star, Avengers 2)}, which shows that \textit{Avengers 2} was produced by the same company as that of \textit{Avengers 4} and also has a common actor in \textit{Avengers 4} that Alice watched in the past, it is deducible that \textit{Avengers 2: age of ultron} would meet Alice's preferences for movies. Apparently, this strategy could equip recommendation with abilities in reasoning and explainability. Methodologically, in order to precisely generate random walks on a knowledge graph which are integrated with a bipartite graph containing user nodes and item nodes, the path-based strategy generally meets the following three issues: first, how to design reasonable meta paths used to rule random walking? Second, how to measure user-item proximity through random walks? Three, how to distinguish the different importance of multiple random walks in order to highlight the most valuable ones? As resorts, for instance, by learning the embeddings of nodes in each random walk, KPRN \cite{wang2019explainable} can represent the hidden sequential patterns between nodes, which will be used to learn the random walk's embedding representing its holistic semantics through a long short-term memory (LSTM) \cite{hochreiter1997long}. These learned embeddings of random walks can be used to measure user-item proximity. Furthermore, by means of a weighted pooling layer, KPRN can also distinguish the different importance of multiple random walks between a user-item pair. However, the pooling layer is built by enumerating all possible random walks between a user-item pair, requiring a high computing complexity, particularly on knowledge graphs with an enormous scale. Alternatively, Xian et al. \cite{xian2019reinforcement} designed a soft reward strategy that can quickly highlight the most valuable random walks between a user-item pair. After taking into account user's history click sequences, Zhu et al. \cite{zhu2020knowledge} further proposed a strategy for weighting random walks. Recently, Chen et al. \cite{chen2021temporal} discovered that ignoring the temporal factors in user-item interactions could result in less convincing and inaccurate recommendation explainability. For that, Chen et al. proposed TMER to construct item-item instance paths between consecutive items for sequential recommendation.

In application, there are two promising research directions of knowledge graph embedding-based recommendation. One is the conversational recommender system (CRS) \cite{Sun2018Conversational, lei2020conversational, jannach2021survey, gao2021advances}, which is equipped with abilities in chatting with users, understanding their expressions (\textit{i.e.}, language) and reasoning their preferences for items on the advice of semantic knowledge or domain knowledge. The motivation of CRS lies in taking actives steps to solve the issue of user privacy infringement encountered by conventional recommendation systems, like those that manage to collect side information from user's social and location information, which is generally protected as privacy. In contrast, CRS can straightforwardly ask needed information for recommendation from users, in which case users have endowed the rights to determine which information they would prefer to share. In general, CRS consists of two components \cite{zhou2020towards}: one is the dialog component designed to interpret human language into proper machine-readable forms and generate responses to users through a multi-round natural language conversational system \cite{li2015diversity,vinyals2015neural}. The other is the recommendation component designed to reason user's preferences for items by analyzing the content of user-machine conversations, based on which recommendations can be generated and returned to users. Based on the two components, the implementation of CRS basically includes the following three stages: initiation, conversation and display \cite{zhang2018towards}. In detail, by raising a suitable question the dialog component can initiate a conversation with users. This conversation will continue to collect needed information for recommendation from users through multi-round questions and answers. After analyzing user's preferences for items from the collected information, the recommendation component can pick up possible candidates which would meet user's interest and then generate the top-N items as recommendations returned to users. The three stages will iteratively implement many rounds till the returned recommendations satisfy users. In addition, by means of knowledge graphs CRS can also reason on user's preferences for items, realizing the explainability of recommendation results. For example, suppose a user expressed his affection on \emph{Avengers 2: age of ultron} and \emph{Avengers 4: the final battle} in conversations with CRS. Combined with the knowledge graph in Fig.~\ref{KG}, CRS can infer that the user may also like other movies produced by TWDC or stared by Robert. Actually, by building user's profiles in the dialog component and matching them with item's attributes (like reviews \cite{zhou2020towards}) in the recommendation component as illustrated above, CRS can be recognized as a content-based recommendation method. Methodologically, item-oriented method \cite{chen2019towards,lei2020estimation, zhang2019toward, he2017distributed, li2018towards, zhang2018towards} is the most common-used implementation for CRS, which aims to uncover hidden relations between items by employing knowledge graphs. Furthermore, Zhou et al. \cite{zhou2020improving} discovered that the word-level enrichment in conversations could reveal user's personal habits in word usage and hence would be valuable to understand user's preferences. In light of that, Zhou et al. proposed a word-oriented method, which can assist the item-oriented models in learning user embeddings. Recently, research into efficiently guiding users from non-recommendation scenarios to some desired ones via proactive conversations has become a crucial direction of CRS \cite{zhou2020towards,tang2019target, kang2019recommendation, liu2020towards}.

The other promising direction of knowledge graph embedding-based recommendation in the application is news recommender systems \cite{karimi2018news, li2019survey, abdollahpouri2021toward, raza2021news}. Compared with recommendation based on non-textual scenarios, since the primary information for recommendation in news recommender systems is textual content \cite{wang2018dkn}, there raise three more questions: first, the news is distinctly time-sensitive. In other words, the relevance to reality makes the news fade away rapidly over time, where out-of-date news is constantly replaced by new ones. Second, user's interests in the news are generally topic-sensitive, diversified and changeable, related to current social focuses and issues. Third, news language is generally comprised of both professional knowledge structured logically and common sense presented causally. To understand, reason and represent it, techniques in NLP combined with graph embedding are both required, which are more sophisticated than previous ones for non-textual recommendation. Equipping content-based recommendation methods with knowledge as external resources enable the learning and representation of relatedness between news words as well as their latent knowledge-level connections more precisely. For instance, by employing an enormous knowledge graph in recommendation DKN \cite{wang2018dkn} can build relations between the word entities identified and extracted from the news, used to construct a small domain knowledge graph for each piece of news consisting of these word entities and relations. Based on the domain knowledge graph, DKN can learn the embeddings of news pieces through a convolutional neural network (CNNs) \cite{albawi2017understanding}.

\subsection{Summary}

\begin{figure}[!ht]
\centering
\includegraphics[scale=0.65]{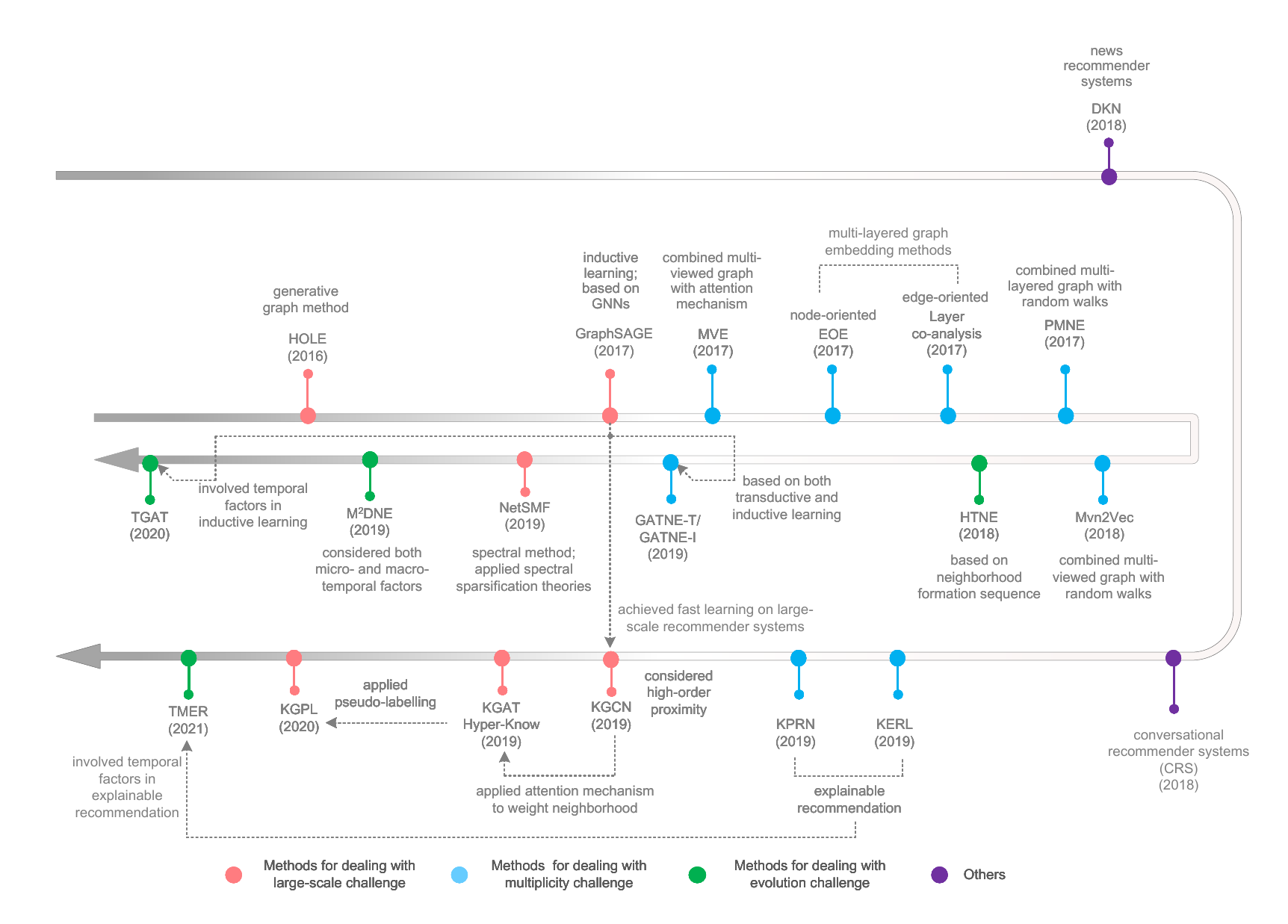}
\hspace{0.1in}
\caption{\textbf{Timeline of key developments in knowledge graph embedding for recommendation.}}
\label{Ch6_Summary}
\end{figure}

Fig.~\ref{Ch6_Summary} concisely summarizes the key developments of knowledge graph embedding techniques (\textit{i.e.}, by the inner timeline in the figure) and recommendation models based on these techniques (\textit{i.e.}, by the outer timeline in the figure), respectively. Conceivably, compared with the developments of bipartite graph-based and general graph embedding-based recommendation shown in Figs.~\ref{Ch4_Summary} and \ref{Ch5_Summary}, the developments of knowledge graph embedding-based one began, for the first time in large numbers, in recent years. its developments have advanced into research into efficiently incorporating multiplex and evolving large-scale knowledge in recommendation.

\section{Performance evaluation}
\label{evaluation}

This section designs and runs control experiments and displays experimental results on representative graph embedding-based recommendation models and the most common-used conventional recommendation models, comparing their pros and cons on six tasks related to different recommendation scenarios, data scales and data sparsity, in order to make a trade-off between them oriented to specific scenarios. In the first place, Sec.~\ref{Experiment_Setup} designs the experiment setups, including data sets, evaluation metrics and evaluated models. Then, Sec.~\ref{Results_and_Analysis} presents experimental results and analysis.

\subsection{Experiment setups}
\label{Experiment_Setup}

Although comprehensive evaluations on conventional recommendation models or graph embedding-based recommendation models did by existed research or published program libraries \cite{zhao2020recbole, hug2020surprise}, there still lacks sufficient comparisons between both of them under a unified experimental framework, providing little prospect on analyzing and utilizing their respective strengths to complement each other. For that gap, this section designs six recommendation tasks for predicting both implicit and explicit user-item interactions, based on which several graph embedding-based and conventional recommendation models will be evaluated on five common-used metrics.

\subsubsection{Data sets}
\label{datasets}

Three data sets from two different recommendation scenarios and with distinctive data scales and sparsity used in experiments are described in Tab.~\ref{Datasets}. Among them, MovieLens 100K\footnote{https://grouplens.org/datasets/movielens/} and MovieLens 1M\footnote{https://grouplens.org/datasets/movielens/} are two data sets from a popular online movie recommender system named MovieLens where users are allowed to explicitly express their preferences for watched movies by giving ratings, which will be used to implement further movie recommendations. As a control experimental group, the two data sets record the ratings of movies given by users while being different from scales and sparsity, used to evaluate the performance of recommendation models in the same recommendation scenario but with different data scales and sparsity. In addition, another data set is Jester 617K\footnote{http://eigentaste.berkeley.edu/dataset/} recording the ratings of jokes given by users, which is from a popular joke review platform where users are allowed to explicitly express their preferences for jokes.

\begin{table}[!ht]
\setlength{\abovecaptionskip}{0.3cm}
\setlength{\belowcaptionskip}{0.3cm}
\caption{\textbf{Descriptions of data sets.} Density, defined as the ratio of the observed number of interactions to the maximum possible number of interactions between all users and items, is used to quantify data sparsity. $|U|$ and $V$ denote the number of users and items, respectively.}
\label{Datasets}
\begin{adjustbox}{center}
\footnotesize
\begin{tabular}{ccllccc}
\toprule
\textbf{Datasets}       & \multicolumn{3}{c}{\textbf{$|U|$}} & \textbf{$|I|$} & \textbf{Observed interactions} & \textbf{Density(\%)}  \\ \hline
MovieLens 100K & \multicolumn{3}{c}{610}            & 9724           & 100836           & 1.7                            \\
MovieLens 1M   & \multicolumn{3}{c}{6040}           & 3682           & 1000209          & 4.497                         \\
Jester 617K    & \multicolumn{3}{c}{24938}          & 100            & 616913           & 24.738                           \\
\bottomrule
\end{tabular}
\end{adjustbox}
\end{table}

As explicit user-item interactions, the ratings recorded in the three data sets are slightly different in value: in MovieLens 100K, user $i$ gave a rating of item $j$ on a scale from $0.5$ through $5$, \textit{i.e.},$\operatorname{r_{ij}^{(100K)}} \in [0.5, 5]$, with step $0.5$, while in MovieLens 1M it is $\operatorname{r_{ij}^{(1M)}} \in [1,5]$ with step $1$, and in Jester 617K it is $\operatorname{r_{ij}^{(617K)}} \in [-10, 10]$ with step $0.0005$. For unity, $\operatorname{r_{ij}^{(617K)} \in [-10, 10]}$ is normalized to $\operatorname{\overline{r}_{ij}^{(617K)} \in [0, 5]}$ by $\displaystyle \operatorname{\overline{r}_{ij}^{(617K)}= 5 \times \frac{r_{ij}^{(617K)} - (-10)}{10 - ( - 10)}}$ in experimental settings. Furthermore, by converting the ratings no less than $3$ into value $1$ while the rest of the ratings less than $3$ into value $0$, the corresponding implicit user-item interactions can be constructed. In this way, six different recommendation tasks are designed for evaluations of recommendation models.

Based on the above settings to implement model evaluations, each of the data sets is organized into tuples (user, item, rating/$0$/$1$) and is randomly split into a training set by $80$\% of the data set and a test set by $20$\% of the data set. Following that, recommendation models can implement based on the observed interactions in the training set, generating a recommendation list for each user. Then, the hit rate of the recommendation list can be calculated by checking how many items recommended to each user meet the observed interacted ones according to the test set. However, such a split strategy inevitably yields deviations of the recommendation accuracy based on different realizations (\textit{i.e.}, different randomly split training-test sets). To assure the reliability of results, $20$ realizations are randomly generated, based on which each of the six recommendation tasks repeatedly and independently implement in order to obtain the average and standard deviation of results.

\subsubsection{Evaluation metrics}

Evaluation metrics \cite{gunawardana2009survey, krichene2020sampled} used in experiments include \textit{mean absolute error (MAE)} \cite{willmott2005advantages} and \textit{root mean squared error (RMSE)} \cite{willmott2005advantages, zhao2017meta} for evaluations on recommendation accuracy in predicting explicit user-item interactions, and \textit{Precision} \cite{wang2013location}, \textit{Recall} \cite{wang2013location, yin2013lcars} and \textit{Rank} \cite{wang2013theoretical} for those in predicting implicit user-item interactions.

Specifically, denote the training sets and test sets as $\mathcal{T_{\operatorname{r}}}$ and $\mathcal{T_{\operatorname{e}}}$, respectively, in which only the users and items that appear both in the training sets and test sets remain while the ones not satisfying the condition as well as the corresponding interactions between them are removed. Denote the observed rating on item $j$ by user $i$ in the test set as $r_{ij}$ and the predicted one as $\hat{r}_{ij}$. Then, MAE and RMSE are defined as
\begin{equation}
\operatorname{MAE}=\frac{1}{|\mathcal{T_{\operatorname{e}}}|}\sum_{(i,j)\in \mathcal{T_{\operatorname{e}}}} |r_{ij}-\hat{r}_{ij}| ,\\
\end{equation}
\begin{equation}
\operatorname{RMSE}=\bigg(\frac{1}{|\mathcal{T_{\operatorname{e}}}|}\sum_{(i,j)\in \mathcal{T_{\operatorname{e}}}}(r_{ij}-\hat{r}_{ij})^2\bigg)^{\frac{1}{2}}.
\end{equation}
Intuitively, MAE and RMSE can quantify the overall deviation between $r_{ij}$ and $\hat{r}_{ij}$, where a smaller value indicates a better recommendation accuracy.

For a user $i$, denote the generated recommendation list containing the top $L$ predicted items in which the user might be most likely to be interested as $\mathcal{O}_i$, and denote the items that the user has interacted with within the test set as $\mathcal{S}_i$. Suppose there are $M$ users in the test set. Then Precision and Recall are defined as
\begin{equation}
\operatorname{Precision}=\frac{\sum_{i=1}^M |\mathcal{O}_i \bigcap \mathcal{S}_i|}{ML},
\end{equation}
\begin{equation}
\operatorname{Recall}=\frac{\sum_{i=1}^M |\mathcal{O}_i \bigcap \mathcal{S}_i|}{|\mathcal{T_{\operatorname{e}}}|},
\end{equation}
where $L=50$ in experiments. Intuitively, Precision and Recall can quantify the hit rate of a recommendation list (\textit{i.e.}, in a recommendation list the proportion of items meeting a user's interacted items within the test set) in two different methods. A higher Precision or Recall indicates a better recommendation accuracy.

Rank is a more fine-grain metric to evaluate the recommendation accuracy. For a user $i$, denote the Rank of each of his interacted item $j \in \mathcal{S}_i$ by $\displaystyle \operatorname{R_{ij}}=\frac{1}{\log_2 (p_j+2)}$ if item $j$ appears in $\mathcal{O}_i$, where $p_j$ is the ranking position of item $j$ in $\mathcal{O}_i$ that ranging from $0$ to $L$, and $0$ otherwise. Then, the overall Rank of items in $\mathcal{T_{\operatorname{e}}}$ is defined as
\begin{equation}
\operatorname{Rank}= \frac{1}{|\mathcal{T_{\operatorname{e}}}|} \sum_{(i,j) \in \mathcal{T_{\operatorname{e}}}} \operatorname{R_{ij}}.
\end{equation}
A higher overall Rank indicates a better recommendation accuracy.

\subsubsection{Evaluated models}

The representative graph embedding-based recommendation models pioneering subsequent variants in each category from Secs.~\ref{BiGE}, \ref{GGE} and \ref{KGBE} are partially selected as evaluated models in experiments. As benchmarks, some of the most commonly used conventional recommendation models are selected. Tab.~\ref{Benchmarks} gives an overview of them.

\begin{table}[!ht]
\setlength{\abovecaptionskip}{0.3cm}
\setlength{\belowcaptionskip}{0.3cm}
\caption{\textbf{Evaluated models.} Corresponding to the data sets, recommendation tasks in experiments are divided into two aspects: predicting explicit and implicit user-item interactions, respectively.}
\label{Benchmarks}
\begin{adjustbox}{center}
\footnotesize
\begin{tabular}{@{}ccc@{}}
\toprule
\textbf{Recommendation tasks}                                                             & \textbf{Graph embedding-based models} & \textbf{Conventional models (benchmarks)}                                                   \\ \midrule
\begin{tabular}[c]{@{}c@{}} Predicting explicit  \\  user-item interactions\end{tabular} & FunkSVD, PMF, FM, AutoRec             & Overall Average, UserKNN, ItemKNN                                                           \\
\begin{tabular}[c]{@{}c@{}} Predicting implicit \\ user-item interactions\end{tabular} & GMF, MLP, NCF, TransE                 & \begin{tabular}[c]{@{}c@{}}UserKNN, ItemKNN, HybridKNN\\ HeatS, ProbS, HybridS\end{tabular} \\ \bottomrule
\end{tabular}
\end{adjustbox}
\end{table}

In detail, as for recommendation tasks for predicting explicit user-item interactions, three graph embedding-based recommendation models are selected from Secs.~\ref{sec:SVD}, \ref{Bayesian_Sec} and \ref{DL-KG}, including FunkSVD \cite{koren2009matrix}, probabilistic matrix factorization (PMF) \cite{mnih2007probabilistic} and AutoRec \cite{sedhain2015autorec}, respectively. In addition, factorization machine (FM) \cite{rendle2010factorization} selected from Sec.\ref{sec:SVD} is a representative model used to tackle the sparsity problem in recommendation. Meanwhile, as benchmarks, three conventional recommendation models including Overall Average, user-based k-nearest neighbor collaborative filtering (UserKNN) \cite{zhao2010user} and item-based k-nearest neighbor collaborative filtering (ItemKNN) \cite{sarwar2001item} are selected, among which Overall Average \cite{lu2012recommender} takes a user's average rating on items in the training set as the predicted rating on his non-interacted items. UserKNN is illustrated in Sec.~\ref{conventional_models} and ItemKNN is an item-oriented version of UserKNN. HybridKNN is a combination of UserKNN and ItemKNN. The recommendation accuracy of these models in predicting explicit user-item interactions is evaluated by metrics MAE and RMSE.

As for recommendation tasks for predicting implicit user-item interactions, three graph embedding-based recommendation models are selected from Sec.~\ref{DL-Bi}, including generalized matrix factorization (GMF) \cite{he2017neural}, multi-layer perceptron (MLP) \cite{srivastava2012multimodal,he2017neural} and neural collaborative filtering (NCF) \cite{he2017neural}. In addition, another evaluated model is TransE \cite{bordes2013translating} selected from Sec.~\ref{Trans-section}, used to explore the performance of the application of models oriented to general graphs on bipartite graphs. Meanwhile, six conventional recommendation models, including UserKNN \cite{zhao2010user}, ItemKNN \cite{sarwar2001item}, hybrid k-nearest neighbor collaborative filtering (HybridKNN) \cite{wang2006unifying}, heat spreading algorithm (HeatS) \cite{zhang2007heat}, probabilistic spreading algorithm (ProbS) \cite{zhang2007recommendation} and hybrid spreading (HybridS) \cite{zhou2010solving}, are selected as benchmarks, among which HeatS, ProbS and HybridS are models established on theories of physical dynamics, and HybridS combines the advantages of ProbS oriented to high recommendation accuracy and HeatS oriented to high recommendation diversity \cite{zhou2013power}. The recommendation accuracy of these models in predicting implicit user-item interactions is evaluated by metrics Precision, Recall and Rank.

\subsection{Results and analysis}
\label{Results_and_Analysis}

On the basis of the above experiment setups, experimental results of evaluated models in predicting explicit and implicit user-item interactions are shown in Tabs. \ref{explicit_result} and \ref{implicit_result}, respectively. The hyper-parameter settings in experiments of evaluated models are given in Tab.~\ref{hyper-parameter_settings} in Appendix A. These hyper-parameter settings were determined by tuning on each of the first randomly split realization of the three data sets, respectively, where the hyper-parameter tuning results in experiments are presented with the tables in Appendix B. Codes and data are available\footnote{https://github.com/pitteryue/Recommender-Systems}.

\subsubsection{Predicting explicit user-item interactions}
\label{explicit_results}

In predicting explicit user-item interactions (\textit{i.e.}, ratings), graph embedding-based recommendation models overall outperform conventional recommendation models in recommendation accuracy, in which case such the advantage of graph embedding-based models becomes more salient with the increase of data scales while data sparsity seems not to be a decisive factor. However, graph embedding-based models are overall less stable than conventional models.

\begin{table}[h]
\setlength{\abovecaptionskip}{0.3cm}
\setlength{\belowcaptionskip}{0.3cm}
\caption{\textbf{Results on explicit user-item interactions.} All results to the nearest $0.001$ are averaged over $20$ realizations, and the value in brackets is the standard deviation to the nearest $0.001$. To make the comparison clearer, the comparatively lower average MAE and average RMSE of models on each data set are in bold. Note that n.a. indicates that FunkSVD failed on Jester617K as a consequence of the gradient disappearance and explosion problem \cite{pascanu2013difficulty}.}
\label{explicit_result}
\begin{adjustbox}{center}
\footnotesize
\begin{tabular}{cccccccc}
\toprule
\textbf{Data sets}        & \multicolumn{2}{c}{\textbf{MovieLens 100K}}                          & \multicolumn{2}{c}{\textbf{MovieLens 1M}}                      & \multicolumn{2}{c}{\textbf{Jester 617K}}              \\ \hline
\textbf{Metrics}         & \textbf{MAE}                 & \textbf{RMSE}                         & \textbf{MAE}                 & \textbf{RMSE}                   & \textbf{MAE}                 & \textbf{RMSE}          \\ \hline
Overall Average &0.827 (0.005)                 &0.909 (0.003)                          &0.934 (0.001)                 &0.966 (0.001)                    &1.144 (0.031)                 &1.070 (0.015)           \\
UserKNN         &0.697 (0.004)                 &0.906 (0.005)                          &0.772 (0.001)                 &0.977 (0.001)                    &0.960 (0.024)                 &1.185 (0.030)           \\
ItemKNN         &0.717 (0.004)                 &0.906 (0.005)                          &0.719 (0.001)                 &0.908 (0.001)                    &0.915 (0.021)                 &1.131 (0.027)  \\ \hline
\textbf{Avg.}            &0.747 (0.003)                 &\textbf{0.907 (0.004)}                 &0.808 (0.001)                 &0.950 (0.001)                    &1.006 (0.025)                 &\textbf{1.129 (0.024)}   \\
\hline
FunkSVD         &0.705 (0.005)                 &0.918 (0.006)                          &0.693 (0.002)                 &0.887 (0.002)                    &n.a.                          &n.a.                    \\
PMF             &0.740 (0.006)                 &0.969 (0.008)                          &0.694 (0.006)                 &0.882 (0.007)                    &0.897 (0.019)                 &1.132 (0.025)  \\
FM              &0.678 (0.009)                 &0.881 (0.006)                          &0.682 (0.005)                 &0.870 (0.003)                    &0.892 (0.019)                 &1.123 (0.027)  \\
AutoRec         &0.742 (0.006)                 &0.959 (0.007)                          &0.724 (0.003)                 &0.919 (0.004)                    &1.053 (0.046)                 &1.304 (0.067)           \\
\hline
\textbf{Avg.}            & \textbf{0.716 (0.007)}       & 0.932 (0.007)                         &\textbf{0.698 (0.004)}        &\textbf{0.890 (0.004)}           &\textbf{0.947 (0.028)}        &1.186 (0.119)            \\

\bottomrule
\end{tabular}
\end{adjustbox}
\end{table}

As shown in Tab.~\ref{explicit_result}, the average MAE of graph embedding-based recommendation models on MovieLens 100K, Jester 617K and MovieLens 1M, which are sorted by data scales in ascending order, are 4.15\% less, 5.86\% less and 13.61\% less than that of conventional recommendation models, respectively. At the same time, in terms of the average RMSE, it is 2.76\% greater, 5.05\% greater and 6.32\% less than that of conventional recommendation models, respectively. Apparently, with the increase of data scales, the average MAE and RMSE of graph embedding-based recommendation models are getting overall lower than those of conventional ones. It appears that the machine learning methodology adopted by graph embedding-based recommendation models plays a large role in those experimental results, benefiting from the better data fitting performance based on a larger data scale. However, when concerning data sparsity the results in Tab.~\ref{explicit_result} seem to show nothing of its role. Although the increase of density from MovieLens 100K to MovieLens 1M brings overall greater advantages in MAE and RMSE of graph embedding-based models beyond conventional models, it is the opposite as the density continues to increase from MovieLens 1M to Jester 617K. Admittedly, these experimental results are insufficient to prove the existence of an optimal data sparsity making the overall recommendation accuracy of graph embedding-based recommendation models better than that of conventional ones. On the other hand, Tab.~\ref{explicit_result} shows that graph embedding-based recommendation models have greater values of standard deviations of the average MAE and average RMSE than those of conventional one on all three data sets, manifesting the overall lower stability of graph embedding-based recommendation models compared with conventional ones.

In addition, it can be observed in Tab.~\ref{explicit_result} that FM outperforms all other models in recommendation accuracy on the three data sets. It appears that the rationale for modeling all interactions between variables (\textit{i.e.}, feature intersection) behind FM helps uncover richer patterns hidden in sparse data for a more accurate recommendation, particularly on MovieLens 100K with the lowest data density. This observation encourages the research into models like FM that can base on small and sparse data sets to achieve high enough recommendation accuracy. Similar to few-shot learning \cite{wang2020generalizing, bendre2020learning}, FM and its variants might be a potential direction for the future breakthrough.

\begin{table}[h]
\setlength{\abovecaptionskip}{0.3cm}
\setlength{\belowcaptionskip}{0.3cm}
\caption{\textbf{Results on implicit user-item interactions.} The length of recommendation list is set to be $50$ (denoted by @$50$). All results to the nearest $0.001$ are averaged over $20$ realizations, and the value in brackets is the standard deviation to the nearest $0.001$. To make the comparison clearer, the comparatively higher average Precision, average Recall and average Rank of models on each data set are in bold. Note that as for conventional models HeatS merely oriented to recommendation diversity is not taken into account for the average results of recommendation accuracy.}
\label{implicit_result}
\begin{adjustbox}{center}
\footnotesize
\begin{tabular}{cccc}
\toprule
\multicolumn{4}{c}{\textbf{MovieLens 100K}}                                                               \\ \hline
\textbf{Metrics}      &\textbf{Precision@50}   & \textbf{Recall@50}     & \textbf{Rank@50}               \\ \hline
UserKNN      &0.126 (0.002)   &0.247 (0.003)  &0.078 (0.001)          \\
ItemKNN      &0.125 (0.002)   &0.245 (0.003)  &0.079 (0.001)          \\
HybridKNN    &0.141 (0.002)   &0.274 (0.003)  &0.089 (0.001)          \\
HeatS        &0.013 (0.001)   &0.024 (0.001)  &0.017 (0.051)                   \\
ProbS        &0.120 (0.002)   &0.235 (0.003)  &0.075 (0.001)          \\
HybridS      &0.128 (0.002)   &0.251 (0.004)  &0.080 (0.001)          \\
\hline
\textbf{Avg}.         &\textbf{0.128 (0.002)}   &\textbf{0.250 (0.003)}  &\textbf{0.080 (0.001)}          \\
\hline
GMF          &0.099 (0.003)   &0.192 (0.005)  &0.057 (0.002)                   \\
MLP          &0.072 (0.004)   &0.140 (0.009)  &0.040 (0.003)                   \\
NCF          &0.117 (0.006)   &0.228 (0.011)  &0.072 (0.004)                   \\
TransE       &0.055 (0.002)   &0.107 (0.004)  &0.032 (0.001)                   \\
\hline
\textbf{Avg.}         &0.086 (0.004)   &0.167 (0.007)  &0.050 (0.003)\\
\bottomrule
\multicolumn{4}{c}{\textbf{MovieLens 1M}}                                                                 \\ \hline
\textbf{Metrics}      &\textbf{Precision@50}  & \textbf{Recall@50}     & \textbf{Rank@50}                \\ \hline
UserKNN      &0.155 (0.000)  &0.280 (0.001)  &0.086 (0.000)           \\
ItemKNN      &0.157 (0.000)  &0.283 (0.001)  &0.086 (0.000)           \\
HybridKNN    &0.178 (0.000)  &0.322 (0.001)  &0.100 (0.000)           \\
HeatS        &0.058 (0.002)  &0.104 (0.004)  &0.021 (0.001)                    \\
ProbS        &0.127 (0.001)  &0.230 (0.001)  &0.072 (0.000)           \\
HybridS      &0.143 (0.000)  &0.259 (0.001)  &0.082 (0.000)           \\
\hline
\textbf{Avg.}         &\textbf{0.152 (0.000)}  &\textbf{0.275 (0.001)}  &\textbf{0.085 (0.000)}\\
\hline
GMF          &0.137 (0.004)  &0.248 (0.007)  &0.071 (0.003)                    \\
MLP          &0.100 (0.014)  &0.181 (0.026)  &0.041 (0.008)                    \\
NCF          &0.120 (0.015)  &0.216 (0.027)  &0.050 (0.009)                    \\
TransE       &0.050 (0.006)  &0.089 (0.011)  &0.024 (0.003)                    \\
\hline
\textbf{Avg.}         &0.102 (0.010)  &0.184 (0.018)  &0.047 (0.006)\\
\bottomrule
\multicolumn{4}{c}{\textbf{Jester 617K}}                                                                  \\ \hline
\textbf{Metrics}      & \textbf{Precision@50}  & \textbf{Recall@50}     & \textbf{Rank@50}                \\ \hline
UserKNN      &0.034 (0.001)           &0.755 (0.020)           &0.331 (0.005)                    \\
ItemKNN      &0.040 (0.000)           &0.906 (0.002)           &0.383 (0.001)           \\
HybridKNN    &0.041 (0.000)           &0.913 (0.002)           &0.389 (0.001)           \\
HeatS        &0.037 (0.000)           &0.836 (0.003)           &0.288 (0.004)                    \\
ProbS        &0.042 (0.000)           &0.948 (0.001)           &0.414 (0.001)           \\
HybridS      &0.042 (0.000)           &0.948 (0.001)           &0.419 (0.001)           \\
\hline
\textbf{Avg.}         &\textbf{0.040 (0.000)}           &\textbf{0.894 (0.005)}           &\textbf{0.387 (0.002)}\\
\hline
GMF          &0.039 (0.006)           &0.879 (0.143)           &0.332 (0.081)                    \\
MLP          &0.042 (0.001)           &0.939 (0.021)           &0.358 (0.013)                    \\
NCF          &0.042 (0.000)           &0.946 (0.002)           &0.365 (0.015)           \\
TransE       &0.031 (0.002)           &0.705 (0.046)           &0.215 (0.025)                    \\
\hline
\textbf{Avg.}         &0.039 (0.002)           &0.867 (0.053)           &0.318 (0.036)\\
\bottomrule
\end{tabular}
\end{adjustbox}
\end{table}

\subsubsection{Predicting implicit user-item interactions}

In predicting implicit user-item interactions (\textit{i.e.}, $0/1$), graph embedding-based recommendation models overall underperform conventional ones in recommendation accuracy on the three data sets, in which case recommendation scenario seems to be a decisive factor. Meanwhile, the lower stability of graph embedding-based recommendation models than conventional ones as manifested in Sec.~\ref{explicit_results} is also embodied in these tasks.

As shown in Tab.~\ref{implicit_result}, graph embedding-based recommendation models underperform conventional ones on MovieLens 100K by 32.81\%, 33.20\% and 37.50\% over average Precision@50, average Recall@50 and average Rank@50, respectively. On MovieLens 1M these results of graph embedding-based recommendation models are 32.89\% less, 33.09\% less and 44.71\% less than those of conventional ones, respectively. On Jester 617K are 2.50\% less, 3.02\% less and 17.83\% less, respectively. As these results put it, neither data scale nor sparsity is a decisive factor in recommendation accuracy. In other words, increasing data scale no longer brings consistent improvement in recommendation accuracy for graph embedding-based recommendation models. From Jester 617K to MovieLens 1M, the average Precision, average Recall and average Rank of graph embedding-based recommendation models are lower than those of conventional ones in a wider gap, while the case from MovieLens 100K to Jester 617K is the opposite. Meanwhile, it can be observed that decreasing data sparsity can shorten the accuracy gap between graph embedding-based recommendation models and conventional ones in terms of their average results on MovieLens 1M and Jester 617K, while this case does not hold on MovieLens 100K and MovieLens 1M. In view of that, the decisive factor influencing recommendation accuracy most likely lies in recommendation scenario distinguished by the difference in the relative number of users and items, for that of Jester 617K, a joke recommendation scenario dominated by the number of users, is distinguished from the cases on MovieLens 100K and 1M about a movie recommendation scenario. When concerning model stability, apparently, conventional recommendation models outperform graph embedding-based ones in terms of the standard deviation of their average results on all three data sets.

\subsection{Summary}

In conclusion, based on the above experimental results, this section provides some constructive suggestions on making a trade-off between graph embedding-based and conventional recommendation in different recommendation tasks, and also proposes some open questions for future research.

The graph embedding-based and conventional recommendation perform distinctively on different tasks. First, as for recommendation accuracy, graph embedding-based recommendation is a prior choice in predicting explicit user-item interactions, especially when the data scale is large. Meanwhile, in predicting implicit user-item interactions, conventional recommendation could still maintain a priority to graph embedding-based recommendation without considering the utilization of side information and knowledge. Second, when concerning stability, conventional recommendation is always a prior choice benefit from its less adjustment for hyper-parameters compared with graph embedding-based recommendation. Third, recommendation efficiency is also a necessity to be taken into account, especially for practical applications with big data. In this regard, graph embedding-based recommendation appears to be always considered as a priority because of its rationale for reusing the embeddings once learned for recommendation.

In practice, recommender systems usually involve both tasks for predicting explicit and implicit user-item interactions. Besides, recommender systems intrinsically develop from small to large in data scale and from sparse to dense in data sparsity, especially after being incorporated with side information and knowledge. Correspondingly, making a trade-off between graph embedding-based and conventional recommendation appears to vary in these different development stages of recommender systems, as the experimental results manifest in this section. As suggestions, in recommender systems' early developing stages conventional recommendation could perform well on a small data scale. At the same time its better explainability compared with graph embedding-based recommendation provides clearer views on user's behaviors, consider as straightforward feedback for guiding a more adaptive recommendation (models') configuration. With the increase of data scale when advancing into the following developing stages of recommender systems, graph embedding-based recommendation is supposed to gradually dominate recommendation configuration for better efficiency and accuracy. In the long run, in order to complement each other with their respective advantages, a versatile recommendation configuration mixing conventional and graph embedding-based recommendation models is an optimal strategy for improving recommender systems.

However, there still remain a few open questions. First, does data sparsity determines a model's recommendation performance? If it does, how? Second, how to select appropriate models for a specific recommendation scenario? Third, when designing a recommendation model, is it better to adopt a task-oriented strategy or a generalization-oriented strategy?

\section{Discussions and outlook}
\label{outlook}

The rapid development of computing resources and machine learning methodology contributes to the prevalence of graph embedding-based recommendation. In summary, as retrospected in this article, bipartite graph embedding-based recommendation can provide an extensible framework based on which the auxiliary information and temporal factors involved in user-item interactions can be incorporated. However, it can not sufficiently and efficiently incorporate side information or knowledge. To fill this gap, general graph embedding-based recommendation was invented. Furthermore, in order to enhance the abilities of recommendation models in representing the multiplicity and evolution of large-scale data, knowledge graph embedding-based recommendation has recently attracted many scholar attention. Nevertheless, among these techniques, flaws and challenges also exist as illustrated in this article. For that, complementing them to each other with their respective advantages is seemly promising, by such as combining embedding-based methods with path-based methods or extending shallow models to deep learning ones like deep matrix factorization \cite{de2021survey}. To the end, this section puts forward some open questions of graph embedding-based recommendation as well as their correspondingly possible solutions as follows.

\subsection{Graph topological analysis for recommendation}

To promote model accuracy, how could graph topological analysis contribute to graph embedding-based recommendation? In making great strides in recommender systems from conventional recommendation to graph embedding-based one, are we really making much progress on its performance? This debate has been raised many times in recent years, where Dacrema et al. \cite{dacrema2019we} threw cold water on DNN-based recommendation (a recently prevalent focus) and proved that conventional recommendation models can still achieve higher accuracy compared to graph embedding-based ones in some tasks, as verified in Sec.~\ref{evaluation}. In view of that, graph topological analysis (the rationale behind conventional recommendation ) seemly still plays a large role in model's accuracy. As yet, recent research into combining graph topological analysis on subgraphs, motifs or neighborhoods with graph embedding-based recommendation has come to see the merit in improving the performance of model's accuracy. For future research, analyzing graph's higher-order topological characteristics \cite{sizemore2018cliques, shi2019totally, battiston2020networks, shi2021computing}, namely network cycles, cliques and cavities, and employing them into graph embedding-based recommendation appears to be a potential direction, for these characteristics can be used to uncover higher-order relations between nodes, which are beyond nodes' pair-wise relations where most current methods concentrated. Methodologically, in order to realize the combination between graph's higher-order characteristics and graph embedding-based recommendation, it could resort to utilizing higher-order graph topology as a guideline for random walking in some graph embedding-based recommendation models. In addition, it is also positive to base more sufficient graph representations (like hypergraphs \cite{ouvrard2020hypergraphs, battiston2020networks}) beyond the conventional three categories in Sec.~\ref{graph_representation} on their higher-order topology. Moreover, designing novel label propagation (or message passing) mechanisms for GNNs based on graph's higher-order topology is also a promising direction.

\subsection{Explainable recommendation}

Explainable recommendation drives recommendation models to evolve from machine perceptual learning based on data fitting to machine cognitive learning for reasoning, long regarded as a necessity for comprehending user's preferences for items, assuring recommendation results' reliability affecting user's trust (\textit{i.e.}, users should know why did them receive these recommendations), and contributing to illuminating the black-box mechanism of graph embedding-based recommendation (at least in a phenomenon-level). For those purposes, promising future research directions might include knowledge-based explainable recommendation \cite{zhang2018explainable, gao2019explainable, palmonari2020knowledge, xie2021explainable}, causal learning (causal inference) \cite{gopnik2007causal, liang2016causal, bonner2018causal, yao2020survey, wang2020causal, xu2021causal, xu2021learning} and neural network interpretability \cite{zhang2021survey}.

\subsection{Protect user's private information in recommendation}

In practice, explicit interactions and side information (like social networks) are common-used information for recommendation. However, one might be loath to share with recommender systems the two kinds of recommendation since his privacy could be leaked or even be infringed by others. From another perspective, perceiving user's preferences for items (\textit{i.e.}, explicit interactions) from the frequency of implicit user-item interactions (\textit{e.g.}, click or play counts and viewing time)  would be a potential solution. On the other hand, instead of escaping from utilizing this privacy-sensitive information, federated learning \cite{aledhari2020federated, mothukuri2021survey, zhang2021survey} is also a promising solution, which can require for user's private data to train several distributed models on user's local devices and then upload the parameters of these learned models to a terminal server to support an aggregated big model. In this process, user's privacy remains on their own devices without being shared with the terminal server. Nevertheless, since user's local devices could lack enough computing efficiency for training a model, efficient embedding techniques for large-scale data are necessary for federated learning.

\subsection{Employ social sciences and NLP techniques in recommendation}

Sufficiently incorporating rich information in recommendation requires for identification and preservation of hidden patterns and multiplicity involved in this information by, for example, identifying different tie strengths in social networks or distinguishing possible abundant semantics of knowledge, in which case techniques of such as community detection or language representation and reasoning play essential roles. In light of that, social sciences and natural language processing (NLP) seemly are two fields most likely overlapped with research into graph embedding-based recommendation. As conversational recommender systems (CRS) has become a promising solution to user privacy protection and explainable recommendation, the techniques of CRS in semantic sentiment analysis and dialogue system are largely based on NLP.

In the future, it is conceivable that recommender systems will be a perpetual theme in the information age, as long as users have the demand to quickly access their preferred or needed items from millions of candidates and commercials have the willingness to efficiently reap profits by their products through Precise Marketing. The rise of graph embedding-based recommendation is just starting, underlying challenges are coming, and promising solutions are sprouting. Expanding the research into graph embedding-based recommendation from the current focus on recommendation accuracy to improving the diversity and novelty of recommendation could also broaden its potential applications. Opening up more applications of graph embedding-based recommendation in practice would give full play to its social and commercial values. It is worth noting that more comprehensive papers about recommender systems can be found and downloaded on public projects\footnote{https://github.com/hongleizhang/RSPapers}.

\addcontentsline{toc}{section}{Acknowledgements}
\section*{Acknowledgements}

The author acknowledges Linyuan L\"u, Shuqi Xu, Xu Na, Hao Wang and Honglei Zhang for their discussions and suggestions.

\newpage
\addcontentsline{toc}{section}{References}
\small

\bibliography{elsarticle-template}

\newpage
\begin{appendices}

\setcounter{table}{0}   
\renewcommand{\thetable}{A\arabic{table}}

\clearpage
\section{Hyper-parameter settings in experiments}
\label{Hyper-parameter_settings_in_experiments}

\begin{table}[!ht]
\setlength{\abovecaptionskip}{0.3cm}
\setlength{\belowcaptionskip}{0.3cm}
\caption{\textbf{Hyper-parameter settings.} The evaluated models' hyper-parameter settings in Sec.~\ref{Results_and_Analysis} follow the principle of making a balance between model's recommendation accuracy and computing efficiency. As for the conventional recommendation models, on the three experimental data sets the k-nearest neighbors ($K$) of UserKNN, ItemKNN and HybridKNN are set to $10$ and the combination ratio ($\alpha$) of HybridKNN and HybridS is set to $0.5$. As for graph embedding-based recommendation models, the optimal embedding dimension (dim) is searched in the space of $[4,8,16,32,64,128]$ and the optimal learning rate (lr) is searched in the space of $[0.1,0.05,0.01,0.005,0.001]$. In addition, as for deep learning-based models, their layers and dropout rate are set by experience. More details are shown below, where n.a. indicates that a model failed to implement on a data set.}
\label{hyper-parameter_settings}
\begin{adjustbox}{center}
\footnotesize
\begin{tabular}{cccc}
\toprule
\textbf{}        & \textbf{MovieLens 100K}                                      & \textbf{MovieLens 1M}                                                                       & \textbf{Jester 617K}   \\ \midrule
\textbf{UserKNN} & K=10                                                         & K=10                                                                                        & K=10                   \\
\textbf{ItemKNN} & K=10                                                         & K=10                                                                                        & K=10                   \\
\textbf{HybridKNN} & K=10, $\alpha=0.5$                                         & K=10, $\alpha=0.5$                                                                          & K=10, $\alpha=0.5$      \\
\textbf{HybridS} & $\alpha=0.5$                                 & $\alpha=0.5$                                                                                & $\alpha=0.5$           \\
\textbf{FunkSVD} & dim=4, lr=0.01                                               & dim=8, lr=0.01                                                                              & n.a.                   \\
\textbf{PMF}     & dim=16, lr=0.01                                              & dim=8, lr=0.01                                                                              & dim=64, lr=0.01        \\
\textbf{FM}      & dim=8, lr=0.01                                               & dim=4, lr=0.01                                                                              & dim=4, lr=0.01         \\
\textbf{AutoRec} & dim=4, lr=0.01                                               & dim=32, lr=0.001                                                                            & dim=8, lr=0.05         \\
\textbf{GMF}     & dim=128, lr=0.01                                             & dim=128, lr=0.005                                                                           & dim=4, lr=0.05         \\
\textbf{MLP}     & \begin{tabular}[c]{@{}c@{}}dim=64, lr=0.05, \\ layers=3, dropout=0.5\end{tabular}             & \begin{tabular}[c]{@{}c@{}}dim=32, lr=0.005, \\ layers=3, dropout=0.5\end{tabular}           & \begin{tabular}[c]{@{}c@{}}dim=16, lr=0.01, \\ layers=3, dropout=0.5\end{tabular}            \\
\textbf{NCF}   & \begin{tabular}[c]{@{}c@{}}dim=128, lr=0.01,0.01,0.005, \\ layers=3, dropout=0.5\end{tabular} & \begin{tabular}[c]{@{}c@{}}dim=64, lr=0.01,0.01,0.005, \\ layers=3, dropout=0.5\end{tabular} & \begin{tabular}[c]{@{}c@{}}dim=128, lr=0.01,0.01,0.01, \\ layers=3, dropout=0.5\end{tabular} \\
\textbf{TransE}  & dim=128, lr=0.01                                            & dim=64, lr=0.005                                                                            & dim=128, lr=0.1        \\
\bottomrule
\end{tabular}
\end{adjustbox}
\end{table}

\section{Hyper-parameter tuning results in experiments}
\label{Hyper-parameter tuning results in experiments}

\begin{table}[h]
\caption{\textbf{Hyper-parameter tuning results on MovieLens 100K for predicting explicit user-item interactions.} The notation n.a. indicates that a model failed to implement on the data set.}
\begin{tabular}{@{}ccccccccccccc@{}}
\toprule
Metric                                              & \multicolumn{6}{c}{MAE}                                                                                                                                                                                                                                                                                              & \multicolumn{6}{c}{RMSE}                                                                                                                                                                                                                                                                                             \\ \midrule
Model                                               & \begin{tabular}[c]{@{}c@{}}dim\\ =4\end{tabular} & \begin{tabular}[c]{@{}c@{}}dim\\ =8\end{tabular} & \begin{tabular}[c]{@{}c@{}}dim\\ =16\end{tabular} & \begin{tabular}[c]{@{}c@{}}dim\\ =32\end{tabular} & \begin{tabular}[c]{@{}c@{}}dim\\ =64\end{tabular} & \begin{tabular}[c]{@{}c@{}}dim\\ =128\end{tabular} & \begin{tabular}[c]{@{}c@{}}dim\\ =4\end{tabular} & \begin{tabular}[c]{@{}c@{}}dim\\ =8\end{tabular} & \begin{tabular}[c]{@{}c@{}}dim\\ =16\end{tabular} & \begin{tabular}[c]{@{}c@{}}dim\\ =32\end{tabular} & \begin{tabular}[c]{@{}c@{}}dim\\ =64\end{tabular} & \begin{tabular}[c]{@{}c@{}}dim\\ =128\end{tabular} \\ \midrule
\begin{tabular}[c]{@{}c@{}}UserKNN\end{tabular} & \multicolumn{6}{c}{0.702}                                                                                                                                                                                                                                                                                            & \multicolumn{6}{c}{0.912}                                                                                                                                                                                                                                                                                            \\
\begin{tabular}[c]{@{}c@{}}ItemKNN\end{tabular} & \multicolumn{6}{c}{0.718}                                                                                                                                                                                                                                                                                            & \multicolumn{6}{c}{0.936}                                                                                                                                                                                                                                                                                            \\
FunkSVD                                             & 0.709                                            & 0.724                                            & 0.736                                             & 0.750                                             & 0.780                                             & 0.813                                              & 0.926                                            & 0.944                                            & 0.961                                             & 0.974                                             & 1.014                                             & 1.052                                              \\
PMF                                                 & 0.759                                            & 0.758                                            & 0.743                                             & 0.785                                             & 0.760                                             & 0.825                                              & 0.998                                            & 0.996                                            & 0.972                                             & 1.034                                             & 0.989                                             & 1.077                                              \\
FM                                                  & 0.692                                            & 0.692                                            & 0.697                                             & 0.714                                             & 0.731                                             & 0.759                                              & 0.890                                            & 0.889                                            & 0.896                                             & 0.915                                             & 0.935                                             & 0.973                                              \\
AutoRec                                             & 0.733                                            & 0.756                                            & 0.737                                             & 0.792                                             & 0.813                                             & 0.787                                              & 0.947                                            & 0.980                                            & 0.955                                             & 1.020                                             & 1.042                                             & 1.009                                              \\ \bottomrule
\end{tabular}
\end{table}

\begin{table}[h]
\caption{\textbf{Hyper-parameter tuning results on MovieLens 1M for predicting explicit user-item interactions.}}
\begin{tabular}{@{}ccccccccccccc@{}}
\toprule
Metric                                              & \multicolumn{6}{c}{MAE}                                                                                                                                                                                                                                                                                              & \multicolumn{6}{c}{RMSE}                                                                                                                                                                                                                                                                                             \\ \midrule
Model                                               & \begin{tabular}[c]{@{}c@{}}dim\\ =4\end{tabular} & \begin{tabular}[c]{@{}c@{}}dim\\ =8\end{tabular} & \begin{tabular}[c]{@{}c@{}}dim\\ =16\end{tabular} & \begin{tabular}[c]{@{}c@{}}dim\\ =32\end{tabular} & \begin{tabular}[c]{@{}c@{}}dim\\ =64\end{tabular} & \begin{tabular}[c]{@{}c@{}}dim\\ =128\end{tabular} & \begin{tabular}[c]{@{}c@{}}dim\\ =4\end{tabular} & \begin{tabular}[c]{@{}c@{}}dim\\ =8\end{tabular} & \begin{tabular}[c]{@{}c@{}}dim\\ =16\end{tabular} & \begin{tabular}[c]{@{}c@{}}dim\\ =32\end{tabular} & \begin{tabular}[c]{@{}c@{}}dim\\ =64\end{tabular} & \begin{tabular}[c]{@{}c@{}}dim\\ =128\end{tabular} \\ \midrule
\begin{tabular}[c]{@{}c@{}}UserKNN\end{tabular} & \multicolumn{6}{c}{0.773}                                                                                                                                                                                                                                                                                            & \multicolumn{6}{c}{0.978}                                                                                                                                                                                                                                                                                            \\
\begin{tabular}[c]{@{}c@{}}ItemKNN\end{tabular} & \multicolumn{6}{c}{0.718}                                                                                                                                                                                                                                                                                            & \multicolumn{6}{c}{0.907}                                                                                                                                                                                                                                                                                            \\
FunkSVD                                             & 0.699                                            & 0.694                                            & 0.700                                             & 0.710                                             & 0.719                                             & 0.731                                              & 0.892                                            & 0.887                                            & 0.894                                             & 0.905                                             & 0.916                                             & 0.930                                              \\
PMF                                                 & 0.695                                            & 0.682                                            & 0.707                                             & 0.699                                             & 0.710                                             & 0.698                                              & 0.883                                            & 0.869                                            & 0.897                                             & 0.888                                             & 0.901                                             & 0.887                                              \\
FM                                                  & 0.685                                            & 0.685                                            & 0.699                                             & 0.715                                             & 0.736                                             & 0.760                                              & 0.869                                            & 0.873                                            & 0.889                                             & 0.907                                             & 0.929                                             & 0.959                                              \\
AutoRec                                             & 0.776                                            & 0.736                                            & 0.732                                             & 0.722                                             & 0.727                                             & 0.758                                              & 0.973                                            & 0.927                                            & 0.923                                             & 0.917                                             & 0.923                                             & 0.960                                              \\ \bottomrule
\end{tabular}
\end{table}

\begin{table}[h]
\caption{\textbf{Hyper-parameter tuning results on Jester 617K for predicting explicit user-item interactions.}}
\begin{tabular}{@{}ccccccccccccc@{}}
\toprule
Metric                                              & \multicolumn{6}{c}{MAE}                                                                                                                                                                                                                                                                                              & \multicolumn{6}{c}{RMSE}                                                                                                                                                                                                                                                                                             \\ \midrule
Model                                               & \begin{tabular}[c]{@{}c@{}}dim\\ =4\end{tabular} & \begin{tabular}[c]{@{}c@{}}dim\\ =8\end{tabular} & \begin{tabular}[c]{@{}c@{}}dim\\ =16\end{tabular} & \begin{tabular}[c]{@{}c@{}}dim\\ =32\end{tabular} & \begin{tabular}[c]{@{}c@{}}dim\\ =64\end{tabular} & \begin{tabular}[c]{@{}c@{}}dim\\ =128\end{tabular} & \begin{tabular}[c]{@{}c@{}}dim\\ =4\end{tabular} & \begin{tabular}[c]{@{}c@{}}dim\\ =8\end{tabular} & \begin{tabular}[c]{@{}c@{}}dim\\ =16\end{tabular} & \begin{tabular}[c]{@{}c@{}}dim\\ =32\end{tabular} & \begin{tabular}[c]{@{}c@{}}dim\\ =64\end{tabular} & \begin{tabular}[c]{@{}c@{}}dim\\ =128\end{tabular} \\ \midrule
\begin{tabular}[c]{@{}c@{}}UserKNN\end{tabular} & \multicolumn{6}{c}{0.857}                                                                                                                                                                                                                                                                                            & \multicolumn{6}{c}{1.053}                                                                                                                                                                                                                                                                                            \\
\begin{tabular}[c]{@{}c@{}}ItemKNN\end{tabular} & \multicolumn{6}{c}{0.823}                                                                                                                                                                                                                                                                                            & \multicolumn{6}{c}{1.015}                                                                                                                                                                                                                                                                                            \\
FunkSVD                                             & \multicolumn{6}{c}{n.a.}                                                                                                                                                                                                                                                                                             & \multicolumn{6}{c}{n.a.}                                                                                                                                                                                                                                                                                             \\
PMF                                                 & 0.828                                            & 0.828                                            & 0.818                                             & 0.819                                             & 0.817                                             & 0.860                                              & 1.041                                            & 1.040                                            & 1.026                                             & 1.027                                             & 1.026                                             & 1.082                                              \\
FM                                                  & 0.813                                            & 0.816                                            & 0.821                                             & 0.827                                             & 0.836                                             & 0.852                                              & 1.008                                            & 1.011                                            & 1.016                                             & 1.022                                             & 1.032                                             & 1.051                                              \\
AutoRec                                             & 0.925                                            & 0.911                                            & 0.912                                             & 1.123                                             & 0.949                                             & 0.970                                              & 1.119                                            & 1.106                                            & 1.107                                             & 1.400                                             & 1.139                                             & 1.162                                              \\ \bottomrule
\end{tabular}
\end{table}

\begin{table}[h]
\caption{\textbf{Hyper-parameter tuning results on MovieLens 100K for predicting implicit user-item interactions.}}
\begin{tabular}{@{}cccccccccccccc@{}}
\toprule
\multicolumn{2}{c}{Metric}                                                                                                     & \multicolumn{6}{c}{Recall}                                                                                                                                                                                                                                                                                           & \multicolumn{6}{c}{NDCG}                                                                                                                                                                                                                                                                                             \\ \midrule
Model                                                                  & \begin{tabular}[c]{@{}c@{}}list\\ length\end{tabular} & \begin{tabular}[c]{@{}c@{}}dim\\ =4\end{tabular} & \begin{tabular}[c]{@{}c@{}}dim\\ =8\end{tabular} & \begin{tabular}[c]{@{}c@{}}dim\\ =16\end{tabular} & \begin{tabular}[c]{@{}c@{}}dim\\ =32\end{tabular} & \begin{tabular}[c]{@{}c@{}}dim\\ =64\end{tabular} & \begin{tabular}[c]{@{}c@{}}dim\\ =128\end{tabular} & \begin{tabular}[c]{@{}c@{}}dim\\ =4\end{tabular} & \begin{tabular}[c]{@{}c@{}}dim\\ =8\end{tabular} & \begin{tabular}[c]{@{}c@{}}dim\\ =16\end{tabular} & \begin{tabular}[c]{@{}c@{}}dim\\ =32\end{tabular} & \begin{tabular}[c]{@{}c@{}}dim\\ =64\end{tabular} & \begin{tabular}[c]{@{}c@{}}dim\\ =128\end{tabular} \\ \midrule
\multirow{5}{*}{\begin{tabular}[c]{@{}c@{}}User-\\ KNN\end{tabular}}   & @10                                                   & \multicolumn{6}{c}{0.090}                                                                                                                                                                                                                                                                                            & \multicolumn{6}{c}{0.044}                                                                                                                                                                                                                                                                                            \\
                                                                       & @20                                                   & \multicolumn{6}{c}{0.147}                                                                                                                                                                                                                                                                                            & \multicolumn{6}{c}{0.059}                                                                                                                                                                                                                                                                                            \\
                                                                       & @30                                                   & \multicolumn{6}{c}{0.188}                                                                                                                                                                                                                                                                                            & \multicolumn{6}{c}{0.067}                                                                                                                                                                                                                                                                                            \\
                                                                       & @40                                                   & \multicolumn{6}{c}{0.222}                                                                                                                                                                                                                                                                                            & \multicolumn{6}{c}{0.074}                                                                                                                                                                                                                                                                                            \\
                                                                       & @50                                                   & \multicolumn{6}{c}{0.252}                                                                                                                                                                                                                                                                                            & \multicolumn{6}{c}{0.079}                                                                                                                                                                                                                                                                                            \\ \midrule
\multirow{5}{*}{\begin{tabular}[c]{@{}c@{}}Item-\\ KNN\end{tabular}}   & @10                                                   & \multicolumn{6}{c}{0.092}                                                                                                                                                                                                                                                                                            & \multicolumn{6}{c}{0.046}                                                                                                                                                                                                                                                                                            \\
                                                                       & @20                                                   & \multicolumn{6}{c}{0.146}                                                                                                                                                                                                                                                                                            & \multicolumn{6}{c}{0.060}                                                                                                                                                                                                                                                                                            \\
                                                                       & @30                                                   & \multicolumn{6}{c}{0.185}                                                                                                                                                                                                                                                                                            & \multicolumn{6}{c}{0.068}                                                                                                                                                                                                                                                                                            \\
                                                                       & @40                                                   & \multicolumn{6}{c}{0.220}                                                                                                                                                                                                                                                                                            & \multicolumn{6}{c}{0.075}                                                                                                                                                                                                                                                                                            \\
                                                                       & @50                                                   & \multicolumn{6}{c}{0.245}                                                                                                                                                                                                                                                                                            & \multicolumn{6}{c}{0.079}                                                                                                                                                                                                                                                                                            \\ \midrule
\multirow{5}{*}{\begin{tabular}[c]{@{}c@{}}Hybrid-\\ KNN\end{tabular}} & @10                                                   & \multicolumn{6}{c}{0.105}                                                                                                                                                                                                                                                                                            & \multicolumn{6}{c}{0.053}                                                                                                                                                                                                                                                                                            \\
                                                                       & @20                                                   & \multicolumn{6}{c}{0.168}                                                                                                                                                                                                                                                                                            & \multicolumn{6}{c}{0.068}                                                                                                                                                                                                                                                                                            \\
                                                                       & @30                                                   & \multicolumn{6}{c}{0.212}                                                                                                                                                                                                                                                                                            & \multicolumn{6}{c}{0.078}                                                                                                                                                                                                                                                                                            \\
                                                                       & @40                                                   & \multicolumn{6}{c}{0.247}                                                                                                                                                                                                                                                                                            & \multicolumn{6}{c}{0.084}                                                                                                                                                                                                                                                                                            \\
                                                                       & @50                                                   & \multicolumn{6}{c}{0.276}                                                                                                                                                                                                                                                                                            & \multicolumn{6}{c}{0.090}                                                                                                                                                                                                                                                                                            \\ \midrule
\multirow{5}{*}{ProbS}                                                 & @10                                                   & \multicolumn{6}{c}{0.087}                                                                                                                                                                                                                                                                                            & \multicolumn{6}{c}{0.043}                                                                                                                                                                                                                                                                                            \\
                                                                       & @20                                                   & \multicolumn{6}{c}{0.139}                                                                                                                                                                                                                                                                                            & \multicolumn{6}{c}{0.056}                                                                                                                                                                                                                                                                                            \\
                                                                       & @30                                                   & \multicolumn{6}{c}{0.177}                                                                                                                                                                                                                                                                                            & \multicolumn{6}{c}{0.065}                                                                                                                                                                                                                                                                                            \\
                                                                       & @40                                                   & \multicolumn{6}{c}{0.210}                                                                                                                                                                                                                                                                                            & \multicolumn{6}{c}{0.071}                                                                                                                                                                                                                                                                                            \\
                                                                       & @50                                                   & \multicolumn{6}{c}{0.236}                                                                                                                                                                                                                                                                                            & \multicolumn{6}{c}{0.076}                                                                                                                                                                                                                                                                                            \\ \midrule
\multirow{5}{*}{HybridS}                                               & @10                                                   & \multicolumn{6}{c}{0.092}                                                                                                                                                                                                                                                                                            & \multicolumn{6}{c}{0.046}                                                                                                                                                                                                                                                                                            \\
                                                                       & @20                                                   & \multicolumn{6}{c}{0.149}                                                                                                                                                                                                                                                                                            & \multicolumn{6}{c}{0.060}                                                                                                                                                                                                                                                                                            \\
                                                                       & @30                                                   & \multicolumn{6}{c}{0.191}                                                                                                                                                                                                                                                                                            & \multicolumn{6}{c}{0.069}                                                                                                                                                                                                                                                                                            \\
                                                                       & @40                                                   & \multicolumn{6}{c}{0.223}                                                                                                                                                                                                                                                                                            & \multicolumn{6}{c}{0.075}                                                                                                                                                                                                                                                                                            \\
                                                                       & @50                                                   & \multicolumn{6}{c}{0.254}                                                                                                                                                                                                                                                                                            & \multicolumn{6}{c}{0.081}                                                                                                                                                                                                                                                                                            \\ \midrule
GMF                                                                    & @10                                                   & 0.023                                            & 0.029                                            & 0.037                                             & 0.044                                             & 0.055                                             & 0.060                                              & 0.011                                            & 0.014                                            & 0.018                                             & 0.020                                             & 0.025                                             & 0.029                                              \\
                                                                       & @20                                                   & 0.046                                            & 0.056                                            & 0.066                                             & 0.080                                             & 0.095                                             & 0.103                                              & 0.017                                            & 0.021                                            & 0.025                                             & 0.029                                             & 0.035                                             & 0.040                                              \\
                                                                       & @30                                                   & 0.067                                            & 0.079                                            & 0.091                                             & 0.109                                             & 0.128                                             & 0.139                                              & 0.021                                            & 0.026                                            & 0.030                                             & 0.036                                             & 0.043                                             & 0.047                                              \\
                                                                       & @40                                                   & 0.091                                            & 0.100                                            & 0.116                                             & 0.136                                             & 0.152                                             & 0.172                                              & 0.026                                            & 0.030                                            & 0.035                                             & 0.041                                             & 0.047                                             & 0.054                                              \\
                                                                       & @50                                                   & 0.113                                            & 0.121                                            & 0.136                                             & 0.159                                             & 0.176                                             & 0.199                                              & 0.030                                            & 0.034                                            & 0.039                                             & 0.045                                             & 0.051                                             & 0.058                                              \\ \midrule
\multirow{5}{*}{MLP}                                                   & @10                                                   & 0.002                                            & 0.018                                            & 0.035                                             & 0.030                                             & 0.042                                             & 0.042                                              & 0.001                                            & 0.008                                            & 0.016                                             & 0.015                                             & 0.020                                             & 0.021                                              \\
                                                                       & @20                                                   & 0.003                                            & 0.039                                            & 0.066                                             & 0.055                                             & 0.077                                             & 0.078                                              & 0.001                                            & 0.013                                            & 0.024                                             & 0.022                                             & 0.029                                             & 0.030                                              \\
                                                                       & @30                                                   & 0.006                                            & 0.057                                            & 0.094                                             & 0.082                                             & 0.104                                             & 0.103                                              & 0.002                                            & 0.017                                            & 0.030                                             & 0.027                                             & 0.035                                             & 0.035                                              \\
                                                                       & @40                                                   & 0.009                                            & 0.076                                            & 0.121                                             & 0.104                                             & 0.128                                             & 0.128                                              & 0.002                                            & 0.020                                            & 0.035                                             & 0.032                                             & 0.040                                             & 0.040                                              \\
                                                                       & @50                                                   & 0.011                                            & 0.096                                            & 0.143                                             & 0.128                                             & 0.149                                             & 0.148                                              & 0.003                                            & 0.024                                            & 0.039                                             & 0.036                                             & 0.043                                             & 0.043                                              \\ \midrule
\multirow{5}{*}{NCF}                                                   & @10                                                   & 0.036                                            & 0.031                                            & 0.037                                             & 0.046                                             & 0.057                                             & 0.061                                              & 0.016                                            & 0.014                                            & 0.018                                             & 0.022                                             & 0.027                                             & 0.029                                              \\
                                                                       & @20                                                   & 0.066                                            & 0.056                                            & 0.068                                             & 0.081                                             & 0.099                                             & 0.104                                              & 0.024                                            & 0.021                                            & 0.025                                             & 0.031                                             & 0.038                                             & 0.040                                              \\
                                                                       & @30                                                   & 0.091                                            & 0.076                                            & 0.093                                             & 0.109                                             & 0.136                                             & 0.141                                              & 0.029                                            & 0.025                                            & 0.031                                             & 0.037                                             & 0.046                                             & 0.048                                              \\
                                                                       & @40                                                   & 0.117                                            & 0.093                                            & 0.117                                             & 0.135                                             & 0.164                                             & 0.169                                              & 0.034                                            & 0.028                                            & 0.035                                             & 0.042                                             & 0.051                                             & 0.053                                              \\
                                                                       & @50                                                   & 0.139                                            & 0.113                                            & 0.136                                             & 0.158                                             & 0.189                                             & 0.195                                              & 0.038                                            & 0.032                                            & 0.039                                             & 0.046                                             & 0.056                                             & 0.058                                              \\ \midrule
\multirow{5}{*}{TransE}                                                & @10                                                   & 0.001                                            & 0.002                                            & 0.009                                             & 0.014                                             & 0.026                                             & 0.034                                              & 0.001                                            & 0.001                                            & 0.004                                             & 0.007                                             & 0.013                                             & 0.016                                              \\
                                                                       & @20                                                   & 0.003                                            & 0.003                                            & 0.017                                             & 0.025                                             & 0.044                                             & 0.057                                              & 0.001                                            & 0.001                                            & 0.006                                             & 0.010                                             & 0.018                                             & 0.022                                              \\
                                                                       & @30                                                   & 0.005                                            & 0.005                                            & 0.024                                             & 0.036                                             & 0.060                                             & 0.076                                              & 0.002                                            & 0.002                                            & 0.008                                             & 0.012                                             & 0.021                                             & 0.026                                              \\
                                                                       & @40                                                   & 0.007                                            & 0.006                                            & 0.030                                             & 0.047                                             & 0.075                                             & 0.094                                              & 0.002                                            & 0.002                                            & 0.009                                             & 0.014                                             & 0.024                                             & 0.029                                              \\
                                                                       & @50                                                   & 0.008                                            & 0.007                                            & 0.036                                             & 0.055                                             & 0.089                                             & 0.108                                              & 0.002                                            & 0.002                                            & 0.010                                             & 0.015                                             & 0.026                                             & 0.032                                              \\ \bottomrule
\end{tabular}
\end{table}

\begin{table}[h]
\caption{\textbf{Hyper-parameter tuning results on MovieLens 1M for predicting implicit user-item interactions.} The notation `-' means that a model's requirement for computing resources on the data set was beyond that of the author's experimental devices as a consequence of embedding vector's large dimension.}
\begin{tabular}{@{}cccccccccccccc@{}}
\toprule
\multicolumn{2}{c}{Metric}                                                                                                     & \multicolumn{6}{c}{Recall}                                                                                                                                                                                                                                                                                           & \multicolumn{6}{c}{NDCG}                                                                                                                                                                                                                                                                                             \\ \midrule
Model                                                                  & \begin{tabular}[c]{@{}c@{}}list\\ length\end{tabular} & \begin{tabular}[c]{@{}c@{}}dim\\ =4\end{tabular} & \begin{tabular}[c]{@{}c@{}}dim\\ =8\end{tabular} & \begin{tabular}[c]{@{}c@{}}dim\\ =16\end{tabular} & \begin{tabular}[c]{@{}c@{}}dim\\ =32\end{tabular} & \begin{tabular}[c]{@{}c@{}}dim\\ =64\end{tabular} & \begin{tabular}[c]{@{}c@{}}dim\\ =128\end{tabular} & \begin{tabular}[c]{@{}c@{}}dim\\ =4\end{tabular} & \begin{tabular}[c]{@{}c@{}}dim\\ =8\end{tabular} & \begin{tabular}[c]{@{}c@{}}dim\\ =16\end{tabular} & \begin{tabular}[c]{@{}c@{}}dim\\ =32\end{tabular} & \begin{tabular}[c]{@{}c@{}}dim\\ =64\end{tabular} & \begin{tabular}[c]{@{}c@{}}dim\\ =128\end{tabular} \\ \midrule
\multirow{5}{*}{\begin{tabular}[c]{@{}c@{}}User-\\ KNN\end{tabular}}   & @10                                                   & \multicolumn{6}{c}{0.095}                                                                                                                                                                                                                                                                                            & \multicolumn{6}{c}{0.047}                                                                                                                                                                                                                                                                                            \\
                                                                       & @20                                                   & \multicolumn{6}{c}{0.157}                                                                                                                                                                                                                                                                                            & \multicolumn{6}{c}{0.062}                                                                                                                                                                                                                                                                                            \\
                                                                       & @30                                                   & \multicolumn{6}{c}{0.205}                                                                                                                                                                                                                                                                                            & \multicolumn{6}{c}{0.073}                                                                                                                                                                                                                                                                                            \\
                                                                       & @40                                                   & \multicolumn{6}{c}{0.246}                                                                                                                                                                                                                                                                                            & \multicolumn{6}{c}{0.080}                                                                                                                                                                                                                                                                                            \\
                                                                       & @50                                                   & \multicolumn{6}{c}{0.281}                                                                                                                                                                                                                                                                                            & \multicolumn{6}{c}{0.087}                                                                                                                                                                                                                                                                                            \\ \midrule
\multirow{5}{*}{\begin{tabular}[c]{@{}c@{}}Item-\\ KNN\end{tabular}}   & @10                                                   & \multicolumn{6}{c}{0.092}                                                                                                                                                                                                                                                                                            & \multicolumn{6}{c}{0.046}                                                                                                                                                                                                                                                                                            \\
                                                                       & @20                                                   & \multicolumn{6}{c}{0.154}                                                                                                                                                                                                                                                                                            & \multicolumn{6}{c}{0.061}                                                                                                                                                                                                                                                                                            \\
                                                                       & @30                                                   & \multicolumn{6}{c}{0.204}                                                                                                                                                                                                                                                                                            & \multicolumn{6}{c}{0.072}                                                                                                                                                                                                                                                                                            \\
                                                                       & @40                                                   & \multicolumn{6}{c}{0.246}                                                                                                                                                                                                                                                                                            & \multicolumn{6}{c}{0.080}                                                                                                                                                                                                                                                                                            \\
                                                                       & @50                                                   & \multicolumn{6}{c}{0.283}                                                                                                                                                                                                                                                                                            & \multicolumn{6}{c}{0.086}                                                                                                                                                                                                                                                                                            \\ \midrule
\multirow{5}{*}{\begin{tabular}[c]{@{}c@{}}Hybrid-\\ KNN\end{tabular}} & @10                                                   & \multicolumn{6}{c}{0.109}                                                                                                                                                                                                                                                                                            & \multicolumn{6}{c}{0.054}                                                                                                                                                                                                                                                                                            \\
                                                                       & @20                                                   & \multicolumn{6}{c}{0.181}                                                                                                                                                                                                                                                                                            & \multicolumn{6}{c}{0.072}                                                                                                                                                                                                                                                                                            \\
                                                                       & @30                                                   & \multicolumn{6}{c}{0.236}                                                                                                                                                                                                                                                                                            & \multicolumn{6}{c}{0.084}                                                                                                                                                                                                                                                                                            \\
                                                                       & @40                                                   & \multicolumn{6}{c}{0.283}                                                                                                                                                                                                                                                                                            & \multicolumn{6}{c}{0.093}                                                                                                                                                                                                                                                                                            \\
                                                                       & @50                                                   & \multicolumn{6}{c}{0.323}                                                                                                                                                                                                                                                                                            & \multicolumn{6}{c}{0.100}                                                                                                                                                                                                                                                                                            \\ \midrule
\multirow{5}{*}{ProbS}                                                 & @10                                                   & \multicolumn{6}{c}{0.077}                                                                                                                                                                                                                                                                                            & \multicolumn{6}{c}{0.040}                                                                                                                                                                                                                                                                                            \\
                                                                       & @20                                                   & \multicolumn{6}{c}{0.124}                                                                                                                                                                                                                                                                                            & \multicolumn{6}{c}{0.051}                                                                                                                                                                                                                                                                                            \\
                                                                       & @30                                                   & \multicolumn{6}{c}{0.164}                                                                                                                                                                                                                                                                                            & \multicolumn{6}{c}{0.060}                                                                                                                                                                                                                                                                                            \\
                                                                       & @40                                                   & \multicolumn{6}{c}{0.199}                                                                                                                                                                                                                                                                                            & \multicolumn{6}{c}{0.067}                                                                                                                                                                                                                                                                                            \\
                                                                       & @50                                                   & \multicolumn{6}{c}{0.231}                                                                                                                                                                                                                                                                                            & \multicolumn{6}{c}{0.072}                                                                                                                                                                                                                                                                                            \\ \midrule
\multirow{5}{*}{HybridS}                                               & @10                                                   & \multicolumn{6}{c}{0.089}                                                                                                                                                                                                                                                                                            & \multicolumn{6}{c}{0.046}                                                                                                                                                                                                                                                                                            \\
                                                                       & @20                                                   & \multicolumn{6}{c}{0.142}                                                                                                                                                                                                                                                                                            & \multicolumn{6}{c}{0.059}                                                                                                                                                                                                                                                                                            \\
                                                                       & @30                                                   & \multicolumn{6}{c}{0.187}                                                                                                                                                                                                                                                                                            & \multicolumn{6}{c}{0.069}                                                                                                                                                                                                                                                                                            \\
                                                                       & @40                                                   & \multicolumn{6}{c}{0.225}                                                                                                                                                                                                                                                                                            & \multicolumn{6}{c}{0.076}                                                                                                                                                                                                                                                                                            \\
                                                                       & @50                                                   & \multicolumn{6}{c}{0.260}                                                                                                                                                                                                                                                                                            & \multicolumn{6}{c}{0.083}                                                                                                                                                                                                                                                                                            \\ \midrule
GMF                                                                    & @10                                                   & 0.075                                            & 0.071                                            & 0.071                                             & 0.074                                             & 0.078                                             & 0.079                                              & 0.037                                            & 0.034                                            & 0.033                                             & 0.035                                             & 0.038                                             & 0.038                                              \\
                                                                       & @20                                                   & 0.126                                            & 0.124                                            & 0.124                                             & 0.130                                             & 0.136                                             & 0.138                                              & 0.050                                            & 0.047                                            & 0.047                                             & 0.049                                             & 0.052                                             & 0.052                                              \\
                                                                       & @30                                                   & 0.167                                            & 0.168                                            & 0.168                                             & 0.176                                             & 0.183                                             & 0.186                                              & 0.058                                            & 0.057                                            & 0.056                                             & 0.059                                             & 0.062                                             & 0.063                                              \\
                                                                       & @40                                                   & 0.203                                            & 0.207                                            & 0.207                                             & 0.217                                             & 0.224                                             & 0.228                                              & 0.065                                            & 0.064                                            & 0.064                                             & 0.067                                             & 0.070                                             & 0.071                                              \\
                                                                       & @50                                                   & 0.235                                            & 0.242                                            & 0.242                                             & 0.253                                             & 0.260                                             & 0.265                                              & 0.071                                            & 0.071                                            & 0.070                                             & 0.073                                             & 0.077                                             & 0.078                                              \\ \midrule
\multirow{5}{*}{MLP}                                                   & @10                                                   & 0.039                                            & 0.060                                            & 0.060                                             & 0.074                                             & 0.024                                             & -                                                  & 0.018                                            & 0.029                                            & 0.030                                             & 0.036                                             & 0.011                                             & -                                                  \\
                                                                       & @20                                                   & 0.074                                            & 0.107                                            & 0.107                                             & 0.127                                             & 0.046                                             & -                                                  & 0.027                                            & 0.040                                            & 0.042                                             & 0.049                                             & 0.016                                             & -                                                  \\
                                                                       & @30                                                   & 0.107                                            & 0.146                                            & 0.146                                             & 0.170                                             & 0.066                                             & -                                                  & 0.034                                            & 0.049                                            & 0.051                                             & 0.058                                             & 0.021                                             & -                                                  \\
                                                                       & @40                                                   & 0.139                                            & 0.181                                            & 0.181                                             & 0.207                                             & 0.087                                             & -                                                  & 0.040                                            & 0.056                                            & 0.057                                             & 0.066                                             & 0.025                                             & -                                                  \\
                                                                       & @50                                                   & 0.168                                            & 0.213                                            & 0.213                                             & 0.241                                             & 0.106                                             & -                                                  & 0.045                                            & 0.061                                            & 0.063                                             & 0.072                                             & 0.028                                             & -                                                  \\ \midrule
\multirow{5}{*}{NCF}                                                   & @10                                                   & 0.076                                            & 0.080                                            & 0.080                                             & 0.080                                             & 0.083                                             & -                                                  & 0.037                                            & 0.039                                            & 0.040                                             & 0.039                                             & 0.041                                             & -                                                  \\
                                                                       & @20                                                   & 0.126                                            & 0.135                                            & 0.135                                             & 0.136                                             & 0.141                                             & -                                                  & 0.050                                            & 0.053                                            & 0.054                                             & 0.053                                             & 0.055                                             & -                                                  \\
                                                                       & @30                                                   & 0.168                                            & 0.181                                            & 0.181                                             & 0.184                                             & 0.188                                             & -                                                  & 0.059                                            & 0.063                                            & 0.064                                             & 0.063                                             & 0.065                                             & -                                                  \\
                                                                       & @40                                                   & 0.204                                            & 0.222                                            & 0.222                                             & 0.223                                             & 0.228                                             & -                                                  & 0.066                                            & 0.071                                            & 0.071                                             & 0.071                                             & 0.073                                             & -                                                  \\
                                                                       & @50                                                   & 0.235                                            & 0.257                                            & 0.257                                             & 0.258                                             & 0.264                                             & -                                                  & 0.071                                            & 0.077                                            & 0.078                                             & 0.077                                             & 0.080                                             & -                                                  \\ \midrule
\multirow{5}{*}{TransE}                                                & @10                                                   & 0.003                                            & 0.003                                            & 0.003                                             & 0.003                                             & 0.010                                             & 0.009                                              & 0.001                                            & 0.001                                            & 0.001                                             & 0.001                                             & 0.005                                             & 0.004                                              \\
                                                                       & @20                                                   & 0.006                                            & 0.006                                            & 0.006                                             & 0.006                                             & 0.018                                             & 0.017                                              & 0.002                                            & 0.002                                            & 0.002                                             & 0.002                                             & 0.007                                             & 0.006                                              \\
                                                                       & @30                                                   & 0.009                                            & 0.009                                            & 0.009                                             & 0.009                                             & 0.026                                             & 0.024                                              & 0.003                                            & 0.003                                            & 0.003                                             & 0.003                                             & 0.008                                             & 0.008                                              \\
                                                                       & @40                                                   & 0.012                                            & 0.012                                            & 0.012                                             & 0.012                                             & 0.034                                             & 0.031                                              & 0.003                                            & 0.003                                            & 0.003                                             & 0.003                                             & 0.010                                             & 0.009                                              \\
                                                                       & @50                                                   & 0.015                                            & 0.015                                            & 0.015                                             & 0.015                                             & 0.041                                             & 0.038                                              & 0.004                                            & 0.004                                            & 0.004                                             & 0.004                                             & 0.011                                             & 0.010                                              \\ \bottomrule
\end{tabular}
\end{table}

\begin{table}[]
\caption{\textbf{Hyper-parameter tuning results on Jester 617K for predicting implicit user-item interactions.}}
\begin{tabular}{@{}cccccccccccccc@{}}
\toprule
\multicolumn{2}{c}{Metric}                                                                                                     & \multicolumn{6}{c}{Recall}                                                                                                                                                                                                                                                                                           & \multicolumn{6}{c}{NDCG}                                                                                                                                                                                                                                                                                             \\ \midrule
Model                                                                  & \begin{tabular}[c]{@{}c@{}}list\\ length\end{tabular} & \begin{tabular}[c]{@{}c@{}}dim\\ =4\end{tabular} & \begin{tabular}[c]{@{}c@{}}dim\\ =8\end{tabular} & \begin{tabular}[c]{@{}c@{}}dim\\ =16\end{tabular} & \begin{tabular}[c]{@{}c@{}}dim\\ =32\end{tabular} & \begin{tabular}[c]{@{}c@{}}dim\\ =64\end{tabular} & \begin{tabular}[c]{@{}c@{}}dim\\ =128\end{tabular} & \begin{tabular}[c]{@{}c@{}}dim\\ =4\end{tabular} & \begin{tabular}[c]{@{}c@{}}dim\\ =8\end{tabular} & \begin{tabular}[c]{@{}c@{}}dim\\ =16\end{tabular} & \begin{tabular}[c]{@{}c@{}}dim\\ =32\end{tabular} & \begin{tabular}[c]{@{}c@{}}dim\\ =64\end{tabular} & \begin{tabular}[c]{@{}c@{}}dim\\ =128\end{tabular} \\ \midrule
\multirow{5}{*}{\begin{tabular}[c]{@{}c@{}}User-\\ KNN\end{tabular}}   & @10                                                   & \multicolumn{6}{c}{0.511}                                                                                                                                                                                                                                                                                            & \multicolumn{6}{c}{0.279}                                                                                                                                                                                                                                                                                            \\
                                                                       & @20                                                   & \multicolumn{6}{c}{0.648}                                                                                                                                                                                                                                                                                            & \multicolumn{6}{c}{0.314}                                                                                                                                                                                                                                                                                            \\
                                                                       & @30                                                   & \multicolumn{6}{c}{0.703}                                                                                                                                                                                                                                                                                            & \multicolumn{6}{c}{0.326}                                                                                                                                                                                                                                                                                            \\
                                                                       & @40                                                   & \multicolumn{6}{c}{0.771}                                                                                                                                                                                                                                                                                            & \multicolumn{6}{c}{0.339}                                                                                                                                                                                                                                                                                            \\
                                                                       & @50                                                   & \multicolumn{6}{c}{0.842}                                                                                                                                                                                                                                                                                            & \multicolumn{6}{c}{0.352}                                                                                                                                                                                                                                                                                            \\ \midrule
\multirow{5}{*}{\begin{tabular}[c]{@{}c@{}}Item-\\ KNN\end{tabular}}   & @10                                                   & \multicolumn{6}{c}{0.590}                                                                                                                                                                                                                                                                                            & \multicolumn{6}{c}{0.309}                                                                                                                                                                                                                                                                                            \\
                                                                       & @20                                                   & \multicolumn{6}{c}{0.784}                                                                                                                                                                                                                                                                                            & \multicolumn{6}{c}{0.358}                                                                                                                                                                                                                                                                                            \\
                                                                       & @30                                                   & \multicolumn{6}{c}{0.843}                                                                                                                                                                                                                                                                                            & \multicolumn{6}{c}{0.371}                                                                                                                                                                                                                                                                                            \\
                                                                       & @40                                                   & \multicolumn{6}{c}{0.879}                                                                                                                                                                                                                                                                                            & \multicolumn{6}{c}{0.378}                                                                                                                                                                                                                                                                                            \\
                                                                       & @50                                                   & \multicolumn{6}{c}{0.911}                                                                                                                                                                                                                                                                                            & \multicolumn{6}{c}{0.384}                                                                                                                                                                                                                                                                                            \\ \midrule
\multirow{5}{*}{\begin{tabular}[c]{@{}c@{}}Hybrid-\\ KNN\end{tabular}} & @10                                                   & \multicolumn{6}{c}{0.591}                                                                                                                                                                                                                                                                                            & \multicolumn{6}{c}{0.316}                                                                                                                                                                                                                                                                                            \\
                                                                       & @20                                                   & \multicolumn{6}{c}{0.768}                                                                                                                                                                                                                                                                                            & \multicolumn{6}{c}{0.361}                                                                                                                                                                                                                                                                                            \\
                                                                       & @30                                                   & \multicolumn{6}{c}{0.850}                                                                                                                                                                                                                                                                                            & \multicolumn{6}{c}{0.378}                                                                                                                                                                                                                                                                                            \\
                                                                       & @40                                                   & \multicolumn{6}{c}{0.886}                                                                                                                                                                                                                                                                                            & \multicolumn{6}{c}{0.385}                                                                                                                                                                                                                                                                                            \\
                                                                       & @50                                                   & \multicolumn{6}{c}{0.918}                                                                                                                                                                                                                                                                                            & \multicolumn{6}{c}{0.391}                                                                                                                                                                                                                                                                                            \\ \midrule
\multirow{5}{*}{ProbS}                                                 & @10                                                   & \multicolumn{6}{c}{0.628}                                                                                                                                                                                                                                                                                            & \multicolumn{6}{c}{0.338}                                                                                                                                                                                                                                                                                            \\
                                                                       & @20                                                   & \multicolumn{6}{c}{0.831}                                                                                                                                                                                                                                                                                            & \multicolumn{6}{c}{0.390}                                                                                                                                                                                                                                                                                            \\
                                                                       & @30                                                   & \multicolumn{6}{c}{0.881}                                                                                                                                                                                                                                                                                            & \multicolumn{6}{c}{0.401}                                                                                                                                                                                                                                                                                            \\
                                                                       & @40                                                   & \multicolumn{6}{c}{0.916}                                                                                                                                                                                                                                                                                            & \multicolumn{6}{c}{0.407}                                                                                                                                                                                                                                                                                            \\
                                                                       & @50                                                   & \multicolumn{6}{c}{0.948}                                                                                                                                                                                                                                                                                            & \multicolumn{6}{c}{0.413}                                                                                                                                                                                                                                                                                            \\ \midrule
\multirow{5}{*}{HybridS}                                               & @10                                                   & \multicolumn{6}{c}{0.641}                                                                                                                                                                                                                                                                                            & \multicolumn{6}{c}{0.347}                                                                                                                                                                                                                                                                                            \\
                                                                       & @20                                                   & \multicolumn{6}{c}{0.831}                                                                                                                                                                                                                                                                                            & \multicolumn{6}{c}{0.395}                                                                                                                                                                                                                                                                                            \\
                                                                       & @30                                                   & \multicolumn{6}{c}{0.881}                                                                                                                                                                                                                                                                                            & \multicolumn{6}{c}{0.406}                                                                                                                                                                                                                                                                                            \\
                                                                       & @40                                                   & \multicolumn{6}{c}{0.917}                                                                                                                                                                                                                                                                                            & \multicolumn{6}{c}{0.413}                                                                                                                                                                                                                                                                                            \\
                                                                       & @50                                                   & \multicolumn{6}{c}{0.949}                                                                                                                                                                                                                                                                                            & \multicolumn{6}{c}{0.419}                                                                                                                                                                                                                                                                                            \\ \midrule
GMF                                                                    & @10                                                   & 0.568                                            & 0.567                                            & 0.555                                             & 0.575                                             & 0.578                                             & 0.574                                              & 0.285                                            & 0.275                                            & 0.263                                             & 0.278                                             & 0.276                                             & 0.289                                              \\
                                                                       & @20                                                   & 0.816                                            & 0.803                                            & 0.824                                             & 0.823                                             & 0.824                                             & 0.825                                              & 0.349                                            & 0.335                                            & 0.331                                             & 0.342                                             & 0.339                                             & 0.353                                              \\
                                                                       & @30                                                   & 0.873                                            & 0.862                                            & 0.871                                             & 0.867                                             & 0.875                                             & 0.872                                              & 0.361                                            & 0.348                                            & 0.342                                             & 0.351                                             & 0.350                                             & 0.363                                              \\
                                                                       & @40                                                   & 0.910                                            & 0.902                                            & 0.899                                             & 0.891                                             & 0.906                                             & 0.900                                              & 0.368                                            & 0.356                                            & 0.347                                             & 0.356                                             & 0.356                                             & 0.369                                              \\
                                                                       & @50                                                   & 0.943                                            & 0.937                                            & 0.923                                             & 0.914                                             & 0.932                                             & 0.925                                              & 0.374                                            & 0.362                                            & 0.351                                             & 0.360                                             & 0.360                                             & 0.374                                              \\ \midrule
\multirow{5}{*}{MLP}                                                   & @10                                                   & 0.564                                            & 0.551                                            & 0.571                                             & 0.569                                             & 0.564                                             & 0.527                                              & 0.266                                            & 0.268                                            & 0.287                                             & 0.282                                             & 0.257                                             & 0.259                                              \\
                                                                       & @20                                                   & 0.827                                            & 0.828                                            & 0.829                                             & 0.828                                             & 0.778                                             & 0.819                                              & 0.333                                            & 0.340                                            & 0.353                                             & 0.349                                             & 0.312                                             & 0.333                                              \\
                                                                       & @30                                                   & 0.880                                            & 0.879                                            & 0.879                                             & 0.879                                             & 0.823                                             & 0.879                                              & 0.344                                            & 0.351                                            & 0.364                                             & 0.360                                             & 0.322                                             & 0.346                                              \\
                                                                       & @40                                                   & 0.915                                            & 0.915                                            & 0.915                                             & 0.915                                             & 0.858                                             & 0.915                                              & 0.351                                            & 0.358                                            & 0.371                                             & 0.367                                             & 0.328                                             & 0.353                                              \\
                                                                       & @50                                                   & 0.947                                            & 0.948                                            & 0.947                                             & 0.947                                             & 0.889                                             & 0.945                                              & 0.357                                            & 0.363                                            & 0.377                                             & 0.372                                             & 0.334                                             & 0.358                                              \\ \midrule
\multirow{5}{*}{NCF}                                                   & @10                                                   & 0.570                                            & 0.585                                            & 0.567                                             & 0.581                                             & 0.581                                             & 0.580                                              & 0.269                                            & 0.288                                            & 0.277                                             & 0.275                                             & 0.303                                             & 0.293                                              \\
                                                                       & @20                                                   & 0.827                                            & 0.825                                            & 0.828                                             & 0.824                                             & 0.824                                             & 0.827                                              & 0.335                                            & 0.350                                            & 0.344                                             & 0.337                                             & 0.365                                             & 0.357                                              \\
                                                                       & @30                                                   & 0.876                                            & 0.876                                            & 0.878                                             & 0.874                                             & 0.872                                             & 0.879                                              & 0.346                                            & 0.360                                            & 0.355                                             & 0.348                                             & 0.375                                             & 0.368                                              \\
                                                                       & @40                                                   & 0.909                                            & 0.908                                            & 0.914                                             & 0.905                                             & 0.902                                             & 0.914                                              & 0.352                                            & 0.367                                            & 0.362                                             & 0.354                                             & 0.381                                             & 0.375                                              \\
                                                                       & @50                                                   & 0.937                                            & 0.936                                            & 0.945                                             & 0.929                                             & 0.928                                             & 0.947                                              & 0.357                                            & 0.372                                            & 0.368                                             & 0.358                                             & 0.386                                             & 0.380                                              \\ \midrule
\multirow{5}{*}{TransE}                                                & @10                                                   & 0.134                                            & 0.122                                            & 0.162                                             & 0.137                                             & 0.166                                             & 0.255                                              & 0.062                                            & 0.056                                            & 0.077                                             & 0.064                                             & 0.078                                             & 0.120                                              \\
                                                                       & @20                                                   & 0.254                                            & 0.238                                            & 0.287                                             & 0.257                                             & 0.297                                             & 0.426                                              & 0.092                                            & 0.085                                            & 0.109                                             & 0.094                                             & 0.111                                             & 0.163                                              \\
                                                                       & @30                                                   & 0.361                                            & 0.352                                            & 0.402                                             & 0.373                                             & 0.417                                             & 0.546                                              & 0.115                                            & 0.109                                            & 0.133                                             & 0.118                                             & 0.137                                             & 0.189                                              \\
                                                                       & @40                                                   & 0.468                                            & 0.465                                            & 0.508                                             & 0.486                                             & 0.528                                             & 0.643                                              & 0.135                                            & 0.131                                            & 0.153                                             & 0.140                                             & 0.158                                             & 0.207                                              \\
                                                                       & @50                                                   & 0.575                                            & 0.577                                            & 0.608                                             & 0.593                                             & 0.629                                             & 0.726                                              & 0.155                                            & 0.151                                            & 0.171                                             & 0.160                                             & 0.176                                             & 0.222                                              \\ \bottomrule
\end{tabular}
\end{table}

\begin{table}[h]
\caption{\textbf{The convergence steps of the machine learning-based baselines on the three data sets under specific hyper-parameters as shown in Tab.~\ref{hyper-parameter_settings}.} Among them, since NCF consists of three components, their convergence steps with different embedding dimensions on the three data sets are presented by three rows, respectively. Note that n.a. indicates that a model failed to implement on the data set, and `-' means that a model's requirement for computing resources on the data set was beyond that of the author's experimental devices as a consequence of embedding vector's large dimension.}
\begin{adjustbox}{center}
\begin{tabular}{@{}ccccccc@{}}

\toprule
\multicolumn{7}{c}{\textbf{MovieLens 100K}}                               \\ \midrule
Model                & dim=4 & dim=8 & dim=16 & dim=32 & dim=64 & dim=128 \\ \midrule
FunkSVD              & 16    & 12    & 11     & 10     & 9      & 9       \\
PMF                  & 6     & 6     & 7      & 5      & 8      & 5       \\
FM                   & 7     & 7     & 5      & 4      & 3      & 3       \\
AutoRec              & 1845  & 2375  & 1537   & 1558   & 1042   & 928     \\
GMF                  & 68    & 59    & 49     & 45     & 40     & 33      \\
MLP                  & 0     & 79    & 28     & 19     & 19     & 17      \\
\multirow{3}{*}{NCF} & 80    & 61    & 51     & 44     & 37     & 36      \\
                     & 32    & 0     & 0      & 14     & 16     & 1       \\
                     & 0     & 3766  & 5758   & 0      & 0      & 0       \\
TransE               & 32    & 14    & 812    & 689    & 716    & 669     \\ \midrule
\multicolumn{7}{c}{\textbf{MovieLens 1M}}                                 \\ \midrule
Model                & dim=4 & dim=8 & dim=16 & dim=32 & dim=64 & dim=128 \\ \midrule
FunkSVD              & 16    & 15    & 7      & 5      & 4      & 3       \\
PMF                  & 10    & 18    & 4      & 5      & 3      & 5       \\
FM                   & 26    & 33    & 12     & 6      & 4      & 3       \\
AutoRec              & 3698  & 2161  & 3323   & 2951   & 1983   & 1451    \\
GMF                  & 507   & 442   & 251    & 213    & 175    & 150     \\
MLP                  & 167   & 176   & 141    & 139    & 232    & -       \\
\multirow{3}{*}{NCF} & 0     & 0     & 0      & 0      & 0      & -       \\
                     & 317   & 259   & 154    & 130    & 106    & -       \\
                     & 17    & 131   & 122    & 128    & 106    & -       \\
TransE               & 32    & 0     & 1      & 569    & 1987   & 1345    \\ \midrule
\multicolumn{7}{c}{\textbf{Jester 617K}}                                  \\ \midrule
Model                & dim=4 & dim=8 & dim=16 & dim=32 & dim=64 & dim=128 \\ \midrule
FunkSVD              & n.a.  & n.a.  & n.a.   & n.a.   & n.a.   & n.a.    \\
PMF                  & 1     & 1     & 2      & 2      & 2      & 0       \\
FM                   & 8     & 7     & 6      & 5      & 4      & 3       \\
AutoRec              & 408   & 303   & 413    & 1124   & 81     & 117     \\
GMF                  & 40    & 36    & 11     & 9      & 8      & 7       \\
MLP                  & 25    & 30    & 24     & 24     & 0      & 3       \\
\multirow{3}{*}{NCF} & 61    & 47    & 39     & 34     & 28     & 24      \\
                     & 20    & 0     & 27     & 0      & 0      & 3       \\
                     & 0     & 0     & 0      & 0      & 0      & 0       \\
TransE               & 160   & 175   & 291    & 195    & 223    & 270     \\ \bottomrule

\end{tabular}
\end{adjustbox}
\end{table}

\clearpage
\section{Indexing}

\begin{table}[!ht]
\setlength{\abovecaptionskip}{0.3cm}
\setlength{\belowcaptionskip}{0.3cm}
\caption{\textbf{Indexing.}}
\label{Framework_and_literature_technologies_and_models}
\begin{adjustbox}{center}
\footnotesize

\begin{tabular}{ccccc}
\toprule
\textbf{\begin{tabular}[c]{@{}c@{}}Recommendation\\ (information or data)\end{tabular}}                                                                         & \textbf{Method}                                                                 & \textbf{Model}                                                                                      & \textbf{\begin{tabular}[c]{@{}c@{}}Indexing \end{tabular}} & \textbf{\begin{tabular}[c]{@{}c@{}}Typical works \end{tabular}} \\ \midrule
\multirow{14}{*}{\begin{tabular}[c]{@{}c@{}}Bipartite graph \\ embedding-based\\ recommendation\\ (static\\ user-item\\ interactions)\end{tabular}}  & \multirow{5}{*}{\begin{tabular}[c]{@{}c@{}}Matrix \\ factorization \\(MF)\end{tabular}} & \begin{tabular}[c]{@{}c@{}}FunkSVD and its variants\end{tabular}                             & Sec.~\ref{sec:SVD}                                                                  &\cite{koren2009matrix}, \cite{koren2009matrix}, \cite{hu2014your}
                                                                         \\ \cmidrule(l){3-5}
                                                                                             &                                                                                  & Combinations with KNN                                                                                          & Sec.~\ref{sec:SVD}                                                                  &\cite{ning2011slim}, \cite{koren2008factorization}, \cite{kabbur2013fism}                                                                         \\ \cmidrule(l){3-5}
                                                                                             &                                                                                  & \begin{tabular}[c]{@{}c@{}}Non-negative MF \end{tabular}                                 & Sec.~\ref{sec:SVD}                                                                  &\cite{lee1999learning}, \cite{zhang2006learning}, \cite{xu2012alternating}, \cite{hernando2016non}, \cite{luo2015nonnegative}                                                                         \\ \cmidrule(l){3-5}
                                                                                             &                                                                                  & \begin{tabular}[c]{@{}c@{}}Based on metric learning\end{tabular}                                        & Sec.~\ref{sec:SVD}                                                                  &\cite{hsieh2017collaborative}, \cite{zhang2018metric}, \cite{vinh2020hyperml}                                                                         \\ \cmidrule(l){3-5}
                                                                                             &                                                                                  & \begin{tabular}[c]{@{}c@{}}Other factorization frameworks\end{tabular}                         & Sec.~\ref{sec:SVD}                                                                  &\cite{lee2013local}, \cite{aharon2006k}, \cite{halko2011finding}                                                                         \\ \cmidrule(l){2-5}
                                                                                             & \multirow{2}{*}{\begin{tabular}[c]{@{}c@{}}Bayesian \\ analysis\end{tabular}}    & \begin{tabular}[c]{@{}c@{}}Automatic hyper-parameter \\adjustment\end{tabular}                              & Sec.~\ref{Bayesian_Sec}                                                                 &\cite{mnih2007probabilistic}, \cite{salakhutdinov2008bayesian}, \cite{peng2016n}, \cite{acharya2015nonparametric}, \cite{gopalan2014content}, \cite{gouvert2020ordinal}                                                                          \\ \cmidrule(l){3-5}
                                                                                             &                                                                                  & \begin{tabular}[c]{@{}c@{}}Pair-wise ranking\end{tabular}                                           & Sec.~\ref{Bayesian_Sec}                                                                 &\cite{rendle2012bpr}                                                                         \\ \cmidrule(l){2-5}
                                                                                             & \multirow{3}{*}{\begin{tabular}[c]{@{}c@{}}Deep\\ learning\end{tabular}}         & For learning embeddings                                                                                             & Sec.~\ref{DL-Bi}                                                                  &\cite{covington2016deep}, \cite{yi2019deep}, \cite{kim2016convolutional}, \cite{cheng2016wide}                                                                         \\ \cmidrule(l){3-5}
                                                                                             &                                                                                  & For proximity measurement                                                                                             & Sec.~\ref{DL-Bi}                                                                   &\cite{he2017neural}, \cite{xue2017deep}                                                                         \\ \cmidrule(l){3-5}
                                                                                             &                                                                                  & Based on causal learning                                                                                        & Sec.~\ref{DL-Bi}                                                                   &\cite{salakhutdinov2007restricted}, \cite{rendle2010factorization}                                                                         \\ \cmidrule(l){2-5}
                                                                                             & \multirow{4}{*}{Others}                                                          & \begin{tabular}[c]{@{}c@{}}Not missing at random \\ assumption\end{tabular}                       & Sec.~\ref{Other_Models}                                                                  &\cite{devooght2015dynamic}, \cite{liang2016modeling}                                                                         \\ \cmidrule(l){3-5}
                                                                                             &                                                                                  & \begin{tabular}[c]{@{}c@{}} Positive-unlabeled problem\end{tabular}                            & Sec.~\ref{Other_Models}                                                                 &\cite{saito2020unbiased}, \cite{he2016fast}                                                                         \\ \cmidrule(l){3-5}
                                                                                             &                                                                                  & \begin{tabular}[c]{@{}c@{}}Transfer learning\end{tabular}                                      & Sec.~\ref{Other_Models}                                                                  &\cite{man2017cross}, \cite{li2020ddtcdr}                                                                         \\ \cmidrule(l){3-5}
                                                                                             &                                                                                  & Fast online learning                                                                                          & Sec.~\ref{Other_Models}                                                                  &\cite{he2016fast}, \cite{steck2020admm}, \cite{boyd2011distributed}, \cite{yuan2020generalized}                                                                         \\ \midrule
\multirow{3}{*}{\begin{tabular}[c]{@{}c@{}}Bipartite graph \\ embedding-based\\ recommendation\\(temporal\\ user-item\\ interactions)\end{tabular}} & \begin{tabular}[c]{@{}c@{}}Matrix\\ factorization\end{tabular}                   & \begin{tabular}[c]{@{}c@{}}User's long-term preferences\\ (online learning)\end{tabular}      & Sec.~\ref{MF-based_online_learning}                                                                  &\cite{ling2012online}, \cite{koren2009collaborative}, \cite{vidmer2016essential}, \cite{yong2015distance}                                                                         \\ \cmidrule(l){2-5}
                                                                                             & \multirow{2}{*}{\begin{tabular}[c]{@{}c@{}}Markov\\ processes\end{tabular}}      & \begin{tabular}[c]{@{}c@{}}User's short-term preferences\\ (sequential learning)\end{tabular} & Sec.~\ref{Markov_processes}                                                                   &\cite{rendle2010factorizing}, \cite{zhao2017sequential}, \cite{rendle2010factorizing}, \cite{kim2007nonnegative},  \cite{wu2013personalized}, \cite{wang2015learning}                                                                         \\ \cmidrule(l){3-5}
                                                                                             &                                                                                  & \begin{tabular}[c]{@{}c@{}}Automatic hyper-parameter \\ adjustment\end{tabular}                              & Sec.~\ref{Markov_processes}                                                                  &\cite{zhang2019movie}                                                                         \\ \midrule
\multirow{7}{*}{\begin{tabular}[c]{@{}c@{}}General graph \\ embedding-based\\ recommendation\\(side \\ information)\end{tabular}}                 & \multirow{2}{*}{Translation}                                                     & Techniques                                                                                               & Sec.~\ref{Trans-section}                                                                  &\cite{bordes2013translating}, \cite{wang2014knowledge}, \cite{lin2015learning}, \cite{lin2015learning}, \cite{ji2015knowledge}, \cite{ji2016knowledge}, \cite{xiao2015transg}, \cite{lin2019relation}                                                                         \\ \cmidrule(l){3-5}
                                                                                             &                                                                                  & Recommendation models                                                                                      & Sec.~\ref{RSwithSideInfo}                                                                 &\cite{he2017translation}, \cite{li2019translation}                                                                         \\ \cmidrule(l){2-5}
                                                                                             & \multirow{2}{*}{Meta path}                                                       & Techniques                                                                                               & Sec.~\ref{Meta-path-section}                                                                 &\begin{tabular}[c]{@{}c@{}}\cite{perozzi2014deepwalk}, \cite{tang2015line}, \cite{tang2015pte}, \cite{grover2016node2vec}, \cite{dong2017metapath2vec}, \cite{qiu2018network}, \cite{gui2016large}, \cite{shang2016meta}\\ \cite{chang2015heterogeneous}, \cite{fu2017hin2vec}, \cite{gao2018bine}\end{tabular}                                                                          \\ \cmidrule(l){3-5}
                                                                                             &                                                                                  & Recommendation models                                                                                      & Sec.~\ref{RSwithSideInfo}                                                                  &\cite{yu2014personalized}, \cite{zhao2017meta}, \cite{shi2018heterogeneous}, \cite{wang2019explainable}, \cite{barkan2016item2vec}                                                                         \\ \cmidrule(l){2-5}
                                                                                             & \multirow{3}{*}{\begin{tabular}[c]{@{}c@{}}Deep\\ learning\end{tabular}}         & \begin{tabular}[c]{@{}c@{}}Techniques of AutoEncoder\end{tabular}                               & Sec.~\ref{DL-KG}                                                                  &\cite{rumelhart1985learning}, \cite{vincent2010stacked}, \cite{masci2011stacked}, \cite{wang2014generalized}, \cite{rezende2014stochastic}, \cite{wang2016structural}                                                                         \\ \cmidrule(l){3-5}
                                                                                             &                                                                                  & \begin{tabular}[c]{@{}c@{}}Techniques of attention  \\ mechanism\end{tabular}                       & Sec.~\ref{DL-KG}                                                                   &\cite{cho2015describing}, \cite{bahdanau2014neural}, \cite{kim2017structured}, \cite{vaswani2017attention}, \cite{velivckovic2017graph}                                                                         \\ \cmidrule(l){3-5}
                                                                                             &                                                                                  & Recommendation models                                                                                      & Sec.~\ref{RSwithSideInfo}                                                                 &\cite{sedhain2015autorec}, \cite{wu2016collaborative}, \cite{liang2018variational}, \cite{li2017collaborative}, \cite{xiao2017attentional}, \cite{wang2018billion}, \cite{wang2018shine}                                                                         \\ \midrule
\multirow{6}{*}{\begin{tabular}[c]{@{}c@{}}Knowledge graph \\ embedding-based\\ recommendation\\(knowledge)\end{tabular}}                                                                   & \multirow{3}{*}{\begin{tabular}[c]{@{}c@{}}Large-scale\\ learning\end{tabular}}  & Techniques                                                                                               & Sec.~\ref{LEMGE}                                                                  &\cite{joachims1999transductive}, \cite{belkin2005manifold}, \cite{hamilton2017inductive},\cite{qiu2019netsmf}, \cite{qiu2018network}, \cite{cheng2015efficient}, \cite{bahdanau2014neural}, \cite{liu2017analogical}, \cite{nickel2016holographic}                                                                         \\ \cmidrule(l){3-5}
                                                                                             &                                                                                  & \begin{tabular}[c]{@{}c@{}}Recommendation models \end{tabular}                         & Sec.~\ref{KGERS}                                                                  &\cite{anelli2019make}, \cite{zhang2016collaborative}, \cite{wang2019knowledge}, \cite{wang2019kgat}, \cite{ma2021knowledge}, \cite{togashi2020alleviating}, \cite{togashi2020alleviating}, \cite{wang2020reinforced},\cite{chen2020jointly}                                                                         \\ \cmidrule(l){3-5}
                                                                                             &                                                                                  & \begin{tabular}[c]{@{}c@{}}Recommendation models\\ (meta path-based)\end{tabular}                         & Sec.~\ref{KGERS}                                                                   &\cite{xian2019reinforcement}, \cite{wang2019explainable}, \cite{xian2019reinforcement}, \cite{zhu2020knowledge}, \cite{chen2021temporal}                                                                         \\ \cmidrule(l){2-5}
                                                                                             & Multi-viewed graphs                                                               & Techniques                                                                                               & Sec.~\ref{LEMGE}                                                                  &\cite{chaudhuri2009multi}, \cite{kumar2011co}, \cite{xia2010multiview}, \cite{shi2018mvn2vec}, \cite{zhang2018scalable}, \cite{qu2017attention}                                                                         \\ \cmidrule(l){2-5}
                                                                                             & Multi-layered graphs                                                              & Techniques                                                                                               & Sec.~\ref{LEMGE}                                                                  &\cite{perozzi2017don}, \cite{ribeiro2017struc2vec}, \cite{xu2017embedding}, \cite{chang2015heterogeneous}, \cite{li2018multi}, \cite{liu2017principled}, \cite{cen2019representation}, \cite{ni2018co}                                                                         \\ \cmidrule(l){2-5}
                                                                                             & Evolution                                                                        & Techniques                                                                                               & Sec.~\ref{LEMGE}                                                                  &\cite{li2014deep}, \cite{lu2019temporal}, \cite{zuo2018embedding}, \cite{kazemi2020representation}                                                                         \\ \bottomrule
\end{tabular}
\end{adjustbox}
\end{table}

\end{appendices}

\end{document}